\documentclass[conference]{IEEEtran}
\usepackage{blindtext, graphicx}
\usepackage{cite}
\usepackage{array}
\usepackage{multirow}
\usepackage{amssymb,mathtools,bm}
\usepackage{color}
\usepackage{algorithm}
\usepackage[algo2e]{algorithm2e}
\usepackage{subfigure}
\usepackage{amsmath}
\usepackage{lipsum, cuted}
\usepackage{breqn}
\usepackage[table,xcdraw]{xcolor}
\usepackage{multirow}
\usepackage[table,xcdraw]{xcolor}
\usepackage{rotating}
\usepackage{lscape}
\usepackage{bbding}
\usepackage{url}
\usepackage[margin=0.7in]{geometry}
\usepackage[T1]{fontenc}

\usepackage{amssymb}
\usepackage{pifont}
\newcommand{\cmark}{\ding{51}}%
\newcommand{\xmark}{\ding{55}}%

\newcolumntype{P}[1]{>{\centering\arraybackslash}p{#1}}
\def\BibTeX{{\rm B\kern-.05em{\sc i\kern-.025em b}\kern-.08em
    T\kern-.1667em\lower.7ex\hbox{E}\kern-.125emX}}
\begin{document}

\title{From Simulators to Digital Twins for Enabling Emerging Cellular Networks: A Tutorial and Survey}
\author{\IEEEauthorblockN{Marvin Manalastas\IEEEauthorrefmark{1},
Muhammad Umar Bin Farooq\IEEEauthorrefmark{1},
Syed Muhammad Asad Zaidi\IEEEauthorrefmark{1}, 
\\Haneya Naeem Qureshi\IEEEauthorrefmark{1}, Yusuf Sambo\IEEEauthorrefmark{2}, and Ali Imran\IEEEauthorrefmark{1}\IEEEauthorrefmark{2}}          
\IEEEauthorblockA{\IEEEauthorrefmark{1}AI4Networks Research Center, School of Electrical and Computer Engineering, University of Oklahoma, USA}
\IEEEauthorblockA{\IEEEauthorrefmark{2}James Watt School of Engineering, University of Glasgow, G12 8QQ Glasgow, UK\\
Email: \{marvin, umar.farooq, asad, haneya, ali.imran\}@ou.edu, \{ali.imran, yusuf.sambo\} @glasgow.ac.uk}}

\maketitle

\begin{abstract}

Simulators are indispensable parts of the research and development necessary to advance countless industries, including cellular networks. With simulators, the evaluation, analysis, testing, and experimentation of novel designs and algorithms can be executed in a more cost-effective and convenient manner without the risk of real network service disruption. Additionally, recent trends indicate that the advancement of these Digital System Models (DSM), such as system-level simulators, will hold a pivotal role in advancing cellular networks by facilitating the development of digital twins. Given this growing significance, in this survey and tutorial paper, we present an extensive review of the currently available DSMs for 5G and beyond (5G\&B) networks. Specifically, we begin with a tutorial on the fundamental concepts of 5G\&B network simulations, followed by an identification of the essential design requirements needed to model the key features of these networks. We also devised a taxonomy of different types of 5G\&B network simulators. In contrast to existing simulator surveys, which mostly leverage traditional metrics applicable to legacy networks, we devise and use 5G-specific evaluation metrics that capture three key facets of a network simulator, namely realism, completeness, and computational efficiency. We evaluate each simulator according to the devised metrics to generate an applicability matrix that maps different 5G\&B simulators vis-à-vis the different research themes they can potentially enable. We also present the current challenges in developing 5G\&B simulators while laying out several potential solutions to address the issues. Finally, we discuss the future challenges related to simulator design provisions that will arise with the emergence of 6G networks.

\end{abstract}


\IEEEpeerreviewmaketitle

\section{Introduction}
\label{sec:intro}

\begin{figure}[]
\centerline{\fbox{\includegraphics[width=0.46\textwidth,height=2.3in]{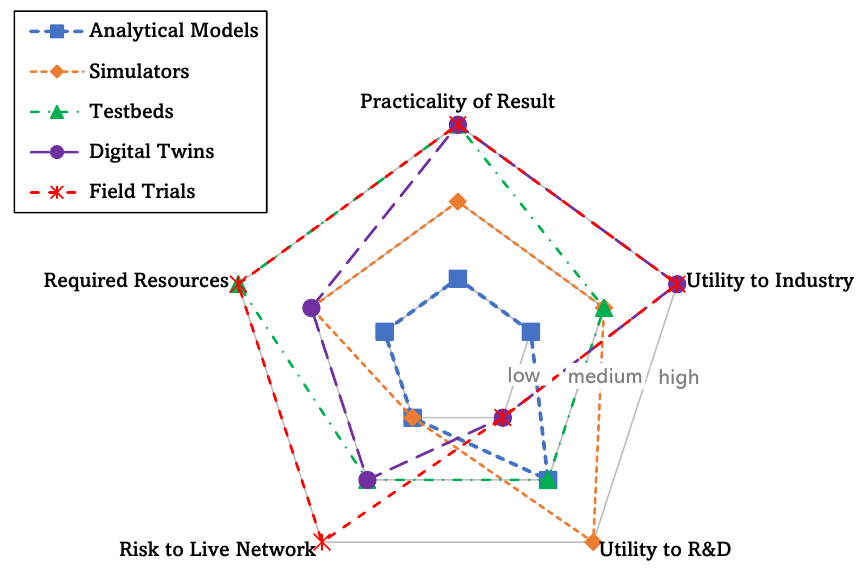}}}
\caption{Spider web diagram of different cellular network evaluation tools using a penta-prong metrics including practicality of the results, utility to industry, utility to academic R\&D, risk to the live network, and required resources.}
\label{fig:risk}
\vspace{-0.1in}
\end{figure}

Computer-aided numerical simulations, or simulators, are used as a first-tier assessment tool to evaluate the diverse features of cellular networks \cite{9084113}. For instance, cellular network operators traditionally rely on cellular network simulators such as Atoll \cite{mathworks}, Planet \cite{planet}, and Asset \cite{asset} to assist in the design, planning, and optimization stages of network rollout. Academic researchers, on the other hand, use simulators such as MATLAB \cite{5Gmatlab}, Vienna \cite{mehlfuhrer2011vienna}, and ns-3 \cite{10.5555/2151054.2151129} to design, analyze, and test new protocols, architectures, and features. Compared to alternative strategies such as analytical models, testbeds, and field trials, simulators are more practical when considering factors such as risk to real networks, benefits to industry, utilities to research and development, and resources needed to develop and perform, as summarized in Fig. \ref{fig:risk}. Compared to analytical modeling, simulators have the ability to generate results even for non-tractable mathematical problems. Although testbeds and field trials can provide a more realistic and practical assessment of wireless networks, simulators pose minimal risk to live cellular networks and require fewer resources to implement and develop. As cellular network technology evolves to enable an expanding number of emerging use cases and as the demand for mobile data grows at an unprecedented magnitude, simulators have become increasingly important for both industry and academia. For instance, simulators are being considered as a promising solution to address the data scarcity challenge in the cellular network domain \cite{sparsity}. However, amidst the increasing complexity and stringent requirements of future cellular networks, simulators must continue to evolve to remain relevant.

\begin{figure*}[t]
	\centering
	\fbox{\includegraphics[width=0.75\textwidth, height=6.5cm]{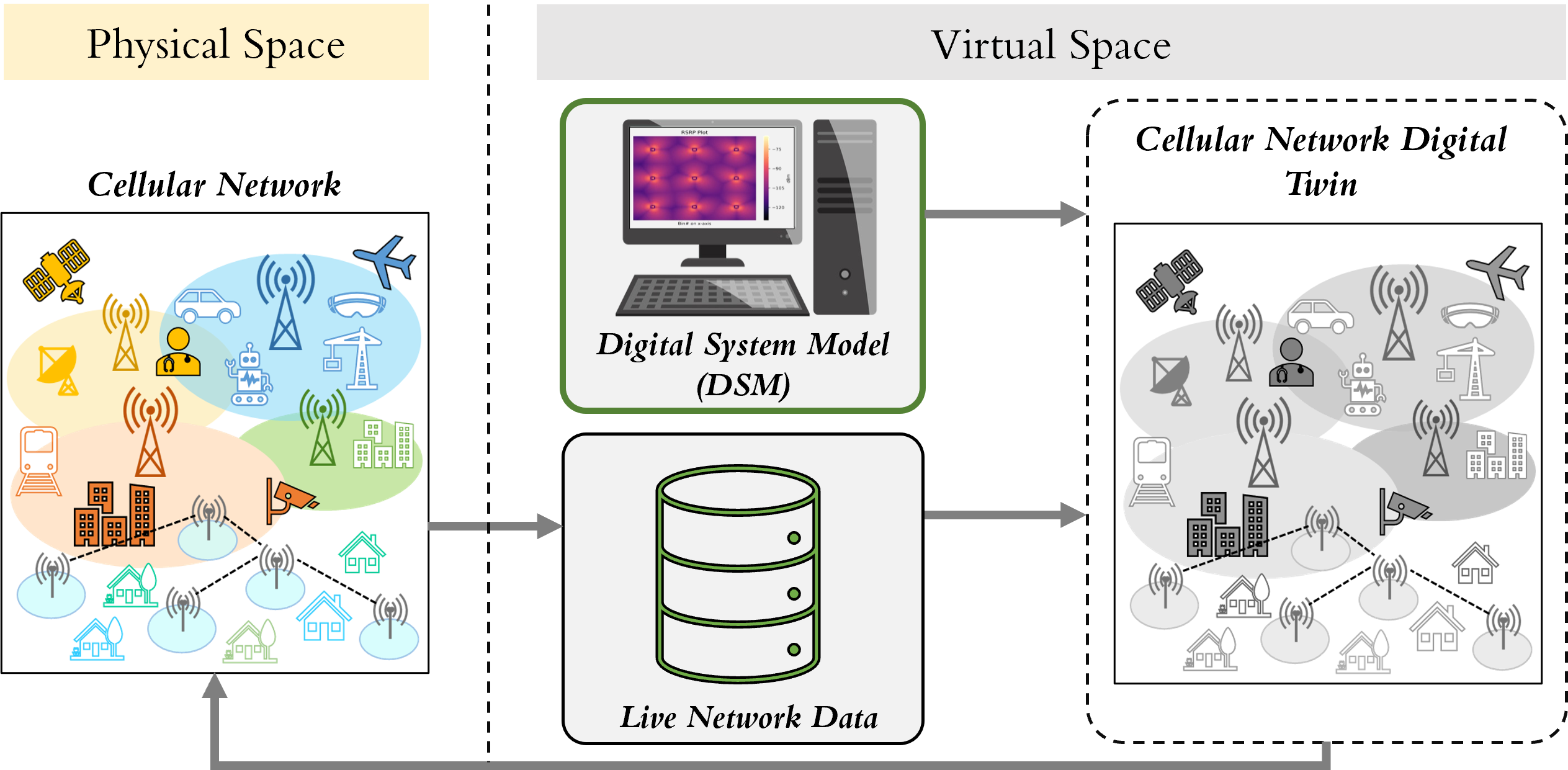}}
	\caption[]{Digital system model utilized as digital twin kernel.}
	\label{fig:simulator_dt}
\end{figure*}

Recent trends suggest that to provide a more advanced method of network modeling, the cellular network industry is moving towards digital twins (DTs), which are virtual and up-to-date representations of a physical system (i.e., a cellular network) \cite{9711524,9374645}. This emphasis on DTs has gained traction within the cellular network community, particularly in shaping the foundation of 6G networks, as highlighted by several studies \cite{9711524,9374645,kuruvatti2022empowering,khan2022digital,masaracchia2022digital,alkhateeb2023real,tang2022survey,ahmadi2021networked,guo2023five,lin20236g,masaracchia2023digital,wang20236g,wu2021digital,bariah2023digital}. These collective endeavors signify a substantial shift towards leveraging DTs as a fundamental enabler in the evolution towards 6G networks. DTs are increasingly recognized to be valuable assets in addressing diverse challenges encountered within cellular networks, including resource allocation~\cite{he2022resource}, energy efficiency~\cite{dong2019deep}, cellular edge networks~\cite{sun2020reducing,fan2021digital,tang2022survey}, optimizing communication and computation costs associated with DTs~\cite{yang2023joint} and the deployment of Open Radio Access Network (O-RAN)~\cite{masaracchia2023digital}. As DT closely mirrors a real cellular network, it serves as a reliable platform for developing and benchmarking new research directions accurately. Furthermore, DTs can serve as platforms for network management and optimization, mitigating the risk of service degradation. Leveraging DTs as the testing ground for modifications, such as parameter adjustments, provides a safeguard against potential disruptions, as changes are only implemented in the real network after successful validation within the digital twin.

DTs are composed of three unique components: the physical product, the digital or virtual model of the product, and the interconnections represented by the data that reflects the current state of a live system, as shown in Fig. \ref{fig:simulator_dt}. While creating a digital copy of a real system is relatively straightforward for simple systems with low dynamicity, such as static machinery, the process becomes exceptionally challenging when dealing with complex and highly dynamic systems like cellular networks. In such cases, cellular network digital system models (DSMs) play a pivotal role. These DSMs take on a critical function in modeling the intricacies of cellular networks, including different network components (i.e., base station and antenna), radio propagation, protocols, user mobility, traffic patterns, and network performance, to name a few. The concept of evolving DSMs into comprehensive digital twins is exemplified in \cite{villa2024colosseum}, wherein the authors showcase the potential of a state-of-the-art emulator to transition into a fully realized digital twin. A DSM is a versatile term that can encompass various simulation types, such as system-level or link-level simulators, as well as digital twins, depending on the specific features. For instance, DSM with high fidelity can become a kernel in developing live DT models of cellular networks \cite{9374645,9842839}. The distinction lies in the set of attributes that characterize a DSM as either a simulator or a digital twin.

While there are overlaps between DSMs and DTs in terms of their functionalities, several differences exist between the two. These distinctions are summarized in Table \ref{tab:dsmvsdt}. DSMs, although functional independently, cannot be classified as DTs as they lack connectivity with the real network. The presence of a feedback loop between the digital representation and the real network is one of the major disparities between the two. This connection is pivotal for facilitating seamless data exchange between the DT and the real network. This link is imperative as DTs heavily rely on data-driven models to function \cite{guo2023five} unlike simulators which rely more on pre-defined rules and deterministic models. The reliance of DTs on data-driven modeling is crucial for mitigating the complexity of modeling numerous real-time network functionalities.  \cite{kuruvatti2022empowering}. Consequently, this makes the requirements for efficient predictive modeling capabilities, more stringent in DTs than in simulators. The DTs of 6G wireless systems must rely on efficient AI schemes tailored for handling extensive datasets \cite{9711524}. This underscores AI's pivotal role as an enabler of DTs and is expected to be natively integrated into their design, unlike in simulators where its implementation remains optional \cite{lin20236g}. Finally, in terms of visualization, DTs demand more advanced capabilities to provide richer representations of physical systems, such as 3D maps, buildings, vegetation, etc.

\begin{table}[]
\centering\caption{Distinguishing factors between DSM and DT}
\label{tab:dsmvsdt}
\begin{tabular}{l|l|l|}
\cline{2-3}
 & Digital Twins & \multicolumn{1}{c|}{\begin{tabular}[c]{@{}c@{}}DSMs \\ (i.e., simulators)\end{tabular}} \\ \hline
\multicolumn{1}{|l|}{\begin{tabular}[c]{@{}l@{}}Real-time interaction \\ with real network\end{tabular}} & Required & Nonexistent \\ \hline
\multicolumn{1}{|l|}{\begin{tabular}[c]{@{}l@{}}Reliance on data-driven \\ models\end{tabular}} & High & Low \\ \hline
\multicolumn{1}{|l|}{\begin{tabular}[c]{@{}l@{}}Predictive modeling \\ capabilities\end{tabular}} & High & Low \\ \hline
\multicolumn{1}{|l|}{AI implementation} & Native & Optional \\ \hline
\multicolumn{1}{|l|}{Visualization} & Advanced & Less advanced \\ \hline
\end{tabular}
\end{table}

\begin{figure*}[h]
	\centering
	\fbox{\includegraphics[width=0.98\textwidth, height=8.6cm]{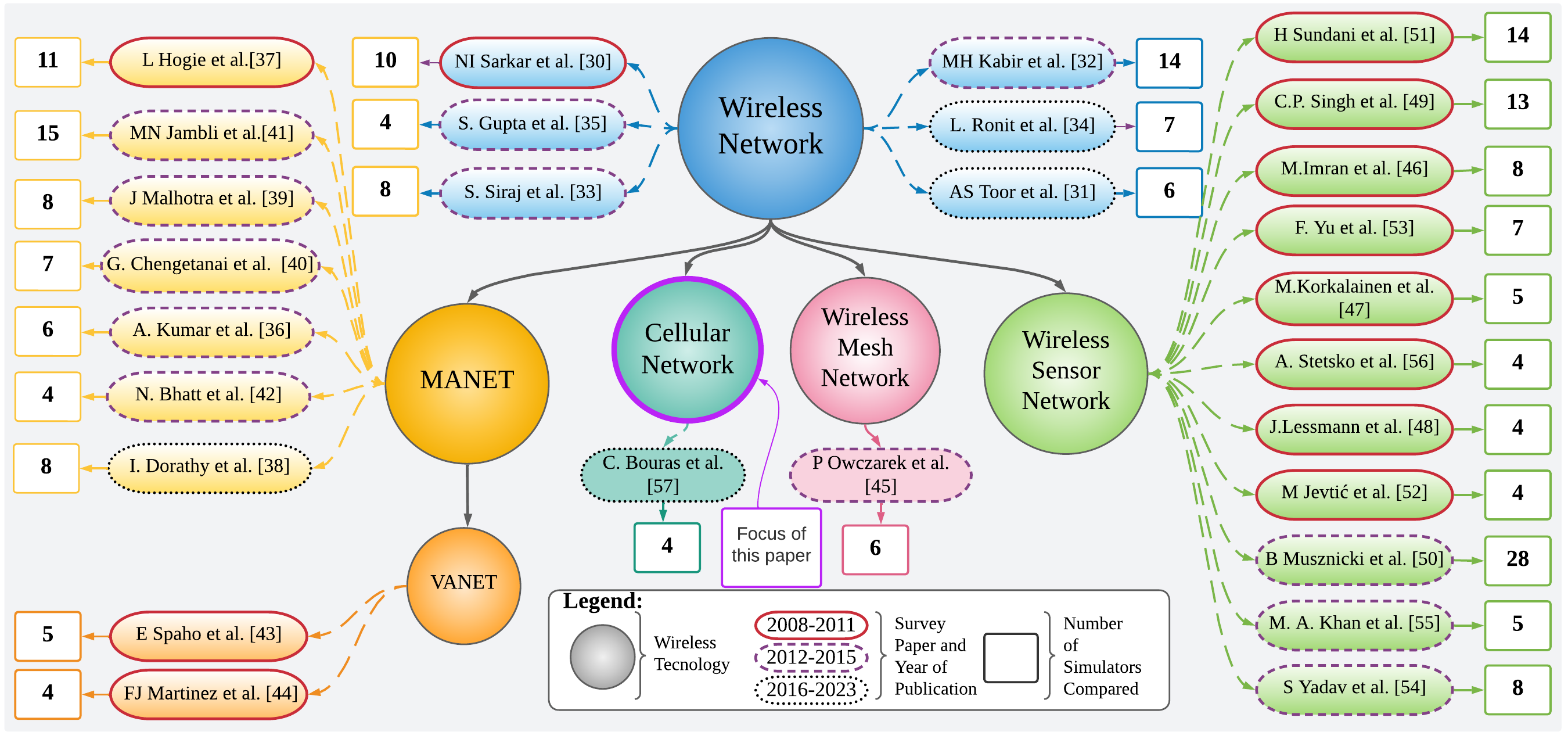}}
	\caption[]{Summary of available literature on simulator comparison in wireless networks. Inside the circles are the different domains of wireless networks. Inside the connected ellipses are the survey papers available in the literature. The outline of these ellipses represents the year of publication. The figures inside the square represent the number of simulators included in each survey paper.}
	\label{fig:lit}
\end{figure*}

In this survey paper, these attributes are presented using a set of criteria that we call the "iron triangle" of cellular networks. The iron triangle encompasses three crucial factors: realism, completeness, and computational efficiency, which together play a critical role in determining whether DSMs can function as simulators or serve as digital twin kernels. By leveraging these criteria, our paper offers valuable insights into the current state of various DSMs and their progress towards becoming digital twin-ready by tackling the challenges posed by the iron triangle. Furthermore, our study aims to raise awareness about the challenges involved in developing these DSMs and explore potential solutions. We also conduct an analysis of future requirements for DSMs to meet the evolving demands of next-generation cellular networks. However, it is essential to clarify that our work does not cover the two-way feedback between the digital model and digital twin, as that aspect falls beyond the scope of this article. Instead, we focus on providing a comprehensive understanding of the key attributes and challenges associated with DSMs, contributing to the broader discourse on digital twin development in the cellular network domain.

\subsection{Related Work}

Wireless networks (WN) have had a colossal impact on various sectors of human society, including communication, education, defense, security, healthcare, agriculture, and manufacturing, among others. It is a broad area of research and can be broken down into several sub-topics, as shown in Fig. \ref{fig:lit}. These specific domains include Wireless Sensor Network (WSN), Wireless Mesh Network (WMN), Cellular Network, Mobile Ad-hoc Network (MANET), and Vehicular Ad-hoc Network (VANET), which is a sub-category of MANET. The overwhelming demand to advance wireless communication has led to the development of several simulators that perform a broad array of functionalities.

There are several comparative studies on simulators for WNs in the literature. Fig. \ref{fig:lit} shows a summary of the related literature we assimilate in this survey paper. Currently, several surveys and comparative studies are focused on WN in general\cite{sarkar2011review, EEI568, kabir2014detail, Siraj2012NetworkST, Ronit2018SurveyON, gupta2013open}. Meanwhile, other works also exist highlighting the simulators for specific sub-categories of WN. For instance, \cite{6391700,hogie2006overview,DORATHY2018,7019120,7226167,jambli2012simulation,Bhatt2013ComarisonAA} are surveys on simulators used for MANETs,\cite{spaho2011vanet,martinez2011survey} are works dedicated to VANET simulators, while \cite{owczarek2014review} confined the simulator discussions for WMN. Moreover, the bulk of the papers on simulator surveys and comparative studies are concentrated on WSN, with 11 papers as of this writing \cite{imran2010survey,korkalainen2009survey,lessmann2008comparative,singh2008survey,musznicki2012survey,sundani2011wireless,jevtic2009evaluation,yu2011survey,yadav2012wireless,6914198,6076678}. While some of these survey papers are comprehensive, including \cite{musznicki2012survey,sundani2011wireless,singh2008survey, kabir2014detail, sarkar2011review, hogie2006overview, jambli2012simulation} wherein the authors evaluated 10 or more simulators, they can be deemed outdated with regards to emerging cellular networks as most were published between 2008 and 2015.

Unlike other branches of WN, such as WSN, WMN, and MANETS, review papers dedicated to simulators pertinent to emerging cellular networks are scarce, as shown in Fig. \ref{fig:lit}. A review paper that is most related to the scope of our paper is \cite{4467}, which offers a comparative study on simulators specific to 4G and 5G cellular networks. However, this review paper has a very limited scope as it considers only four simulators, namely, ns-3, OMNeT++, Riverbed Modeler, and NetSim.

While there are other review papers available on cellular network simulators, such as those in \cite{Gkonis_2020, 7879128, 6994948, 7993797}, they primarily concentrate on the specifications of 5G simulators rather than their evaluation. These papers discuss various requirements for simulators, such as reusability, scalability, flexibility, multiple levels of abstraction, parallel processing, and the integration of link-, system-, an d network-level simulators. However, it is worth noting that most of these requirements are not exclusive to 5G networks, and they provide only a cursory assessment of the challenges in developing 5G simulators. While \cite{7993797} discusses the challenges associated with developing 5G simulators, the study is not comprehensive and does not offer insights into overcoming the challenges inherent in developing 5G\&B simulators.

Along with the aforementioned articles on cellular network simulators, there exist publications that delve into simulator development \cite{9084113, MARTIRADONNA2020107314, TININI2020102030}. In such works, authors not only expound on the features and capabilities of their respective simulators but also perform comparative analyses with other 5G simulators, highlighting the strengths and advantages of their own tools. For instance, the authors in \cite{MARTIRADONNA2020107314} compared their simulator with ten others with respect to supported features such as massive multi-input multi-output (MIMO), flexible numerology, random access procedure, millimeter wave (mmWave) propagation, and general information such as programming language and license type. Similarly, authors in \cite{9084113}  developed a simulator and compared it to existing simulators using almost similar metrics as \cite{MARTIRADONNA2020107314} with the inclusion of cloud computing capabilities. By discussing its primary features, the authors in \cite{TININI2020102030} juxtaposed the simulator they developed with six other simulators.

Table II provides a comparative summary of the existing literature in tabular form, highlighting the distinctions between our work and other studies in this domain. Notably, our study stands out for its comprehensive analysis, which encompasses the broadest range of simulators, including industry-grade tools. Furthermore, our work is distinguished by the diversity of analyses performed, particularly our unique focus on how simulators can facilitate digital twin generation. Unlike previous studies that relied solely on traditional evaluation metrics, our research introduces new metrics, further setting it apart from existing literature.

\subsection{Contributions}

\begin{figure}[]
\centerline{\fbox{\includegraphics[width=0.45\textwidth,height=3.2in]{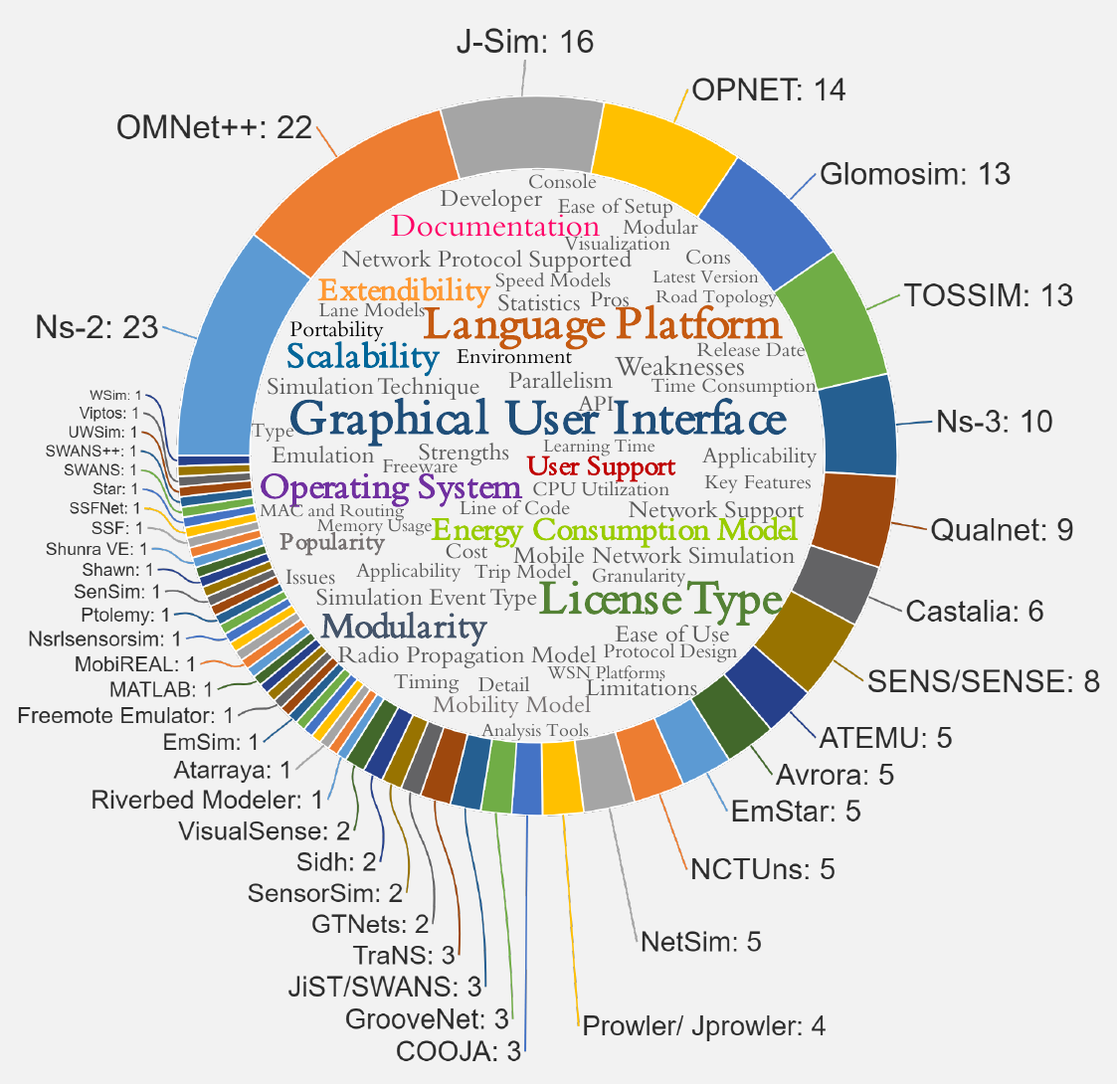}}}
\caption{List of the simulators evaluated and comparison metrics used in relevant literature.}
\label{fig:metrics}
\vspace{-0.1in}
\end{figure}

The examination of the available literature on WN simulators reveals that, in comparison to other domains, cellular network simulators lag in the availability of review papers. Meanwhile, those that are now available are either insufficiently comprehensive, have limited discussion of simulator development challenges and potential solutions, or do not use 5G-specific evaluation measures. Given the shortcomings of the current relevant work on cellular network simulators, there is a dire need for a comprehensive survey that focuses on 5G\&B network simulators. In summary, the main contributions of this work are as follows:

\begin{enumerate}
    
    \item This paper provides a comprehensive survey of the current simulators for 5G\&B networks. In contrast to current literature, which concentrates on general-purpose simulators (shown in Fig. \ref{fig:metrics}), this survey covers simulators targeted for specific 5G use cases. Our discussion includes over 35 link-level, system-level, and network-level simulators. Moreover, this survey is the first to discuss current state-of-the-art, industry-grade commercial simulators. The presented analysis can assist researchers in selecting the proper simulation tool for their needs.
    
    \item We provide potential strategies to reduce the complexity of simulator development by investigating simulator design requirements in light of the nuances and peculiarities of 5G\&B communication. We also present several research topics that can be investigated in greater depth after these design requirements are met.
    
    \item We present a new and insightful evaluation metric tailored specifically for 5G\&B networks, aimed at discerning the operational status, degree of implementation fidelity, and underlying assumptions inherent to each simulator. This metric serves as a decisive benchmark, enabling the categorization of these DSMs into either potential candidates for enabling DTs or retaining their status as conventional simulators. In contrast to conventional and generic assessment methods, which often involve comparisons based on criteria such as graphical user interface, language platform, licensing model, and modularity, as depicted in Figure \ref{fig:metrics}, our innovative metric focuses on a comprehensive evaluation framework composed of three pivotal criteria: realism, comprehensiveness, and computational efficiency. This refined metric is instrumental in not only gauging the capabilities of each simulator but also probing their applicability across different fields of research in 5G\&B.

    \item Simulators meeting the three mentioned criteria can be regarded as possessing high fidelity and good quality, however, they cannot be classified as DTs. Another essential aspect that must be present is the connection between the DSM and the real network. While this interconnection is beyond the scope of this paper, existing literature provides insights into what can be communicated through this link. Hence, we introduce a framework outlining the necessary steps to utilize DSM as a core component for generating DTs. This framework offers readers insights into the required information and the role of DSM in DT generation, while also emphasizing the significance of the three metrics.
    
    \item We outline the challenges that may impede the development of realistic, comprehensive, and computationally efficient DSMs for 5G\&B networks, preventing them from becoming DT kernels. This analysis is guided by insights from 5G\&B network requirements, use cases, and enablers. It is essential to recognize these challenges in order to accelerate research and development toward more accurate and reliable DSMs and transform them into operational DTs rather than just conventional simulators. Furthermore, we identify potential approaches to addressing these challenges based on insights gleaned from academic literature and industrial practice. Finally, based on an array of related articles projecting what 6G will look like, we highlight the upcoming issues of designing 6G-specific DSMs.

\end{enumerate}

\begin{landscape}
\newgeometry{left=0cm,right=0cm,top=4cm,bottom=0cm}
\begin{table*}[ht]
\caption{A Comparison of Existing Simulator Survey Papers}
\label{tab:comparison}
\centering
\begin{tabular}{|l|l|c|c|cc|c|c|ccc|c|}
\hline
\multicolumn{1}{|c|}{\multirow{2}{*}{Related Work}} & \multicolumn{1}{c|}{\multirow{2}{*}{\begin{tabular}[c]{@{}c@{}}Wireless \\ Technology\end{tabular}}} & \multirow{2}{*}{\begin{tabular}[c]{@{}c@{}}Number of \\ Simulators \\ Evaluated\end{tabular}} & \multirow{2}{*}{\begin{tabular}[c]{@{}c@{}}Year of \\ Publication\end{tabular}} & \multicolumn{2}{c|}{\begin{tabular}[c]{@{}c@{}}Comparison \\ Metrics Used\end{tabular}} & \multirow{2}{*}{\begin{tabular}[c]{@{}c@{}}Development \\ Requirements\end{tabular}} & \multirow{2}{*}{\begin{tabular}[c]{@{}c@{}}Development \\ Challenges and \\ Solutions\end{tabular}} & \multicolumn{3}{c|}{\begin{tabular}[c]{@{}c@{}}Type of Simulators \\ (Applicable only to CN \\ Simulators)\end{tabular}} & \multirow{2}{*}{\begin{tabular}[c]{@{}c@{}}Digital Twin \\ Considered\end{tabular}} \\ \cline{5-6} \cline{9-11}
\multicolumn{1}{|c|}{} & \multicolumn{1}{c|}{} &  &  & \multicolumn{1}{c|}{Traditional} & New &  &  & \multicolumn{1}{c|}{Link Level} & \multicolumn{1}{c|}{System Level} & Industry-grade &  \\ \hline
NI Sarkar et al. {\cite{sarkar2011review}} & WN & 10 & 2011 & \multicolumn{1}{c|}{\cmark} & \xmark & \cmark & \xmark & \multicolumn{1}{c|}{N/A} & \multicolumn{1}{c|}{N/A} & N/A & \xmark \\ \hline
S. Gupta et al. {\cite{gupta2013open}} & WN & 4 & 2013 & \multicolumn{1}{c|}{\cmark} & \xmark & \xmark & \xmark & \multicolumn{1}{c|}{N/A} & \multicolumn{1}{c|}{N/A} & N/A & \xmark \\ \hline
S. Siraj et al. {\cite{Siraj2012NetworkST}} & WN & 8 & 2012 & \multicolumn{1}{c|}{\cmark} & \xmark & \xmark & \xmark & \multicolumn{1}{c|}{N/A} & \multicolumn{1}{c|}{N/A} & N/A & \xmark \\ \hline
MH Kabir et al. {\cite{kabir2014detail}} & WN & 14 & 2014 & \multicolumn{1}{c|}{\cmark} & \xmark & \cmark & \xmark & \multicolumn{1}{c|}{N/A} & \multicolumn{1}{c|}{N/A} & N/A & \xmark \\ \hline
L. Ronit et al. {\cite{Ronit2018SurveyON}} & WN & 7 & 2018 & \multicolumn{1}{c|}{\cmark} & \xmark & \xmark & \xmark & \multicolumn{1}{c|}{N/A} & \multicolumn{1}{c|}{N/A} & N/A & \xmark \\ \hline
AS Toor et al. {\cite{EEI568}} & WN & 6 & 2017 & \multicolumn{1}{c|}{\cmark} & \xmark & \xmark & \xmark & \multicolumn{1}{c|}{N/A} & \multicolumn{1}{c|}{N/A} & N/A & \xmark \\ \hline
H Sundani et al. {\cite{sundani2011wireless}} & WSN & 14 & 2011 & \multicolumn{1}{c|}{\cmark} & \xmark & \xmark & \xmark & \multicolumn{1}{c|}{N/A} & \multicolumn{1}{c|}{N/A} & N/A & \xmark \\ \hline
C.P. Singh et al. {\cite{singh2008survey}} & WSN & 13 & 2008 & \multicolumn{1}{c|}{\cmark} & \xmark & \xmark & \xmark & \multicolumn{1}{c|}{N/A} & \multicolumn{1}{c|}{N/A} & N/A & \xmark \\ \hline
M.Imran et al. {\cite{imran2010survey}} & WSN & 8 & 2010 & \multicolumn{1}{c|}{\cmark} & \xmark & \cmark & \xmark & \multicolumn{1}{c|}{N/A} & \multicolumn{1}{c|}{N/A} & N/A & \xmark \\ \hline
F. Yu et al. {\cite{yu2011survey}} & WSN & 7 & 2011 & \multicolumn{1}{c|}{\cmark} & \xmark & \xmark & \xmark & \multicolumn{1}{c|}{N/A} & \multicolumn{1}{c|}{N/A} & N/A & \xmark \\ \hline
M.Korkalainen et al. {\cite{korkalainen2009survey}} & WSN & 5 & 2009 & \multicolumn{1}{c|}{\cmark} & \xmark & \cmark & \xmark & \multicolumn{1}{c|}{N/A} & \multicolumn{1}{c|}{N/A} & N/A & \xmark \\ \hline
A. Stetsko et al. {\cite{6076678}} & WSN & 4 & 2011 & \multicolumn{1}{c|}{\cmark} & \xmark & \xmark & \xmark & \multicolumn{1}{c|}{N/A} & \multicolumn{1}{c|}{N/A} & N/A & \xmark \\ \hline
J.Lessmann et al. {\cite{lessmann2008comparative}} & WSN & 4 & 2008 & \multicolumn{1}{c|}{\cmark} & \xmark & \xmark & \xmark & \multicolumn{1}{c|}{N/A} & \multicolumn{1}{c|}{N/A} & N/A & \xmark \\ \hline
M Jevtic et al. {\cite{jevtic2009evaluation}} & WSN & 4 & 2009 & \multicolumn{1}{c|}{\cmark} & \xmark & \xmark & \xmark & \multicolumn{1}{c|}{N/A} & \multicolumn{1}{c|}{N/A} & N/A & \xmark \\ \hline
B Musznicki et al. {\cite{musznicki2012survey}} & WSN & 28 & 2012 & \multicolumn{1}{c|}{\cmark} & \xmark & \cmark & \xmark & \multicolumn{1}{c|}{N/A} & \multicolumn{1}{c|}{N/A} & N/A & \xmark \\ \hline
M. A. Khan et al. {\cite{6914198}} & WSN & 5 & 2014 & \multicolumn{1}{c|}{\cmark} & \xmark & \xmark & \xmark & \multicolumn{1}{c|}{N/A} & \multicolumn{1}{c|}{N/A} & N/A & \xmark \\ \hline
S Yadav et al. {\cite{yadav2012wireless}} & WSN & 8 & 2012 & \multicolumn{1}{c|}{\cmark} & \xmark & \xmark & \xmark & \multicolumn{1}{c|}{N/A} & \multicolumn{1}{c|}{N/A} & N/A & \xmark \\ \hline
L Hogie et al.{\cite{hogie2006overview}} & MANET & 11 & 2006 & \multicolumn{1}{c|}{\cmark} & \xmark & \xmark & \xmark & \multicolumn{1}{c|}{N/A} & \multicolumn{1}{c|}{N/A} & N/A & \xmark \\ \hline
MN Jambli et al.{\cite{jambli2012simulation}} & MANET & 15 & 2012 & \multicolumn{1}{c|}{\cmark} & \xmark & \cmark & \xmark & \multicolumn{1}{c|}{N/A} & \multicolumn{1}{c|}{N/A} & N/A & \xmark \\ \hline
J Malhotra et al. {\cite{7019120}} & MANET & 8 & 2014 & \multicolumn{1}{c|}{\cmark} & \xmark & \xmark & \xmark & \multicolumn{1}{c|}{N/A} & \multicolumn{1}{c|}{N/A} & N/A & \xmark \\ \hline
G. Chengetanai et al. {\cite{7226167}} & MANET & 7 & 2015 & \multicolumn{1}{c|}{\cmark} & \xmark & \xmark & \xmark & \multicolumn{1}{c|}{N/A} & \multicolumn{1}{c|}{N/A} & N/A & \xmark \\ \hline
A. Kumar et al. {\cite{6391700}} & MANET & 6 & 2012 & \multicolumn{1}{c|}{\cmark} & \xmark & \xmark & \xmark & \multicolumn{1}{c|}{N/A} & \multicolumn{1}{c|}{N/A} & N/A & \xmark \\ \hline
N. Bhatt et al. {\cite{Bhatt2013ComarisonAA}} & MANET & 4 & 2013 & \multicolumn{1}{c|}{\cmark} & \xmark & \xmark & \xmark & \multicolumn{1}{c|}{N/A} & \multicolumn{1}{c|}{N/A} & N/A & \xmark \\ \hline
I. Dorathy et al. {\cite{DORATHY2018}} & MANET & 8 & 2018 & \multicolumn{1}{c|}{\cmark} & \xmark & \xmark & \xmark & \multicolumn{1}{c|}{N/A} & \multicolumn{1}{c|}{N/A} & N/A & \xmark \\ \hline
E Spaho et al. {\cite{spaho2011vanet}} & VANET & 5 & 2011 & \multicolumn{1}{c|}{\cmark} & \xmark & \xmark & \xmark & \multicolumn{1}{c|}{N/A} & \multicolumn{1}{c|}{N/A} & N/A & \xmark \\ \hline
FJ Martinez et al. {\cite{martinez2011survey}} & VANET & 4 & 2011 & \multicolumn{1}{c|}{\cmark} & \xmark & \xmark & \xmark & \multicolumn{1}{c|}{N/A} & \multicolumn{1}{c|}{N/A} & N/A & \xmark \\ \hline
P Owczarek et al. {\cite{owczarek2014review}} & WMN & 6 & 2014 & \multicolumn{1}{c|}{\cmark} & \xmark & \xmark & \xmark & \multicolumn{1}{c|}{N/A} & \multicolumn{1}{c|}{N/A} & N/A & \xmark \\ \hline
C. Bouras et al. {\cite{4467}} & CN & 4 & 2020 & \multicolumn{1}{c|}{\cmark} & \xmark & \xmark & \xmark & \multicolumn{1}{c|}{\cmark} & \multicolumn{1}{c|}{\xmark} & \xmark & \xmark \\ \hline
P. K. Gkonis et al. {\cite{Gkonis_2020}} & CN & 8 & 2020 & \multicolumn{1}{c|}{\xmark} & \xmark & \xmark & \cmark & \multicolumn{1}{c|}{\cmark} & \multicolumn{1}{c|}{\cmark} & \xmark & \xmark \\ \hline
Y. Wang et al. {\cite{6994948}} & CN & \xmark & 2014 & \multicolumn{1}{c|}{\xmark} & \xmark & \xmark & \cmark & \multicolumn{1}{c|}{\xmark} & \multicolumn{1}{c|}{\cmark} & \xmark & \xmark \\ \hline
S. Cho et al. {\cite{7993797}} & CN & \xmark & 2017 & \multicolumn{1}{c|}{\xmark} & \xmark & \cmark & \cmark & \multicolumn{1}{c|}{\cmark} & \multicolumn{1}{c|}{\cmark} & \xmark & \xmark \\ \hline
S. M. A. Zaidi et al. {\cite{9084113}} & CN & 14 & 2020 & \multicolumn{1}{c|}{\cmark} & \xmark & \xmark & \xmark & \multicolumn{1}{c|}{\cmark} & \multicolumn{1}{c|}{\cmark} & \xmark & \xmark \\ \hline
S. Martiradonna et. al {\cite{MARTIRADONNA2020107314}} & CN & 11 & 2020 & \multicolumn{1}{c|}{\cmark} & \xmark & \cmark & \xmark & \multicolumn{1}{c|}{\cmark} & \multicolumn{1}{c|}{\cmark} & \xmark & \xmark \\ \hline
R. I. Tinini et al. {\cite{TININI2020102030}} & CN & 7 & 2020 & \multicolumn{1}{c|}{\cmark} & \xmark & \xmark & \xmark & \multicolumn{1}{c|}{\cmark} & \multicolumn{1}{c|}{\cmark} & \xmark & \xmark \\ \hline
\textbf{This survey \& tutorial paper} & CN & 35 & TBD & \multicolumn{1}{c|}{\cmark} & \cmark & \cmark & \cmark & \multicolumn{1}{c|}{\cmark} & \multicolumn{1}{c|}{\cmark} & \cmark & \cmark \\ \hline
\end{tabular}
\vskip 2ex
\end{table*}
\restoregeometry
\end{landscape}

\subsection{Paper Organization}

\begin{figure}[]
	\centering
	\fbox{\includegraphics[width=0.45\textwidth, height=11.0cm]{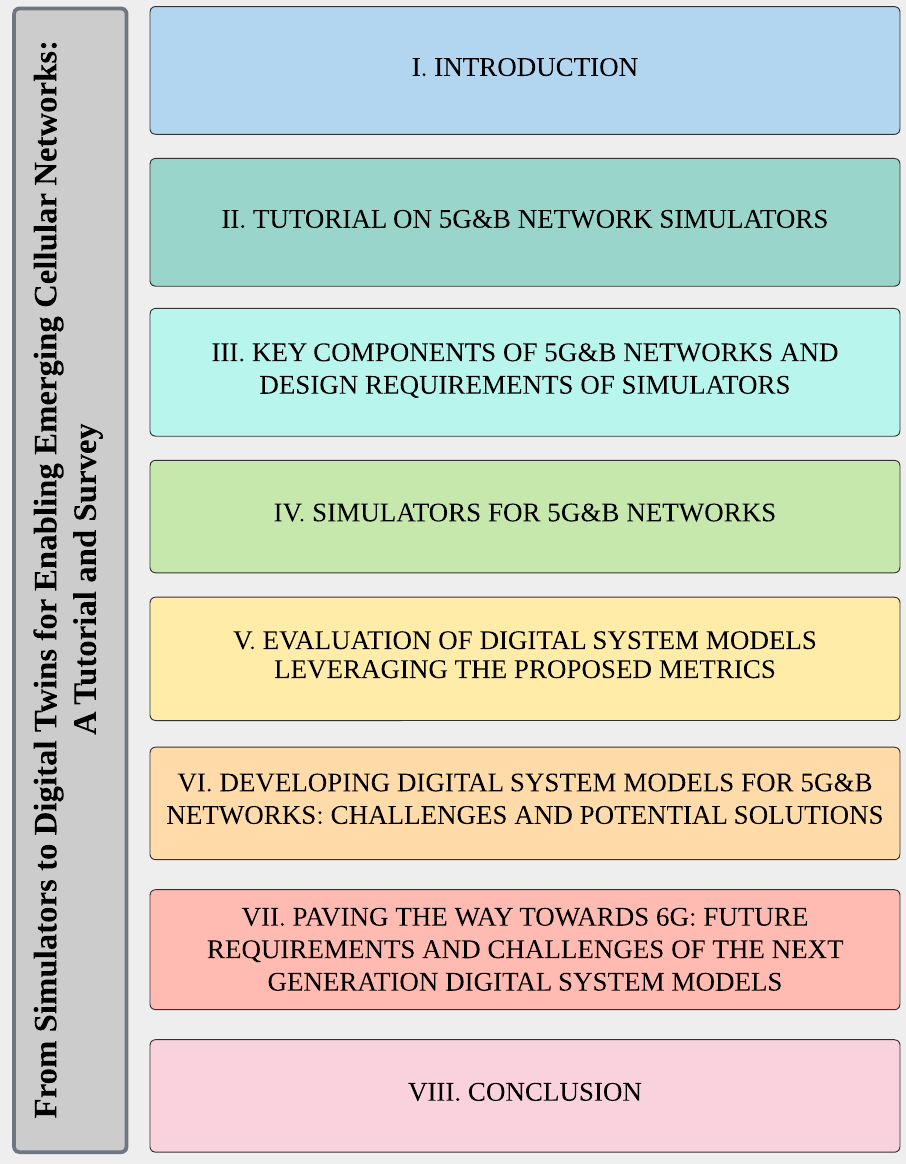}}
	\caption[]{Structure of the paper.}
	\label{fig:structure_1}
\end{figure}

The structure of the paper, visualized in Fig. \ref{fig:structure_1}, is organized as follows: Section \ref{sec:intro} provides the introduction and relevant work. Section \ref{sec:tutorial} presents a tutorial focusing on the roles, strengths, and limitations of wireless network simulators. This section also includes an overview of the different types of simulators used in 5G\&B networks, such as link-, system-, and network-level simulators, as well as the interplay between these simulators. In Section \ref{sec:req}, we discuss the key components of the 5G\&B network and the impact of these components on the design requirements of the simulators. Section \ref{sec:types} presents a taxonomy of the different types of simulators and their evaluation using traditional metrics. This section also includes a brief discussion of around 35 simulators that can be leveraged by both academia and industry for research, network planning, and optimization.

We narrow down the discussion to DSMs consisting of system-level simulators and evaluate them in Section \ref{sec:system}. In this section, we present the new insightful metrics for simulator evaluation based on three factors: realism, completeness, and computational complexity. Section \ref{sec:challenges} deals with the open challenges in the development of DSMs capable of functioning as DT kernels. Additionally, this section also presents several potential solutions to address the challenges. In Section \ref{sec:6G}, we extend the discussion of the challenges specific to 6G. Finally, Section \ref{sec:conclusion} concludes the paper.

To provide assistance to the readers, we provide a list of the acronyms used in this paper in Table \ref{table:acronyms}.

\begin{table}[htbp] 
\centering\caption{List of Acronyms}
\begin{tabular}{m{1.45cm} m{6.3cm}}\hline

\multicolumn{1}{c}{\cellcolor[HTML]{FFFFFF}} & \multicolumn{1}{c}{\cellcolor[HTML]{FFFFFF}} \\
\multicolumn{1}{c}{\multirow{-2}{*}{\cellcolor[HTML]{FFFFFF}\textbf{Acronym}}} & \multicolumn{1}{c}{\multirow{-2}{*}{\cellcolor[HTML]{FFFFFF}\textbf{Description}}} \\ \hline
3GPP & 3rd Generation Partnership   Project \\
4G & Fourth Generation \\
5G & Fifth Generation \\
5G\&B & Fifth Generation and Beyond \\
5GC & 5G Core \\
5G PPP & 5G Infrastructure Public Private Partnership \\
AI & Artificial Intelligence\\
AMF & Access and Mobility Management Function \\
BLER & Block Error Rate \\
BS & Base Station \\
C-RAN & Cloud Radio Access Network\\
CAPEX & Capital Expenditures \\
CQI & Channel Quality Indicator \\
DSM & Digital System Model \\
EPC & Evolved Packet Core \\
FBMC & Filter Bank Multicarrier \\
FSPL & Free Space Pathloss \\
GIS & Geographic Information Systems \\
GW & Gateway \\
HARQ & Hybrid Automatic Repeat Request \\
HetNet & Heterogeneous Network\\
I2I & Indoor–to-indoor \\
InH & Indoor-Hotspot \\
IoT & Internet of Things\\
KPI & Key Performance Indicator\\
LOS & Line-of-Sight \\
LTE & Long Term Evolution \\
LUT & Look-up Tables \\
MAC & Medium Access Control \\
MANET & Mobile  Ad-hoc    Network \\
MCS & Modulation and Coding Scheme \\
ML & Machine Learning \\
MME & Mobility Management Entity\\
MU-MIMO & Multi-user Multiple Input Multiple Output \\
NFV & Network Function Virtualization \\
NLOS & Non Line-of-Sight \\
NR & New Radio \\
NSA & Non-standalone\\
O2I & Outdoor–to-indoor \\
O2O & Outdoor–to-outdoor \\
OFDM & Orthogonal Frequency-division Multiplexing \\
OPEX & Operational Expenditures \\
PDCP & Packet Data Convergence Protocol \\
PHY & Physical Layer\\
RAT & Radio Access Technology\\
RAN & Radio Access Network\\
RLC & Radio Link Control \\
RMa & Rural Macro \\
RRC & Radio Resource Control \\
RW & Random Walk \\
RWP & Randon Waypoint \\
SDN & Software Defined Networking \\
SINR & Signal-to-Interference-plus-Noise Ratio \\
SLAW & Self-similar Least Action Walk \\
SMa & Suburban Macro \\
SMF & Session Management Function \\
SA & Standalone \\
SUMO & Simulation of Urban Mobility \\
TIA & Telecommunications Industry Association \\
UAV & Unmanned Aerial Vehicles \\
UFe & Urban Femto \\
UMa & Urban Macro \\
UMi & Urban Micro \\
UPF & User Plane Function \\
VANET & Vehicular Ad-hoc Network \\
VoIP & Voice Over IP \\
WMN & Wireless  Mesh    Network \\
WSN & Wireless  Sensor Network \\
ZF & Zero Forcing\\
\end{tabular}
\label{table:acronyms}
\end{table}

\section{Tutorial on 5G\&B Network Simulators}
\label{sec:tutorial}

\subsection{Roles, Strengths, and Limitations of Wireless Network Simulators}

\begin{figure*}[]
\centerline{\fbox{\includegraphics[width=.6\textwidth,height=3 in]{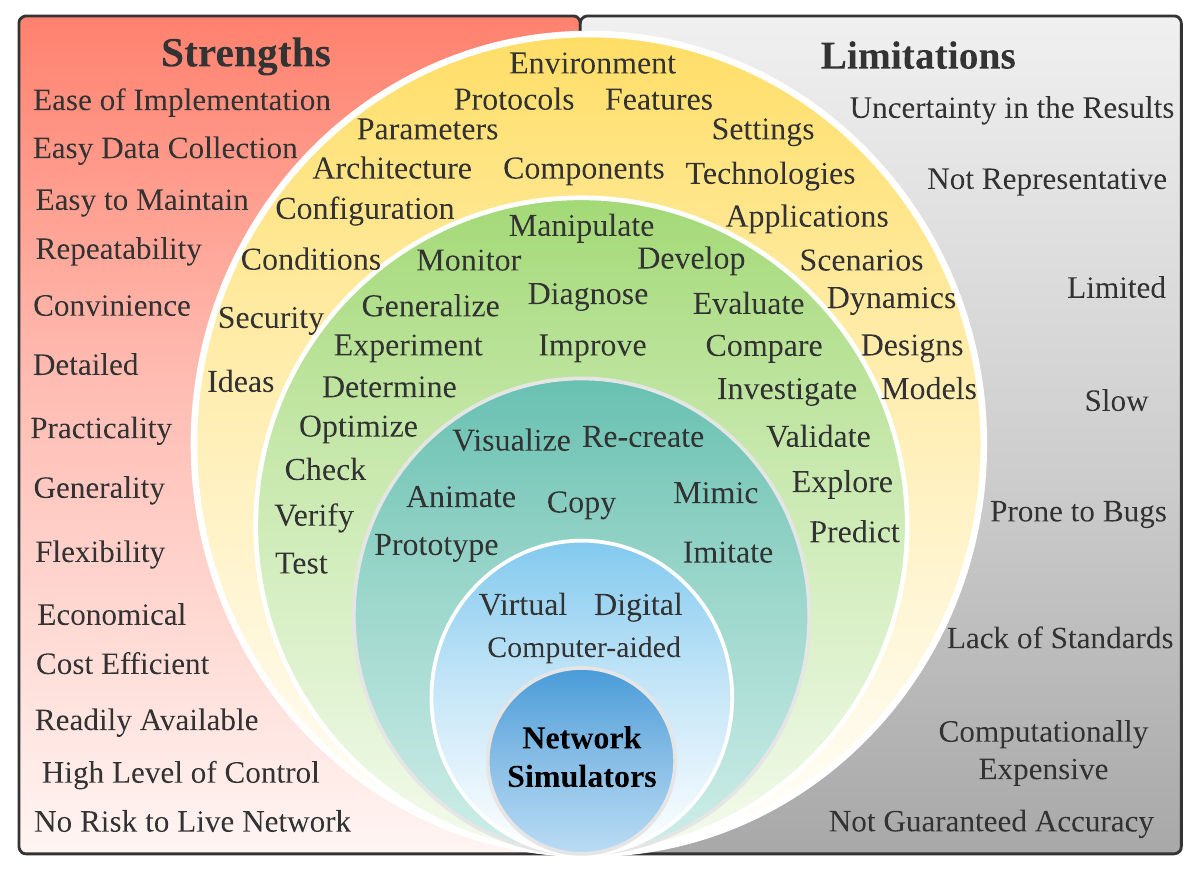}}}
\caption{Summary of network simulator roles from selected sources \cite{singh2008survey,wsn1,nizzi2019role,yu2011survey,jevtic2009evaluation,lessmann2008comparative, korkalainen2009survey,imran2010survey, owczarek2014review,jambli2012simulation,sarkar2011review,5534917}. Starting from the inner circle, this figure should be read as follows: Network simulators are [simulator descriptors (blue circle)] tools which [simulator characteristics (blue-green circle)] real networks and used to [simulator applications (green circle)] existing and new network [simulation scenarios (yellow circle)]. Outside the circles, the strengths and limitations of using a simulator are shown.}
\label{fig:benefits}
\vspace{-0.1in}
\end{figure*}

The role of simulators in wireless network systems has been extensively studied in the literature \cite{singh2008survey,wsn1,nizzi2019role,yu2011survey,jevtic2009evaluation,lessmann2008comparative, korkalainen2009survey,imran2010survey, owczarek2014review,jambli2012simulation,sarkar2011review,5534917}. Fig. \ref{fig:benefits} summarizes the comprehensive role of simulators in wireless networks, including their strengths and limitations. At their core, network simulators are tools that imitate the operation of a real network, predict its behavior, and provide a virtual atmosphere to test existing and new algorithms, architecture, protocols, parameters, and features to allow a cost-efficient and fast way to assess the viability of new solutions, parameter interactions, and network dynamics without requiring actual implementation.

In academia, simulators are commonly used to test key technologies, and to develop, and verify novel solutions using data-driven methods. For instance, the availability of simulators such as ns-2 has been critical in the research and development of 2G and 3G technologies. Meanwhile, Vienna LTE-A \cite{mehlfuhrer2011vienna}, ns-3 LTE \cite{10.5555/2151054.2151129}, and LTE-Sim \cite{5634134} provided much-needed support in experimentation and testing towards maturing the LTE technology. Meanwhile, mobile network operators rely on simulators throughout the process of cellular network deployment, from design, planning, and optimization to network operations and maintenance. Simulators, in particular, lower capital expenditures (CAPEX) by providing the appropriate number and positioning of base stations (BSs), as well as operational expenditures (OPEX) through more effective network monitoring and troubleshooting \cite{9084113}.

The importance of simulators in enabling AI-based zero-touch optimization in cellular networks has recently gained traction. As with any AI-based solution, this technique requires a massive amount of training data to be effective. However, training data for cellular networks is not as widely available as it is in other sectors where AI has had a profound impact (i.e., computer vision, text recognition, and healthcare) \cite{aztek,qureshi2023addressing}. In their study, the authors in \cite{aztek,qureshi2023addressing} brought attention to the detrimental effects of sparse training data from real networks on the efficacy of data-driven AI models for system-level optimization. Furthermore, they emphasized the simulators’ ability to generate synthetic data to supplement sparse real data from live networks, thereby increasing the performance of AI-based solutions.

Fig. \ref{fig:benefits} illustrates the definition of network simulators and provides an overview of the various advantages and disadvantages associated with their use. In the following discussions, we briefly outline these benefits and detriments.

\subsubsection{Strengths}

\begin{itemize} 
    \item \textit{Cost efficiency}: Testing new network protocols and algorithms in a simulator is more cost efficient compared to a real setup \cite{jevtic2009evaluation,singh2008survey}. Furthermore, developing a simulator is typically less expensive than deploying a full-scale testbed as it requires fewer hardware resources and less time for setup and configuration \cite{5534917,yu2011survey}.
    
    \item \textit{Risk reduction}: Simulators reduce the time required to develop new features while eliminating the risk of poor performance on a real network during trials. Moreover, simulators mitigate uncertain outcomes, which can be detrimental in a large network deployment \cite{jevtic2009evaluation}.
    
    \item  \textit{Convenience}: Users can use the simulators at any time and from any location in a time-efficient manner as results are almost immediately available compared to the real network, where observations can take a couple of days or even months in some scenarios \cite{singh2008survey}.
    
    \item \textit{High level of control}: Simulators allow users to create specific scenarios, and freely alter network parameters, which is otherwise impractical in a live network \cite{jevtic2009evaluation}.
    
    \item  \textit{Reproducibility}: A simulator can easily reproduce experimentation results under multiple situations \cite{mehlfuhrer2011vienna}.

\end{itemize}

Despite the aforementioned advantages, it should be highlighted, however, that some precautions must be taken before using any simulators, as they also have limitations. These disadvantages are as follows:

\subsubsection{Limitations}

\begin{itemize}
    \item \textit{Having a tendency to be inaccurate and prone to bugs}: Modeling the cellular network perfectly is an extremely complex task. As a result, the use of simulators can result in less accurate results when compared to real network or testbed evaluations. Additionally, simulators rely heavily on human skills during development, which may lead to defects and errors in the code, ultimately impacting simulation outcomes.

    \item \textit{Complicated structure}: Due to the complex nature of cellular networks, network simulators designed to emulate their behavior can be extremely complex. This is particularly true for simulators that include a wide range of operations and features. The architecture of these simulators can be difficult to design and develop, requiring the integration of multiple software components and algorithms. The code base for these simulators can become fairly extensive, and its maintenance can be challenging, necessitating continuous attention to guarantee that the simulator remains current with evolving network configurations and protocols.
    
    \item \textit{Computationally expensive processing}: Simulators are frequently subjected to computationally intensive operations. The complexity grows with network size, which increases the time complexity of data creation, which, on the other hand, may be readily available in a real network (e.g., data from Minimization of Drive Test (MDT) or operations support systems (OSS)).
    
    \item \textit{Lack of standards}: Currently, there is no widely accepted structure or standard for developing a simulator. As a result, simulators are incompatible with one another, making it difficult to obtain consistent findings from different simulators.
    
\end{itemize}

\subsection{Classifications of Cellular Network Simulators} 

\begin{figure*}[]
	\centering
	\fbox{\includegraphics[width=0.98\textwidth, height=7.6cm]{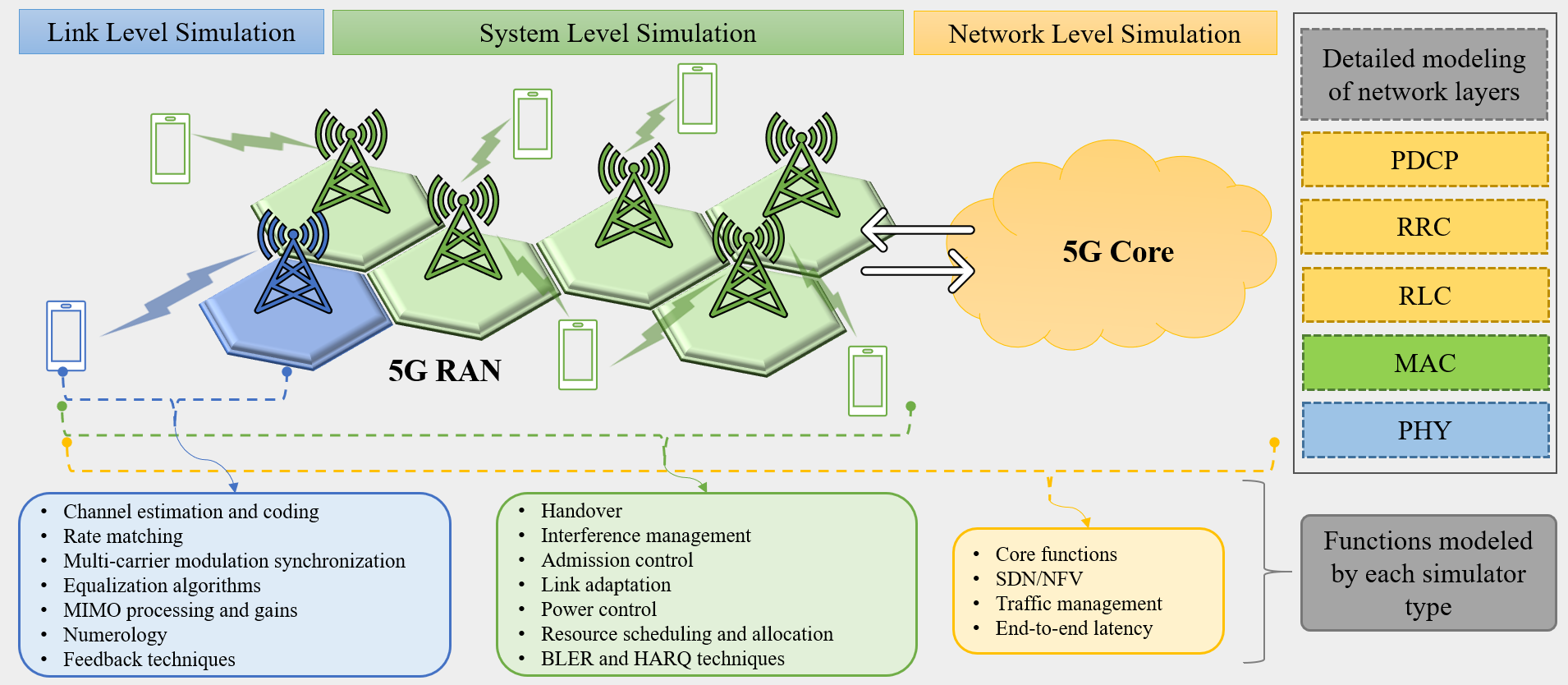}}
	\caption[]{Illustration of different types of simulators used for 5G\&B network.}
	\label{fig:simulator_type}
\end{figure*}

The utilization of link-level, system-level, and network-level simulators enables comprehensive modeling of cellular network dynamics. Fig. \ref{fig:simulator_type} illustrates different classifications of cellular network simulators alongside the layers and functionalities modeled by each simulator type. Despite their close interconnection, each type of simulator can be employed independently, depending on the simulated scenario and intended outcomes. In this subsection, we provide an overview of different types of simulators for 5G\&B networks.

\subsubsection{Link-level simulators}
A link-level simulator is a type of simulator that meticulously and computationally models the radio link between a transmitter and one or multiple receivers. Their primary focus is on modeling the physical layer (PHY) aspects of a communication system \cite{5494007}. In the context of 5G\&B, these simulators are specifically designed to facilitate the evaluation of various components, such as channel estimation, channel coding and decoding, rate matching, multi-carrier modulation, synchronization and equalization algorithms, MIMO processing and gains, numerology, and feedback techniques, among others \cite{vienall,ikuno2010system}.

\subsubsection{System-level simulators}
In cellular networks, numerous interconnected linkages exist between users and BSs. However, representing these connections in detail within simulations can significantly increase computational complexity. To address this challenge and minimize computational costs, system-level simulators employ an abstracted link-level simulation approach \cite{5494007}. While link-level simulators excel at accurately simulating the PHY of a cellular network, they are not as effective when it comes to simulating other layers, such as the Medium Access Control (MAC) layer. System-level simulators play a crucial role in overcoming this limitation by providing detailed modeling of MAC layer functions and incorporating various algorithm designs. These simulators are responsible for simulating complex processes like handover, interference management, admission control, link adaptation, power control, and resource scheduling and allocation \cite{8610404}. By incorporating these functionalities, system-level simulators offer a comprehensive and holistic view of the overall system's behavior and performance.

\subsubsection{Network-level simulators}
Network-level simulators provide a comprehensive evaluation of an entire network's performance at the packet level. These simulators go beyond the scope of link-level and system-level simulators by incorporating the modeling of higher layers in the protocol stack. For instance, network-level simulators encompass the simulation of crucial components such as the radio link control (RLC), radio resource control (RRC), and packet data convergence protocol (PDCP) \cite{8610404}. By including these layers, network-level simulations enable researchers and engineers to assess the network's behavior and performance from an end-to-end perspective. Moreover, network-level simulators also encompass core network modeling, allowing for the measurement of overall communication latency within the network. This capability is essential in evaluating the quality of service and the efficiency of various communication protocols. Similar to system-level simulators, network-level simulators utilize an abstracted PHY model \cite{8651686}.

\begin{figure*}[]
	\centering
	\fbox{\includegraphics[width=.98\textwidth]{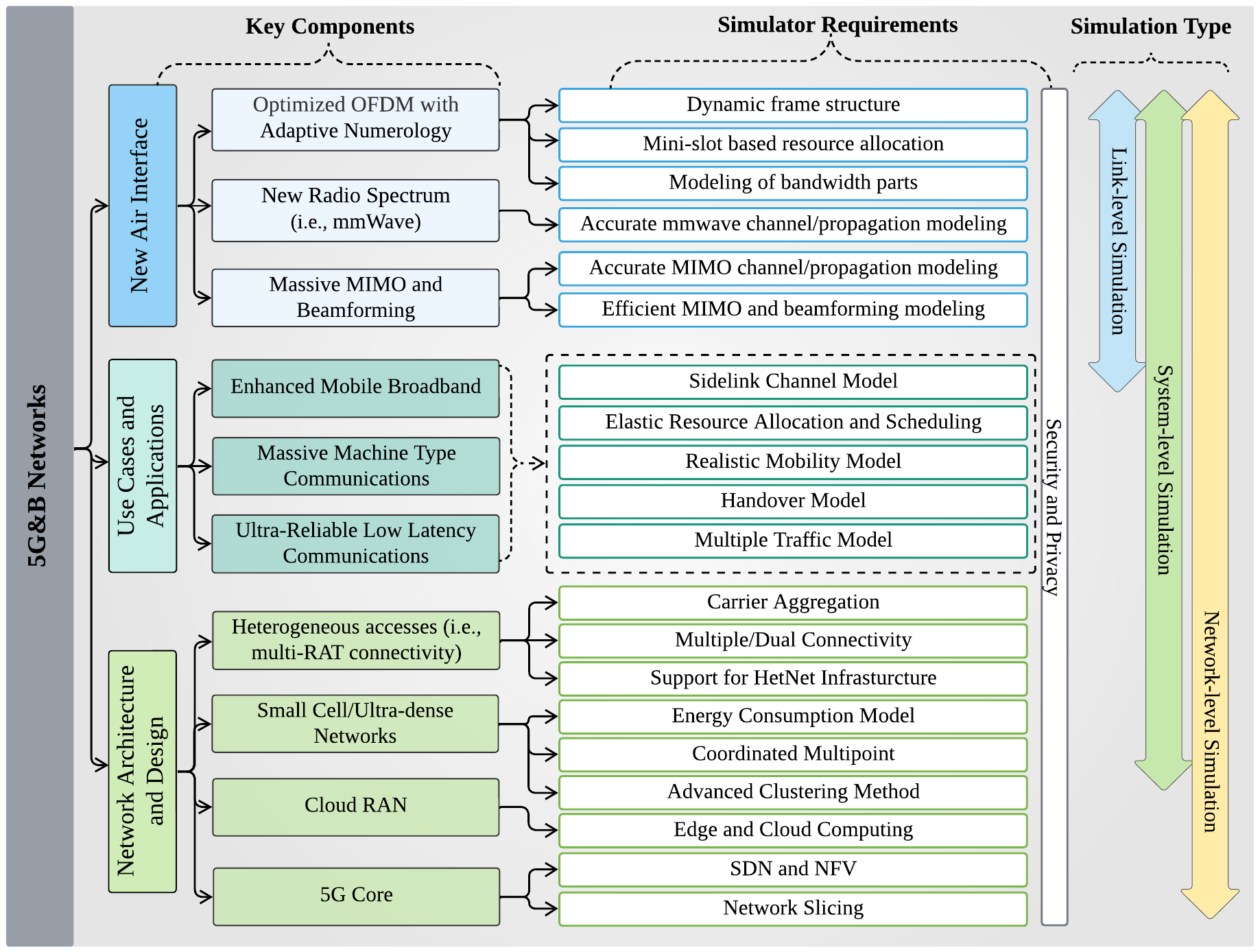}}
	\caption[]{Taxonomy of 5G network key features and components mapped to the resulting 5G\&B simulator requirements.}
	\label{fig:req}
\end{figure*}

\subsection{Interplay Between the Different Types of Cellular Network Simulators}

The interplay between different types of cellular network simulators is a critical aspect of accurately evaluating network performance. The link-level simulator serves as a valuable input source for system-level and network-level simulators through the utilization of link-level look-up tables (LUTs). These tables are generated based on various parameters, including data rate, modulation and coding scheme (MCS), and multiple input multiple output (MIMO) antenna configuration. The LUTs consist of essential information such as channel quality indicator (CQI), signal-to-interference-plus-noise ratio (SINR), and coded block error rate (BLER). System-level simulators rely on these link-level LUTs as inputs for their simulations. By processing this information, system-level simulators generate comprehensive results that provide insights into the overall network performance. These results include metrics such as cell throughput, handover success rate, packet error rate, dropped call rate, radio connection failure, energy efficiency, and spectral efficiency, among others. By evaluating these key performance indicators (KPIs), system-level simulators can offer valuable insights into the network's behavior and enable the assessment of different algorithm designs, protocols, and network configurations.

To achieve the best simulation results, it is crucial to utilize a comprehensive and accurate link-level simulator. It is essential to ensure that the generated LUTs capture the nuances and variations of the radio link accurately. A robust link-level simulator accounts for factors such as fading, interference, and propagation effects, resulting in more realistic and reliable input data for system-level simulations.

\section{Key Components of 5G\&B Networks and Design Requirements of Simulators}
\label{sec:req}

In this section, we discuss some of the key components of the 5G\&B networks, including air interface, architecture, and deployment design to support multiple use cases and applications. We then draw insights from this discussion to highlight the requirements of 5G\&B simulators. We devise a taxonomy that maps the characteristics of 5G\&B networks to the required features and modules of simulators (i.e., traffic, mobility, handover, and propagation), as shown in Fig. \ref{fig:req}.

\subsection{New air interface (New Radio)}

The new air interface, which has several significant distinctions from its predecessors, is one of 5G NR’s cornerstones. The following points discuss three key innovations in the air interface of 5G NR.

\subsubsection{Optimized OFDM with adaptive numerology}

Orthogonal frequency-division multiplexing (OFDM) has been the most widely utilized waveform for 4G LTE and Wi-Fi technology due to its inherent low complexity and efficient hardware implementation \cite{8986659,4607239}. Although several new waveforms have been proposed for 5G \cite{7744813,7469313}, 3GPP standardized OFDM as the waveform for 5G NR. This decision stems from several advantages of OFDM, including high spectral efficiency, low transceiver complexity, and high flexibility \cite{7744816}. The specific OFDM version used in 5G NR is called cyclic prefix OFDM (CP-OFDM). One key distinguishing feature of the optimized CP-OFDM is flexible or adaptive numerology. In adaptive numerology, instead of fixed sub-carrier spacing (i.e., 15 kHz in LTE), several sub-carrier spacings can be used (i.e., 15 kHz, 30 kHz, 60 kHz, 120 kHz, and 240 kHz) depending on the scenario requirements and operating frequency \cite{7883931}.

\subsubsection{New radio spectrum (mmWave)}
Apart from operating at sub-6 GHz frequencies, 5G adds the 34 GHz to 40 GHz mmWave spectrum. In the mmWave spectrum, flexible frequency reuse across a small area allows for more efficient spectrum utilization \cite{6736750}. However, mmWave also faces serious deployment challenges. As the frequency of radio waves increases, the attenuation of signal strength due to path loss and environmental obstructions becomes more pronounced. This phenomenon is especially significant in the higher frequency bands, where signals are more sensitive to absorption, reflection, and dispersion by obstacles such as buildings, vegetation, and atmospheric conditions. Therefore, the utility of mmWave is strongly dependent on line-of-sight (LOS) radio wave propagation \cite{6932503}.

\subsubsection{Massive MIMO and beamforming}

To mitigate the effects of pathloss and non line-of-sight (NLOS) in mmWave, MIMO allows the formation of highly focused beams via a process known as beamforming. Meanwhile, the inherently short wavelengths of the mmWave spectrum allow for fitting multiple antenna arrays on the transmitter and receiver \cite{6824752}. When the number of antennas becomes large enough (in the order of hundreds or thousands), MIMO evolves into massive MIMO. Massive MIMO improves the spectral efficiency of the network by enabling parallel, multiple data stream communication between the BS and multiple users \cite{8371237}.

\subsection{Requirements brought by 5G\&B NR air interface}

\subsubsection{Modeling of dynamic frame structure, mini-slot based resource allocation, and bandwidth parts}

Simulators should be equipped with various features to effectively mimic the optimized OFDM with adaptive numerology. Firstly, simulators should model a dynamic frame structure, which can allocate each symbol for uplink or downlink traffic depending on the requested service \cite{8802028}. Secondly, mini-slot based transmission should be supported, which allows user resource allocation at the symbol level (e.g., 2, 4, or 7 symbols in a mini-slot) \cite{9305243}. Lastly, the implementation of bandwidth parts should be modeled, wherein adjacent physical resource blocks (PRBs) with different numerologies are packed on a given carrier. The presence of these requirements pertaining to 5G frame structure can aid in validating designs such as those presented in \cite{7432148, 8004461, 7805314}.

\subsubsection{Accurate mmWave channel and propagation models} 

Traditional propagation models will not suffice for 5G NR due to the new characteristics of the mmWave frequency. Thus, incorporating extremely accurate and finer resolution mmWave channel and propagation models is crucial in advancing research topics such as coverage evaluation  \cite{7370940}, performance assessment \cite{8611440}, feasibility validation \cite{9366525}, licensed and unlicensed mmWave integration \cite{8643932}, among others.

\subsubsection{Accurate and efficient MIMO channel/propagation models}
Channel models with a high degree of granularity on top of massive MIMO and beamforming techniques to model 5G-NR may result in memory-intensive simulations \cite{7993797}. Thus, computationally efficient MIMO, and beamforming models should be embedded in the simulators to reduce complexity \cite{Gkonis_2020, MARTIRADONNA2020107314}. Incorporating accurate and efficient MIMO channel models into simulators is particularly useful in research topics including massive MIMO solutions \cite{8288009}, performance analysis \cite {9000609}, performance improvement \cite{9256345}, and pilot contamination \cite{8241348,8288009}.

\subsection{Network architecture and design}

The 5G architecture is designed to be service-oriented (i.e., it should be flexible and adaptable to support different services). In the following discussions, we cover the network architecture and design innovations implemented in 5G networks.

\subsubsection{Ultra-dense heterogeneous network deployment}

Employing a very large number of small cells, each with a range of only a few hundred meters, is one of the most effective ways to increase the network capacity, coverage, and energy efficiency, while reducing interference and latency \cite{6824752}. In 5G, BS density is expected to increase to around 40-50 BS per km², as compared to 4-5 BS per km² in 3G and 8-10 BS per km² for 4G \cite{7422408}. Moreover, 5G is envisaged to unify the diversified types of BS (i.e., small cells, macro cells, mmWave cells), paving the way for a heterogeneous system to coexist.

\subsubsection{Multi-RAT connectivity}

5G RAN must coexist with existing technologies such as LTE and Wi-Fi. During the initial phase of the 5G rollout, the 5G RAN is tightly coupled with 4G and makes use of its core via a configuration known as the non-standalone (NSA) operation. In this architecture, users camp traditionally on the LTE network, and later on, if the user requests a 5G service, it will connect to LTE and 5G NR simultaneously. This dual-camping is enabled through a 3GPP-standardized technique known as E-UTRAN new-radio dual-connectivity (EN-DC) \cite{9910474,9322339}. Furthermore, another technology known as carrier aggregation allows the simultaneous utilization of carriers, delivering more capacity for the users \cite{7945855}.

\subsubsection{Cloud radio access network (C-RAN)}

C-RAN leverages a computing-based architecture that integrates several BSs into a centralized base-band processing function entity. The increased energy efficiency and capacity performance of the network are two of the most common benefits of employing this architecture \cite{6897914}.

\subsubsection{5G core (5GC) network}

Apart from the evolution in the RAN, the 5G network core design has also undergone a major overhaul. One of the key distinctions of the 5GC is the inception of control and user plane separation \cite{7140736}. This concept enables a shift from the current node-based to a network function-based communication \cite{7993854}. This new architecture leverages network function virtualization (NFV) and software defined networking (SDN) for flexibility in 5G\&B network operations. This flexibility enables 5GC to support a plethora of new use cases with diverse requirements through network slicing, mobile edge computing and realization of different network functions (NFs), including 3rd party NFs.

\subsection{Requirements brought by 5G network architecture and design}

\subsubsection{Support for HetNet infrastructure}
The ability to support HetNet deployment with coexisting BS types (i.e., macro cell, small cell) and technologies (i.e., 4G, 5G, Wi-Fi) in a single simulation environment is critical, especially for evaluating inter-RAT handovers, quality of service (QoS) management, and network switching design \cite{Gkonis_2020,Vienna5GSLS,wang,liu2016design}.

\subsubsection{Enhanced interference calculation}
The interference distribution varies with a HetNet deployment compared to standard homogeneous networks \cite{6994948}. This makes interference calculation more complicated, necessitating the use of enhanced interference models in 5G\&B simulators. This enables more effective planning and evaluation of BS placement, as well as optimization of antenna parameters such as tilt, azimuth, and transmit power.

\subsubsection{Advanced clustering scheme}
Several studies demonstrate an efficient method for simplifying a HetNet topology by creating several clusters of small cells \cite{8388957,7579583,6717207,6503788}. This technique includes clustering small cells based on their proximity and allocating a macrocell to each cluster. The macrocell then functions as a gateway for all the small cells in that cluster, therefore lowering the network's complexity and increasing its efficiency. Thus, 5G\&B simulators should integrate advanced clustering schemes to evaluate the impact of different clustering procedures on network performance and the trade-off between cluster size and signaling overhead. For instance, in the context of coordinated multi-point, the number of BSs forming clusters should be small; otherwise, the amount of signaling overhead may increase significantly as backhaul traffic increases \cite{6815892}.

\subsubsection{Energy consumption model}
In 5G\&B, there is an expected increase in energy consumption on both the RAN and user sides. The increased energy consumption in RAN is attributed to the increased computational requirement, which is further exacerbated by the expected increase in the number of BSs in ultra-dense heterogeneous networks. Hence, 5G\&B simulators should incorporate elaborate energy consumption models for various BS types and technologies to help in the validation of solutions, including energy-saving schemes \cite{8998281,7904703}, and energy consumption optimization \cite{sachan2016genetic}. Similarly, the need for innovative energy management on the user side stems from the miscellany of connected devices (e.g., smartphones, IoT devices, unmanned aerial vehicles (UAVs), healthcare, industry 4.0 devices, etc.) in the 5G ecosystem. Therefore, having an energy consumption model for various device types will assist in the development of energy management solutions for end users \cite{farooq2018user, farooq2020utilizing}.

\subsubsection{Carrier aggregation and multiple/dual connectivity} 
5G\&B features such as carrier aggregation and multi-technology connectivity should be included in 5G\&B simulators to allow in-depth analysis of their influence on user quality of experience (QoE), intelligent multi-connectivity solutions \cite{9322339}, and various carrier aggregation combinations.

\subsubsection{Support for SDN and NFV}

SDN/NFV-capable simulators can facilitate a thorough investigation of the challenges for practical deployment of SDN/NFV, such as fronthaul, latency of general-purpose platforms, backward compatibility, disruptive deployment, security vulnerabilities, and compelling business cases, as discussed in \cite{8361844}.

\subsubsection{Support for network slicing}
Different resource slices can be allocated to users based on their service demands for network performance optimization and QoS assurance \cite{8802028}. Incorporating network slicing functionality in 5G\&B network simulators can help in research aiming to address the challenges related to network slicing, such as RAN slicing, traffic isolation, slice security, slice optimality, and UE slicing, as outlined in \cite{8320765}. Moreover, experiments related to the impact of slicing on 5G protocol architecture design, network functions (e.g., scheduling and random access), and network management (e.g., slice-based performance monitoring and optimization) can be executed \cite{7561023}.

\subsubsection{Support for edge computing}

Support for edge computing is desirable for 5G\&B simulators to design and evaluate edge computing techniques. More specifically, the optimization of edge latency-related methods can be performed using an edge computing enabled 5G\&B simulator \cite{8802028}. With access to these kinds of simulators, challenges associated with edge computing such as network security, energy utilization, proximity awareness, offloading awareness, network reliability, resource management, intelligent caching, and mobility management, as highlighted in \cite{9206025} can be studied.

\subsection{Use cases and applications}

The radio communication sector of the international telecommunications union (ITU-R) identified three general classifications of 5G use cases, which are later adopted by 3GPP namely eMBB, URLLC, and mMTC \cite{series2015imt}. Each of the use cases demands a unique set of requirements from the 5G network.

\subsubsection{Enhanced mobile broadband (eMBB)}
High throughput, large-volume transmission, and high capacity are required for applications falling under the umbrella of eMBB (e.g., fixed wireless access, ultra-HD video streaming, and in-vehicle entertainment).

\subsubsection{Ultra-reliable and low latency communications (URLLC)}
Under URLLC, the connected devices are provided with low latency and high reliability regardless of the user's speed (e.g., automatic driving, mobile healthcare, remote-controlled vehicles, and V2X).

\subsubsection{Massive machine type communications (mMTC)}
Increase in coverage, high capacity, and the ability to support a large number of devices are the primary requirements of applications under mMTC (e.g., smart cities, home automation, large distributed sensors, and massive IoT).

\subsection{Requirements brought by 5G\&B use cases and applications}

\subsubsection{Elastic resource allocation and scheduling}
Each type of user for different use cases and applications requires different amounts of resources, depending on the application. As a result, simulators require dynamic and elastic resource allocation and scheduling algorithms to assist in addressing associated optimization challenges such as overheads and delay minimization, and the constraint between performance parameters, security, and backhaul/fronthaul \cite{SHARMA2021101415}.

\subsubsection{Multiple traffic models}
In legacy networks, users are homogenized to simplify traffic modeling. For instance, LTE traffic is mainly categorized into voice and data traffic. In comparison to legacy networks, 5G\&B networks have more diverse users with varying service requirements, making traffic modeling complicated. For instance, IMT 2020 defined traffic patterns for several 5G test settings (i.e., full buffer for eMBB and URLLC while Poisson packet arrival for mMTC) \cite{series2015imt}. For 5G\&B simulators to become realistic, a diverse set of traffic models must be included.

\subsubsection{Realistic mobility model}
The introduction of advanced 5G\&B use cases, such as UAVs and self-driving cars, introduces a new dimension in modeling mobility. Currently, some of the most popular mobility models incorporated in cellular network simulators include random walk model, random waypoint (RWP) model, fluid flow model, and Gauss-Markov model \cite {7399689}. These models, however, will not be sufficient to serve a wide range of user types with diverse mobility profiles. Moreover, these 2D models cannot capture users that move in three dimensions (e.g., UAVs). Thus, implementation of a realistic mobility model in a simulator is imperative to drive research pertinent to mobility prediction \cite{8416734,9045127,9117072}, handover management \cite{9206546}, 5G-enabled self-driving cars \cite{8667012}, 5G-enabled mobile healthcare \cite{10032073,9982778}, and flying BSs \cite{8675384,8620550}.

\subsubsection{Handover model}
In the wake of ultra-dense networks and an increasing fraction of high-speed users, handovers are becoming a bottleneck in achieving the user's QoE and throughput requirements. Furthermore, handovers are a major source of signaling overhead for 5G\&B networks. However, most of the current simulators either lack handover models or provide very simplistic and unrealistic models that limit their utility \cite{9084113}. Realistic and detailed handover models should be developed for 5G\&B simulators to enable the design and evaluation of handover optimization solutions \cite{10060203,9730050,9716889,9839024,farooq2020data} and test novel handover algorithms \cite{7502901,8747365,8895796}.

\subsubsection{Accurate channel modeling on top of UE-BS links, such as D2D link, V2X communication, and intermediate relay nodes}
Aside from the traditional uplink and downlink communication, in 5G\&B network, sidelinks also play a vital role. Simulators should have the capability to model sidelink transmissions to support various use cases such as D2D, V2X communication \cite{Gkonis_2020}, and intermediate relay nodes \cite{8013721,7417775}.

\subsection{Enhanced Security and Privacy Across Key Components}

Security and privacy become more crucial as cellular technology advances. Several studies have been conducted to assess the security landscape of 5G networks \cite{9305243, 9972143, article, 8125684, 9813965, 10039654}. The authors of these studies have highlighted several vulnerabilities in 5G networks, including man-in-the-middle attacks, signaling attacks, distributed denial-of-service (DDoS) attacks, and fake base station attacks. These papers highlight that the network architecture and features of 5G systems poses unique cybersecurity challenges. For instance, as networks become more distributed and cloudified, new attack vectors emerge. This observation is exemplified by the current dis-aggregated deployment of 5G into distinct entities such as distributed units (DU) and centralized units (CU). With regards to 5G features, its high connection throughput enables attackers to quickly download large volumes of data, such as private information compromising the users' privacy. Moreover, the low-latency connectivity offered by 5G, facilitated by mobile edge computing infrastructure, is now more susceptible to attacks such as DoS or man-in-the-middle attacks. Aside from 5G network architecture and features, the growing number and heterogeneity of connected devices exacerbate the challenge of identifying potential attackers.

It is important to note that while the potential for attacks may theoretically increase with 5G, several countermeasures have been developed and proposed to address these threats. For example, 5G introduces multiple authentication methods, significantly enhancing security compared to 4G. Whereas 4G relied on a single authentication method (EPS-AKA) \cite{3GPP1}, 5G includes three methods: 5G-AKA, EAP-AKA’, and EAP-TLS \cite{3GPP2}. Additionally, 5G improves network security and privacy by implementing IMSI (International Mobile Subscriber Identity) encryption, ensuring that all user data transmitted through 5G networks is protected both in terms of confidentiality and integrity on a hop-by-hop basis \cite{Ericsson1}. Moreover, ongoing research into security enhancements at the PHY and MAC layers \cite{SHARMA2022101791, sun2017physical, sun2017physical, fang2017security}, as well as at higher layers \cite{sullivan20215g}, further contributes to enhancing 5G security. Another promising approach gaining traction is the use of AI to counter these attacks \cite{lee2022ai}. To effectively evaluate security, simulators must capture these attacks and their corresponding countermeasures.

The authors in \cite{9921600} undertook an extensive evaluation of various wireless network simulators, such as NS-3, OMNeT++, WSNet, TOSSIM, J-Sim, GloMoSim, and others, to assess their performance and scalability in enhancing the security and safety of smart cities. Their analysis highlighted several critical factors that should be considered when selecting a simulator to aid in evaluating network security. These factors include the level of abstraction, processing time (in seconds), memory usage (in kilobytes), CPU utilization (as a percentage), and simulation overhead.

In another study, the authors delved into the role of simulation in cybersecurity, offering valuable insights applicable beyond cellular networks \cite{10.1093/cybsec/tyab005}. They emphasized the necessity for simulation platforms to provide a representative environment for testing various types of cyberattacks. Additionally, they underscored the importance of simulators being capable of testing potential attacks and seamlessly integrating with security-enhancing solutions. With the rising prevalence of AI-based attacks and corresponding defenses, it is imperative for simulators to support these emerging techniques \cite{10082863}.

Moreover, the ability of simulators to generate substantial amounts of data rapidly plays a crucial role in safeguarding privacy. This capability reduces reliance on real data for training ML/DL models, thus mitigating the associated privacy risks.

It is therefore crucial to ensure that the security and privacy aspects of the network can be assessed using simulators. This can be achieved by implementing various threat models, such as those discussed in several studies \cite{10318093, 9825149, electronics11121819, 9604960, 9221122}, which mimic the behavior of real-world attacks, including those previously mentioned.

\section{Simulators for 5G\&B Networks}
\label{sec:types}

This section discusses the different types of 5G\&B simulators that are currently available to the general public, academia, and industry. We begin by developing a taxonomy for categorizing 5G\&B simulators. We then present a quick introduction of the well-known 5G\&B link-level, system-level, and network-level simulators, followed by a discussion of commercially available system-level simulators.

\subsection{Taxonomy of 5G\&B Network Simulators}
\begin{figure*}[]
	\centering
	\includegraphics[width=.98\textwidth, height=13cm]{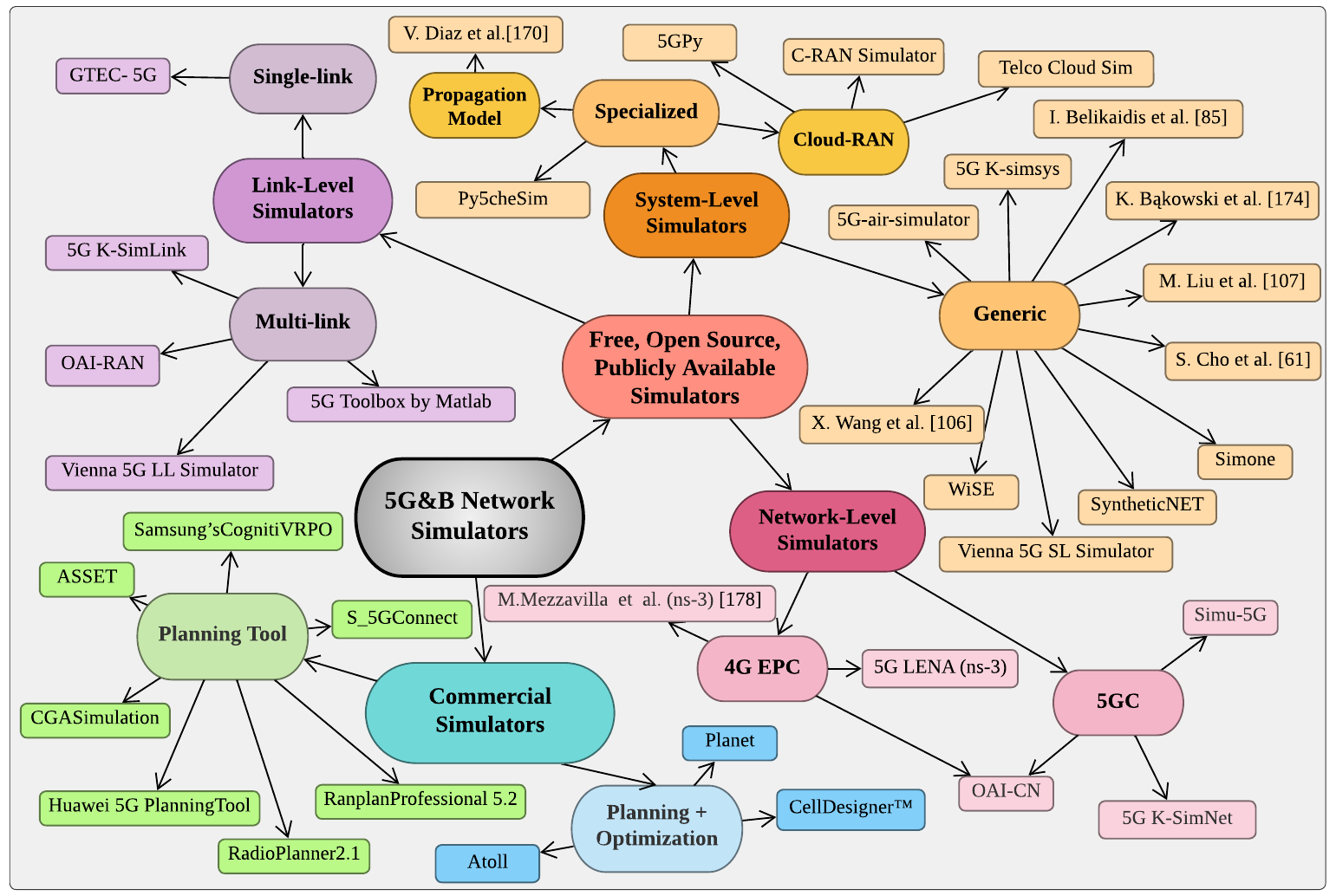}
	\caption[]{Taxonomy of the 5G simulators included in the discussion.}
	\label{fig:taxo}
\end{figure*}

The taxonomy of 5G\&B simulators included in this survey paper is shown in Fig. \ref{fig:taxo}. 5G\&B network simulators are broadly divided into two categories: 1) commercial simulators and 2) free, open-source, or publicly available simulators with published peer-reviewed publications.

The first group includes commercial and industrial-grade simulators used mostly by network operators and equipment vendors. This group is further subdivided into two subgroups based on the functions offered and their utility. A commercial simulator can be used as a planning tool or as a planning and optimization tool in combination. The outermost layer of the taxonomy in Fig. \ref{fig:taxo} shows the names of different simulators belonging to each subgroup. Although these commercial simulators are not generally used in the research community due to their high cost, we explore them in order to gain insight into the state-of-the-art implementation of 5G components and functionalities.

Free, open-source, or publicly available simulators comprise the second group of 5G\&B network simulators. These kinds of simulators are often used by the research community because of their free accessibility. This group is divided into three types of simulators: link-, system-, and network-level simulators. Link-level simulators can be further classified as single-link or multi-link, depending on the simulation capabilities. Similarly, system-level simulators can be further categorized as specialized or generic. The specialized system-level simulators are built to model or serve a specific part of 5G\&B networks (e.g., C-RAN simulations, or propagation modeling), while the generic system-level simulators cover wider aspects of the network and can be used for the evaluation of multiple 5G\&B technologies. Lastly, the network-level simulators can be organized according to the implementation of core networks. The two categories for network-level simulators include 4G evolved packet core (EPC), wherein simulators utilize the LTE core model, and 5GC, in which simulators model the new core model of 5G.

The outer layer of the taxonomy features the names of simulators within each category, setting the stage for our subsequent discussions. Our taxonomy reveals that the majority of these simulators belong to the broader category of generic system-level simulations. Consequently, while we provide a brief overview of each listed simulator, our primary focus is directed towards the generic system-level simulators. Specifically, in Section \ref{sec:system}, our comprehensive examination delves into whether these DSMs can transition from mere simulators to DT models. Meanwhile, the challenges inherent in elevating DSMs to embrace a more substantial DT kernel role, explored in Section \ref{sec:challenges} and Section \ref{sec:6G}, are particularly pertinent to this system-level context.

\subsection{Link-Level Simulators}

In the following discussion, we provide a brief overview of each of the link-level simulators included in this survey paper. 

\begin{table*}[t]
\caption{Comparison Between Link-Level 5G Simulators}
\centering
\label{tab:linklevel}
\begin{tabular}{|c|c|c|c|c|c|c|c|c|c|}
\hline
\rowcolor[HTML]{EFEFEF} 
\textbf{\begin{tabular}[c]{@{}c@{}}Link-Level \\ Simulators\end{tabular}} & \multicolumn{1}{c|}{\cellcolor[HTML]{EFEFEF}\textbf{\begin{tabular}[c]{@{}c@{}}License \\ Type\end{tabular}}} & \multicolumn{1}{c|}{\cellcolor[HTML]{EFEFEF}\textbf{\begin{tabular}[c]{@{}c@{}}Language/\\ Platform\end{tabular}}} & \multicolumn{1}{c|}{\cellcolor[HTML]{EFEFEF}\textbf{\begin{tabular}[c]{@{}c@{}}Link \\ Analysis\end{tabular}}} & \multicolumn{1}{c|}{\cellcolor[HTML]{EFEFEF}\textbf{\begin{tabular}[c]{@{}c@{}}Modu-\\ lation\end{tabular}}} & \multicolumn{1}{c|}{\cellcolor[HTML]{EFEFEF}\textbf{\begin{tabular}[c]{@{}c@{}}Channel \\ Coding\end{tabular}}} & \multicolumn{1}{c|}{\cellcolor[HTML]{EFEFEF}\textbf{\begin{tabular}[c]{@{}c@{}}Adaptive \\ Numerology\end{tabular}}} & \multicolumn{1}{c|}{\cellcolor[HTML]{EFEFEF}\textbf{Channel Models}} & \multicolumn{1}{c|}{\cellcolor[HTML]{EFEFEF}\textbf{\begin{tabular}[c]{@{}c@{}}Supported \\ Frequency \end{tabular}}} \\ \hline

\rowcolor[HTML]{DAE8FC}
\textbf{\begin{tabular}[c]{@{}c@{}}5G \\ K-SimLink \cite{8610463}\end{tabular}} & \begin{tabular}[c]{@{}c@{}}Academic \\ use license\end{tabular} & C++ & Multi-link & \begin{tabular}[c]{@{}c@{}}CP-\\ OFDM\end{tabular} & \begin{tabular}[c]{@{}c@{}}LDPC\\ coding\end{tabular} & Yes & \begin{tabular}[c]{@{}c@{}}AWGN, tapped delay \\ line model, and cluster \\ delay line model\end{tabular} & \begin{tabular}[c]{@{}c@{}}Sub- and\\ above 6 GHz\end{tabular} \\ \hline

\textbf{\begin{tabular}[c]{@{}c@{}}5G Toolbox\\ by MATLAB© \cite{5Gmatlab}\end{tabular}} & \begin{tabular}[c]{@{}c@{}}Total academic \\ headcount \\ license\end{tabular} & MATLAB & Multi-link & \begin{tabular}[c]{@{}c@{}} CP-\\OFDM  \end{tabular} & \begin{tabular}[c]{@{}c@{}}Not \\ specified\end{tabular} & Yes & \begin{tabular}[c]{@{}c@{}}5G channel \\models (TR 38.901)\end{tabular} & \begin{tabular}[c]{@{}c@{}} Sub-6 GHz\\ and mmWave\end{tabular} \\ \hline

\rowcolor[HTML]{DAE8FC}
\textbf{GTEC-5G \cite{GTEC}} & \begin{tabular}[c]{@{}c@{}}General Public \\ License \\ (GPLv3)\end{tabular} & MATLAB & Single-link & \begin{tabular}[c]{@{}c@{}}OFDM \\ and\\ FBMC\end{tabular} & \begin{tabular}[c]{@{}c@{}}Not \\ specified\end{tabular} & No & \begin{tabular}[c]{@{}c@{}}AWGN, flat Rayleigh \\ fading, standardized \\ channel models, and\\ 3GPP typical \\ urban channel model\end{tabular} & Sub-6 GHz \\ \hline

\textbf{\begin{tabular}[c]{@{}c@{}}OpenAir-\\ Interface RAN \\ (OAI-RAN) \cite{OIA-RAN}\end{tabular}} & \begin{tabular}[c]{@{}c@{}}Public License \\ V1.1\end{tabular} & C & Multi-link & OFDM & \begin{tabular}[c]{@{}c@{}}Turbo \\ coding\end{tabular} & No & \begin{tabular}[c]{@{}c@{}}AWGN, standardized \\ channel models,\\ Rayleigh fading, and\\ Rician fading\end{tabular} & Sub-6GHz \\ \hline

\rowcolor[HTML]{DAE8FC} 
\textbf{\begin{tabular}[c]{@{}c@{}}Vienna  5G \\ Link-Level \\ Simulator \cite{vienall}\end{tabular}} & \begin{tabular}[c]{@{}c@{}}Academic \\ use license\end{tabular} & MATLAB & Multi-link & \begin{tabular}[c]{@{}c@{}}OFDM,\\ f-OFDM,\\ WOLA,\\ FBMC, \\ and\\ UFMC\end{tabular} & \begin{tabular}[c]{@{}c@{}}Turbo,\\ Polar, and\\ LDPC \\ coding\end{tabular} & Yes & \begin{tabular}[c]{@{}c@{}}Doubly-fading channel\\ model, and spatial \\ channel model\end{tabular} & \begin{tabular}[c]{@{}c@{}} Sub-6 GHz\\ up to 100 GHz\end{tabular} \\ \hline

\end{tabular}

\end{table*}

\begin{itemize}
    
\item {\textbf{5G K-SimLink} \cite{8610463}}: 
5G K-SimLink is a C++-based multi-link link-level simulator that implements several main features of 5G in compliance with 3GPP Release-15. These features include an evolved frame structure with variable sub-carrier spacing, a CP-OFDMA waveform, the adoption of new reference signals, and LDPC channel coding. Additionally, the developers implemented distinct channel models for the sub-6 GHz and mmWave bands.

\item {\textbf{5G Toolbox by MATLAB©} \cite{5Gmatlab}}: 
Although 5G Toolbox by MATLAB© is a commercial product, it is often free for academic use. This toolbox simulates the operations of the transmitter and receiver, as well as the channel between them, based on 3GPP Release-15. It analyzes the link's performance using several metrics, such as BLER and throughput.

\item{\textbf{GTEC-5G} \cite{GTEC}}:  
GTEC-5G is a MATLAB-based single-link link-level simulator that supports both OFDM and filter bank multicarrier (FBMC) waveform simulations. The modular architecture of GTEC-5G includes a complete implementation of transmitter and receiver with various channel models. Additionally, GTEC-5G’s integration with the GTEC testbed enables over-the-air measurements using diverse scenarios.

\item{\textbf{OpenAirInterface RAN (OAI-RAN)} \cite{OIA-RAN}}: 
OAI-RAN is a component of the broader OpenAirInterface simulator. It is an open and flexible platform for cellular network experimentation and prototyping. It has an over-the-air interface that supports multiple radio access technologies, including LTE, NR, and Wi-Fi. The 5G NR software implementation adheres to 3GPP specifications. OAI-RAN enables the investigation of critical 5G characteristics, such as machine-to-machine (M2M) communication and C-RAN. Although OIA-RAN possesses certain system-level capabilities, it functions more as an emulator than a simulator. Consequently, we have omitted it from our current discussion.

\item{\textbf{Vienna 5G Link-Level Simulator} \cite{vienall}}: 
The Vienna 5G link-level simulator is a flexible MATLAB-based simulation tool that complies with 3GPP standards for both 4G LTE and 5G NR. The popularity of the Vienna 5G link-level simulator is attributed to the high-granularity implementation of the PHY layer and support for 5G features like adaptive numerology and mmWave channel modeling. It leverages Monte Carlo simulations to model and evaluate the performance of the PHY layer.

\end{itemize}

Table \ref{tab:linklevel} compares some of the most important features of link-level 5G simulators. The comparison shows that the Vienna 5G link-level simulator is the most comprehensive, with a rich selection of waveforms and channel coding techniques. Furthermore, it is the only simulator that supports up to 100 GHz carrier frequency. The Vienna 5G link-level simulator, along with 5GK-SimLink and OAI-RAN, supports the simulation of multiple link scenarios. Meanwhile, all of the simulators offer a variety of channel models for varying link-level evaluation scenarios.

\subsection{System-Level Simulators}
The discussion of system-level simulators is divided into two main parts: specialized system-level simulators and general-purpose system-level simulators.

We start by examining specialized system-level simulators, which are specifically designed to simulate particular network functions, use cases, or network architectures. These simulators are tailored to address specific requirements and provide in-depth analysis within their designated domains.

Following that, we delve into the discussion of general-purpose system-level simulators. Unlike their specialized counterparts, these simulators offer a broader range of functions and are not limited to any particular network function or architecture. They provide a more versatile platform that can be applied to various scenarios, offering a wider scope of simulation capabilities.

\subsubsection{Specialized system-level simulators}

\begin{itemize}
    
\item {\textbf{5GPy} \cite{TININI2020102030}}: 
5GPy is a python-based, event-driven simulator developed as a platform to perform C-RAN architecture simulations. 5GPy supports both small- and large-scale simulations for C-RAN.

\item {\textbf{C-RAN Simulator} \cite{7347976}}:
This simulator is designed to provide a platform for system-level evaluation of 5G cloud-based networks. Some of the features implemented in the C-RAN Simulator include centralized user scheduling, joint transmission of edge users, and support for global per-antenna carrier aggregation.

\item {\textbf{Telco Cloud Simulator} \cite{8858483}}:
Telco Cloud Simulator (TCS) is developed to simulate virtualized 5G networks and serves as a platform for conducting 5G cloud-related research.

\item {\textbf{V. Diaz et al.}} \cite{diaz}: 
This specialized simulator is a web-based path loss simulation tool based on 3GPP channel models for 5G networks. Specifically, this simulator implements channel models standardized in 3GPP technical report TR 38.901 (i.e., UMa, RMa, UMi, InH) \cite{zhu20193gpp}.

\item {\textbf{Py5cheSim} \cite{9640086}}:
Py5cheSim is a Python-based, open-source simulator designed to model cell capacity in 5G\&B networks and serves as the first simulator that enables network slicing at the RAN.

\end{itemize}

\begin{table*}[]
\caption{Summary of Comparison Between System-level 5G Simulators using Traditional Metrics}
\centering
\label{tab:traditional}
\begin{tabular}{|c|c|c|c|c|c|c|c|}
\hline

\rowcolor[HTML]{EFEFEF} 
\textbf{Simulator} &  \textbf{\begin{tabular}[c]{@{}c@{}}Simulator \\ Type\end{tabular}} & \textbf{User Interface} & \textbf{\begin{tabular}[c]{@{}c@{}}Language/\\ Platform\end{tabular}} & \textbf{\begin{tabular}[c]{@{}c@{}}Open \\ Source\end{tabular}} & \textbf{\begin{tabular}[c]{@{}c@{}}License \\ Type\end{tabular}} & \textbf{\begin{tabular}[c]{@{}c@{}}Documentation \\ and User Support\end{tabular}} & \textbf{Modular} \\ \hline

\rowcolor[HTML]{FFFFFF} 
\textbf{5GPy \cite{TININI2020102030}}  & Specialized & Not specified & Python & Yes & \begin{tabular}[c]{@{}c@{}}General Public   \\ License\end{tabular} & Limited & Yes \\ \hline

\rowcolor[HTML]{DAE8FC} 
\textbf{\begin{tabular}[c]{@{}c@{}}C-RAN \\ Simulator \cite{7347976}\end{tabular}}  & Specialized & \begin{tabular}[c]{@{}c@{}}Command Line   \\ Interface\end{tabular} & MATLAB & \begin{tabular}[c]{@{}c@{}}Not \\ specified\end{tabular} & \begin{tabular}[c]{@{}c@{}}Academic Use \\ License\end{tabular} & Limited & Yes \\ \hline

\rowcolor[HTML]{FFFFFF} 
\textbf{\begin{tabular}[c]{@{}c@{}}Telco Cloud \\ Simulator \cite{8858483}\end{tabular}}  & Specialized & \begin{tabular}[c]{@{}c@{}}Graphical User   \\ Interface\end{tabular} & Java & \begin{tabular}[c]{@{}c@{}}Not \\ specified\end{tabular} & Not specified & Limited & \begin{tabular}[c]{@{}c@{}}Not \\ specified\end{tabular} \\ \hline

\rowcolor[HTML]{DAE8FC} 
\textbf{V. Diaz et al. \cite{diaz}} & Specialized & \begin{tabular}[c]{@{}c@{}}Menu Driven \\ Interface\end{tabular} & PHP & No & \begin{tabular}[c]{@{}c@{}}General Public   \\ License\end{tabular} & Limited & \begin{tabular}[c]{@{}c@{}}Not \\ specified\end{tabular} \\ \hline

\rowcolor[HTML]{FFFFFF} 
\textbf{Py5cheSim \cite{9640086}} & Specialized & \begin{tabular}[c]{@{}c@{}}Not specified\end{tabular} & Python & Yes & \begin{tabular}[c]{@{}c@{}}General Public \\ License\end{tabular} & Limited & \begin{tabular}[c]{@{}c@{}}Yes \end{tabular} \\ \hline

\rowcolor[HTML]{DAE8FC} 
\textbf{\begin{tabular}[c]{@{}c@{}}5G-air-\\ simulator \cite{MARTIRADONNA2020107314}\end{tabular}} & Generic & \begin{tabular}[c]{@{}c@{}}Command Line\\ Interface\end{tabular} & C++ & Yes & \begin{tabular}[c]{@{}c@{}}General Public   \\ License\end{tabular} & Extensive & Yes \\ \hline

\rowcolor[HTML]{FFFFFF} 
\textbf{5G K-simsys \cite{han20185g}} & Generic & \begin{tabular}[c]{@{}c@{}}Graphical User \\ Interface\end{tabular} & C++ & Yes & \begin{tabular}[c]{@{}c@{}}Academic Use \\ License\end{tabular} & Extensive & Yes \\ \hline

\rowcolor[HTML]{DAE8FC} 
\textbf{\begin{tabular}[c]{@{}c@{}}I. Belikaidis \\ et al. \cite{8802028}\end{tabular}} & Generic & \begin{tabular}[c]{@{}c@{}}Graphical User   \\ Interface\end{tabular} & \begin{tabular}[c]{@{}c@{}}Not \\ specified\end{tabular} & \begin{tabular}[c]{@{}c@{}}Not \\ specified\end{tabular} & Not specified & Limited & Yes \\ \hline

\rowcolor[HTML]{FFFFFF}
\textbf{M. Liu et al. \cite{liu2016design}}  & Generic & \begin{tabular}[c]{@{}c@{}}Graphical User \\ Interface\end{tabular} & Not specified & \begin{tabular}[c]{@{}c@{}}Not \\ specified\end{tabular} & Not specified & Limited & \begin{tabular}[c]{@{}c@{}}Not \\ specified\end{tabular} \\ \hline

\rowcolor[HTML]{DAE8FC} 
\textbf{S. Cho et al. \cite{7993797}} & Generic & Not specified & C++ & \begin{tabular}[c]{@{}c@{}}Not \\ specified\end{tabular} & Not specified & Limited & Yes \\ \hline

\rowcolor[HTML]{FFFFFF} 
\textbf{\begin{tabular}[c]{@{}c@{}}K. Bąkowski \\ et al.\cite{7454442}\end{tabular}} & Generic & \begin{tabular}[c]{@{}c@{}}Graphical User   \\ Interface\end{tabular} & Not specified & \begin{tabular}[c]{@{}c@{}}Not \\ specified\end{tabular} & Not specified & Limited & \begin{tabular}[c]{@{}c@{}}Not \\ specified\end{tabular} \\ \hline

\rowcolor[HTML]{DAE8FC} 
\textbf{SiMoNe \cite{7146084}} &  Generic & \begin{tabular}[c]{@{}c@{}}Graphical User   \\ Interface\end{tabular} & \begin{tabular}[c]{@{}c@{}}Not \\ specified\end{tabular} & \begin{tabular}[c]{@{}c@{}}Not \\ specified\end{tabular} & Not specified & Limited & Yes \\ \hline

\rowcolor[HTML]{FFFFFF}
\textbf{SyntheticNET \cite{9084113}}  & Generic & \begin{tabular}[c]{@{}c@{}}Command Line   \\ Interface\end{tabular} & Python & No & \begin{tabular}[c]{@{}c@{}}General Public   \\ License\end{tabular} & Limited & Yes \\ \hline

\rowcolor[HTML]{DAE8FC}
\textbf{\begin{tabular}[c]{@{}c@{}}Vienna 5G \\ SL Simulator \cite{Vienna5GSLS} \end{tabular}} & Generic & \begin{tabular}[c]{@{}c@{}}Command Line   \\ Interface\end{tabular} & MATLAB & Yes & \begin{tabular}[c]{@{}c@{}}Academic Use \\ License\end{tabular} & Extensive & Yes \\ \hline

\rowcolor[HTML]{FFFFFF}
\textbf{WiSE \cite{8352614}}  & Generic & \begin{tabular}[c]{@{}c@{}}Graphical User   \\ Interface\end{tabular} & C++ & No & Not specified & Limited & \begin{tabular}[c]{@{}c@{}}Not \\ specified\end{tabular} \\ \hline

\rowcolor[HTML]{DAE8FC} 
\textbf{X. Wang et al. \cite{wang}} & Generic & Not specified & Unknown & \begin{tabular}[c]{@{}c@{}}Not \\ specified\end{tabular} & Not specified & Limited & \begin{tabular}[c]{@{}c@{}}Not \\ specified\end{tabular} \\ \hline

\end{tabular}
\end{table*}

\subsubsection{Generic system-level simulators}

\begin{itemize}

\item {\textbf{5G-air-simulator}}\cite{MARTIRADONNA2020107314}:
5G-air-simulator models various critical components designed for the 5G air interface at the system level. These key technical components include massive MIMO, extended multi-cast, and broadcast transmission schemes, predictor antennas, enhanced random access procedures, and NB-IoT. Additionally, it includes a variety of network architectures capable of simulating multiple cells and users, a wide selection of mobility models, and a high-fidelity link-to-system model for the physical and data-link layers.

\item {\textbf{5G K-simsys}} \cite{han20185g}:
It is designed to simulate and evaluate the performance of various aspects of 5G networks, including network architecture, protocols, and deployment scenarios. One key aspect of 5G K-SimSys is its modular and flexible architecture, which enables the reuse of modules across different system configurations. This is achieved by modifying existing modules or incorporating additional ones to accommodate newly introduced algorithms or functionalities.

\item {\textbf{I. Belikaidis et al.}} \cite{8802028}:
This system-level simulator aims to develop a tool for simulating multi-connectivity scenarios. The authors utilized multiple channel models, traffic models, and mobility models for testing different service requirements of 5G networks.

\item {\textbf{K. Bąkowski et al.}} \cite{7454442}: 
The authors presented a system-level simulation tool for evaluating some of the major research directions of the 5G network. These research directions include long-term interference mitigation, D2D resource allocation, and two-way relaying.

\item {\textbf{M. Liu et al.}} \cite{liu2016design}:
The developers of this simulator introduced a novel design for 5G system-level simulations. Since 5G was still nonexistent at the time of its publication, the authors applied the proposed approach to a heterogeneous environment comprised of LTE-A and IEEE 802.11-based Wi-Fi.

\item {\textbf{S. Cho et al.}} \cite{7993797}:
This study demonstrated a work-in-progress simulator that addressed several of the challenges associated with 5G system-level simulators, including scalability, reusability, accuracy in the presence of abstraction, computational efficiency, and tight coupling between link and system-level simulations.

\item {\textbf{SiMoNe}} \cite{7146084}:
Simulator for Mobile Networks, also known as SiMoNe, is a system-level simulator developed primarily to facilitate the testing of novel self-organizing networks (SON) solutions. Through seamless parameter variations, SiMoNe provides a platform for testing a large variety of network configurations.

\item {\textbf{SyntheticNET}} \cite{9084113}:
SyntheticNET is a cellular network simulator built in Python for 4G and 5G\&B networks in compliance with 3GPP Release-15. It is a modular, flexible, and versatile simulator supporting advanced features like adaptive numerology, handover, and futuristic database-aided edge computing, to name a few.

\item {\textbf{Vienna 5G system-level simulator}} \cite{Vienna5GSLS}:
Vienna system-level simulator is built in MATLAB and is one of the most popular simulators in the field of mobile network communications. The Vienna system-level simulator allows large-scale multi-tier network performance evaluation and supports various types of network nodes. A modular and flexible architecture combined with an efficient object-oriented programming (OOP)-based implementation has enabled the Vienna system-level simulator to support large-scale simulations.

\item {\textbf{WiSE}} \cite{8352614}:
Wireless Simulator Evolution (WiSE) is a powerful system-level simulator originally developed for 4G environments. Recently, the scope has been extended to support 5G by adding vital aspects of 5G NR, such as new channel models, and 5G NR RAT features including scalable numerologies, flexible duplex, and code block group-based transmission.

\item {\textbf{X.Wang et al.}} \cite{wang}:
The authors proposed a new design for a system-level simulator specifically for heterogeneous networks. This simulator decouples the radio resource heads (RRHs) from eNodeB by assigning a cell ID to each RRH. This enabled more comprehensive testing and analysis of algorithms such as load balancing, intra-site coordinated beamforming, and scheduling.

\end{itemize}

In Table \ref{tab:traditional}, we evaluate the different system-level simulators based on general information such as user interface type, language, and traditional metrics such as documentation, user support, and modularity. The evaluation shows that the majority of system-level simulators employ a graphical user interface. Meanwhile, the most frequently used programming language or platform for developing these simulators is C++, followed by MATLAB. SyntheticNET and 5Gpy are Python-based simulators, while TCS and V. Diaz et al. have leveraged Java and PHP, respectively. While the majority of simulators have limited documentation and user support, 5G-air-simulator, 5G K-SimSys, and Vienna 5G all have detailed documentation and user guides. Additionally, the majority of simulators are modular in design. Meanwhile, the licensing type for a large number of simulators is not indicated in the referenced manuscripts.

Additional crucial aspects to consider when evaluating simulators include extensibility, ease of use, and interoperability. While details about these aspects may not be explicitly mentioned in the surveyed literature, we can deduce the state of the simulators in terms of extensibility, ease of use, and interoperability using the metrics presented in Table \ref{tab:traditional}. For example, the openness of the code implementation can significantly impact the extensibility of a simulator. Simulators that are open source, such as 5GPy, Py5cheSim, and 5G K-SimSys, among others, offer users the ability to extend their functionalities. Additionally, the modularity of these simulators also contributes to their extensibility by facilitating the incorporation of additional features or functionalities without necessitating extensive modifications to the existing codebase. Moreover, detailed documentation and user guides, as seen in 5G K-SimSys and Vienna 5G, make learning to use these simulators relatively easy compared to others. Furthermore, ease of use can be attributed to the nature of the user interface, with simulators equipped with GUIs, such as Telco Cloud Simulator, 5G K-SimSys, and SiMoNe, being more user-friendly compared to those using a command-line interface. Finally, interoperability with other simulators may be influenced by the programming language used for developing the simulator, with simulators built in similar languages having greater ease in enabling interoperability.

\begin{table*}[]
\caption{Comparison Between Network-Level 5G Simulators}
\centering
\label{tab:networklevel}
\begin{tabular}{|c|c|c|c|c|c|}
\hline

\rowcolor[HTML]{EFEFEF} 
\textbf{Network-Level Simulator} & \textbf{License Type} & \textbf{\begin{tabular}[c]{@{}c@{}}Language/\\ Platform\end{tabular}} & \textbf{Core Network Model}  \\ \hline 

\rowcolor[HTML]{DAE8FC} 
\textbf{5G LENA \cite{patriciello2019e2e}} & \begin{tabular}[c]{@{}c@{}}General Public \\ License\end{tabular} & C++ & Simplified EPC model consisting of one GW and MME \\ \hline

\textbf{5G K-SimNet \cite{8651686}} & \begin{tabular}[c]{@{}c@{}}Academic Use \\ License\end{tabular} & C++ & SDN/NFV,  5GC (AMF, SMF, UPF) \\ \hline

\rowcolor[HTML]{DAE8FC} 
\textbf{\begin{tabular}[c]{@{}c@{}}M.Mezzavilla et al. \cite{8344116}\end{tabular}} & \begin{tabular}[c]{@{}c@{}}General Public \\ License\end{tabular} & C++ & Utilized 5G LENA's core network model \\ \hline

\rowcolor[HTML]{FFFFFF} 
\textbf{Simu-5G \cite{9211504}} & \begin{tabular}[c]{@{}c@{}}Not \\ specified\end{tabular} & C++ & Simplified   UPF model for 5GC of 5G networks \\ \hline

\rowcolor[HTML]{DAE8FC} 
\textbf {\begin{tabular}[c]{@{}c@{}}OpenAirInterface \\ Core Network (OAI-CN) \cite{OIA-RAN}\end{tabular} } & \begin{tabular}[c]{@{}c@{}}General Public \\ License\end{tabular} & C & 4G EPC and 5GC (AMF, SMF, UPF, UDM, AUSF) \\ \hline

\end{tabular}
\end{table*}

\subsection{Network-Level Simulators}

In the following discussion, we provide a brief overview of each of the network-level simulators included in this survey paper.

\begin{itemize}

\item {\textbf{5G-LENA (ns-3)}} \cite{patriciello2019e2e}:
5G-LENA is a 5G NR pluggable module for ns-3. It is developed to cater to the needs of the research community for an efficient, well-documented, and easy-to-use tool for 5G simulations. 5G-LENA includes the implementation of NR features including, but not limited to, mmWave, MIMO, beamforming, and 3GPP-compliant channel models. The implementation of fundamental PHY and MAC features is based on 3GPP NR Release-15.

\item {\textbf{5G K-SimNet}} \cite{8651686}: 
5G K-SimNet is developed to provide an end-to-end performance evaluation of the 5G cellular network. This simulator integrates features like 5G NR standards, 5GC implementation, multi-connection traffic management, and SDN/NFV design.

\item {\textbf{M. Mezzavilla et al. (ns-3)}} \cite{8344116}:
Built on top of ns-3, authors in \cite{8344116} presented a module for end-to-end simulation of 5G mmWave networks. This module includes various statistical channels and MIMO models. In addition, it also provides the ability to incorporate actual measurements or ray-tracing data. Similar to \cite{patriciello2019e2e}, this module, although abstracted, has a high-fidelity PHY and MAC layer implementation.

\item {\textbf{Simu5G}} \cite{9211504}:
Simu5G is an OMNET++ library built to simulate and evaluate 5G network performance. This network-level simulator provides an end-to-end perspective of the network, including all fundamental protocol layers. It models the 5G data plane and incorporates modeling of the core network based on 3GPP Release-16. Some of the most notable features include frequency division duplex (FDD) and time division duplex (TDD) modes of communication, HetNet BS, handover support, and dual connectivity between 4G and 5G (EN-DC).

\item {\textbf{OpenAirInterface Core Network (OAI-CN)}} \cite{OIA-RAN}:
OIA-CN is the network-level counterpart of the OpenAir-Interface system-level simulator and models the core side of 5G. It models several network functionalities, such as AMF, SMF, UPF, UDM, and AUSF.

\end{itemize}

Table \ref{tab:networklevel} outlines the key features of each network-level simulator. It is notable that the majority of these simulators are accessible to the general public. Additionally, these simulators are primarily built in the C++ programming language. We focus our comparison on the core network implementation of these simulators. Currently, the most advanced simulators, in terms of 5GC implementation, are 5G K-SimNet and OAI-CN. They currently model several new 5G core entities, such as AMF, SMF, and UPF. In addition, 5G K-SimNet is the only network-level simulator that incorporates SDN/NFV functionality. Simu-5G models a simplified model of UPF. Lastly, 5G LENA currently employs the EPC model for the core consisting of a single gateway (GW) and mobility management entity (MME). This same model is utilized in simulators developed by \emph{M.Mezzavilla et al.} \cite{8344116}.

\subsection{Commercial Network Planning and Optimization Tools}

This sub-section examines commercial simulators to create awareness among the research community about recent advancements in industrial simulation. We review 10 simulators that are most popular in the industry and highlight the 5G features available in each simulator, as well as the propagation model, automation features, and some state-of-the-art innovations. These tools are mostly used to plan and optimize networks prior to deployment. To remain competitive, network operators rely on these simulators to optimize the cost and efficiency of their 5G deployments. However, due to the high licensing costs associated with these commercial simulators, their use in the research community is limited.

\begin{table*}[]
\caption{Summary of Comparison Between Commercial 5G Simulators}
\centering
\label{tab:commercial}
\begin{tabular}{|c|c|c|c|c|c|c|c|c|}
\hline

\rowcolor[HTML]{EFEFEF} 
\multicolumn{1}{|c|}{\cellcolor[HTML]{EFEFEF}} & \cellcolor[HTML]{EFEFEF} &  \multicolumn{5}{c|}{\cellcolor[HTML]{EFEFEF}\textbf{Implemented 5G Features}} & \multicolumn{1}{c|}{\cellcolor[HTML]{EFEFEF}} \\ \cline{4-8}

\rowcolor[HTML]{EFEFEF} 
\multicolumn{1}{|c|}{\multirow{-2}{*}{\cellcolor[HTML]{EFEFEF}\textbf{\begin{tabular}[c]{@{}c@{}}Commercial \\ Simulators\end{tabular}}}} & \multicolumn{1}{c|}{\multirow{-2}{*}{\cellcolor[HTML]{EFEFEF}\textbf{Usage}}} & \textbf{\begin{tabular}[c]{@{}c@{}}mmWave \\ Propagation\end{tabular}} & \textbf{\begin{tabular}[c]{@{}c@{}}Massive\\ MIMO\end{tabular}} & \textbf{\begin{tabular}[c]{@{}c@{}}Beamfor-\\ ming\end{tabular}} & \textbf{\begin{tabular}[c]{@{}c@{}}Scalable\\ Numerology\end{tabular}} & \textbf{\begin{tabular}[c]{@{}c@{}}Dual \\ Connectivity\end{tabular}} & \multicolumn{1}{c|}{\multirow{-2}{*}{\cellcolor[HTML]{EFEFEF}\textbf{\begin{tabular}[c]{@{}c@{}}Principal Propagation \\ Model\end{tabular}}}} \\ \hline

\rowcolor[HTML]{DAE8FC} 
\textbf{Atoll \cite{mathworks}} & \begin{tabular}[c]{@{}c@{}}Network Planning\\ and Optimization\end{tabular} & \checkmark & \checkmark & \checkmark & \xmark & \checkmark & \begin{tabular}[c]{@{}c@{}}3D ray tracing \\ propagation model\end{tabular} \\ \hline

\rowcolor[HTML]{FFFFFF} 
\textbf{CellDesigner™ \cite{celldesigner}}  & \begin{tabular}[c]{@{}c@{}}Network Planning \\ and Optimization\end{tabular} & \checkmark & \xmark & \xmark & \xmark & \checkmark & \begin{tabular}[c]{@{}c@{}}Korowajczuk 3D \\ propagation model\end{tabular} \\ \hline

\rowcolor[HTML]{DAE8FC} 
\textbf{Planet \cite{planet}}  & \begin{tabular}[c]{@{}c@{}}Network Planning \\ and Optimization\end{tabular} & \checkmark & \checkmark & \checkmark & \checkmark & \checkmark & \begin{tabular}[c]{@{}c@{}}CRC-Predict, Planet 3D \\ model, Universal model\end{tabular} \\ \hline

\rowcolor[HTML]{FFFFFF} 
\textbf{ASSET \cite{asset}} & Network Planning & \checkmark & \checkmark & \checkmark & \xmark & \checkmark & \begin{tabular}[c]{@{}c@{}}MYRIAD, Volcano \\ models\end{tabular} \\ \hline

\rowcolor[HTML]{DAE8FC} 
\textbf{CGA Simulation \cite{cga}}  & Network Planning & \checkmark & \xmark & \xmark & \xmark & \xmark & \begin{tabular}[c]{@{}c@{}}Not \\ specified\end{tabular} \\ \hline

\rowcolor[HTML]{FFFFFF} 
\textbf{\begin{tabular}[c]{@{}c@{}}Huawei 5G \\ Planning Tool \cite{huawei}\end{tabular}}  & Network Planning & \checkmark & \checkmark & \checkmark & \xmark & \xmark & \begin{tabular}[c]{@{}c@{}}3D ray tracing \\ propagation model\end{tabular} \\ \hline

\rowcolor[HTML]{DAE8FC} 
\textbf{RadioPlanner 2.1 \cite{radioplanner}} & Network Planning & \checkmark & \xmark & \xmark & \xmark & \xmark & \begin{tabular}[c]{@{}c@{}}ITU-R P.1812-4 \\ model\end{tabular} \\ \hline

\rowcolor[HTML]{FFFFFF} 
\textbf{\begin{tabular}[c]{@{}c@{}}RanPlan  \\ Professional 5.2 \cite{ranplan}\end{tabular}} & Network Planning & \checkmark & \checkmark & \checkmark & \checkmark & \checkmark & \begin{tabular}[c]{@{}c@{}}3D ray tracing \\ propagation model\end{tabular} \\ \hline

\rowcolor[HTML]{DAE8FC} 
\textbf{S\_5GConnect \cite{s_5g}} & Network Planning & \checkmark & \xmark & \xmark & \xmark & \xmark & \begin{tabular}[c]{@{}c@{}}Volcano ray tracing \\ propagation model\end{tabular} \\ \hline

\rowcolor[HTML]{FFFFFF} 
\textbf{\begin{tabular}[c]{@{}c@{}}Samsung’s \\ CognitiV RPO \cite{samsung}\end{tabular}} & Network Planning & \checkmark & \checkmark & \xmark & \xmark & \xmark & \begin{tabular}[c]{@{}c@{}}3D ray tracing \\ propagation model\end{tabular} \\ \hline
\end{tabular}
\end{table*}

\begin{table*}[h!]
  \centering
  \caption{Proposed 5G-specific Evaluation Metrics for Digital System Models.}
  \label{tab:metric}
  \fbox{\includegraphics[width=0.9\textwidth, height=16.5cm]{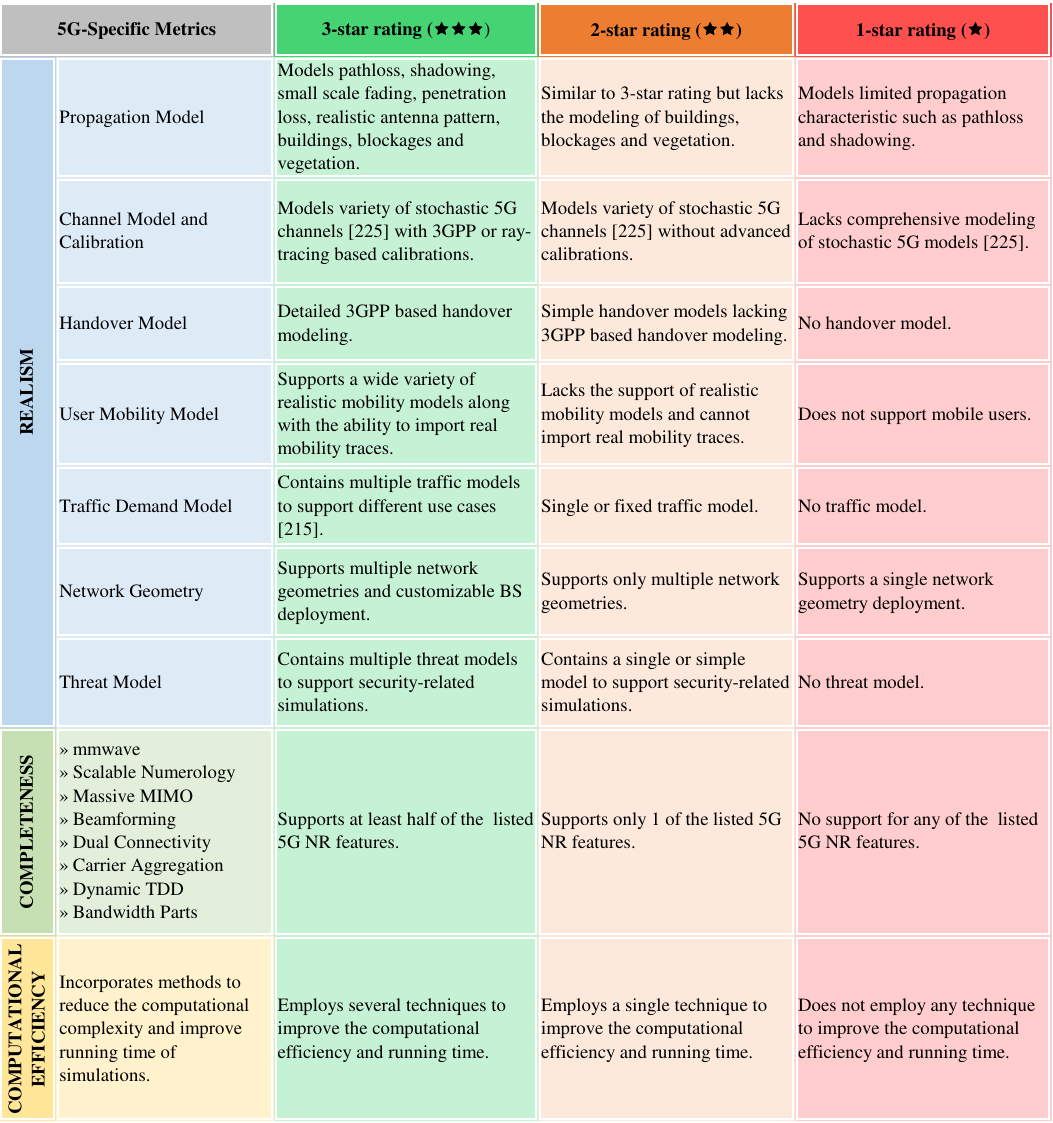}}
\end{table*}

\begin{itemize}

\item {\textbf{Atoll}} \cite{mathworks}:
Atoll’s modular architecture, advanced radio technology modeling capabilities, and support for high-frequency propagation models provide network operators with flexibility in designing and deploying their network. By adding real network data, Atoll elevates its radio signal propagation and traffic prediction capabilities. Additionally, it incorporates modeling of both indoor and outdoor environments. Atoll is one of the only commercial tools that supports open-loop SON.

\item {\textbf{CellDesigner}} \cite{celldesigner}:
CellDesigner is a planning and optimization tool with an integrated Geographic Information System (GIS) platform. It utilizes the Korowajczuk 3D (K3D) propagation model. K3D is a proprietary model that accurately predicts the signal propagation in three dimensions by considering the effect of diverse morphologies in the propagation path, such as varied clutter types, heights, and resolution layers.

\item {\textbf{Planet}} \cite{planet}:
Planet includes a wide range of realistic propagation models, including the Communications Research Centre (CRC) predict model \cite{995510}, the planet 3D model (P3M), and the universal model. All of these propagation models are calibrated and validated for 5G. It also leverages crowd-sourced data to improve the fidelity of the network design. Additionally, Planet offers increased flexibility and better utilization of computing resources by allowing decentralized radio prediction calculations using the Planet distributed radio propagation engine (D-RPE).

\item {\textbf{ASSET}} \cite{asset}:
ASSET offers improved 5G network modeling by combining real-world data, such as 3D building data, with a powerful multi-height prediction model and deterministic propagation models. Apart from the advanced propagation model, the software also supports complex antenna arrays and simulations of multi-technology 3D coverage and capacity.

\item {\textbf{CGASimulation}} \cite{cga}:
This tool leverages digital twin technology to create a digital copy of the environment and accurately model the finest details of a city. This provides a realistic simulation environment and a cost-effective method for large-scale 5G deployment. However, CGASimulation is still in its infancy and currently contains only one city model.

\item {\textbf{Huawei 5G Planning Tool}} \cite{huawei}:
This tool features a high precision 5G propagation model based on ray tracing. It enables both static and dynamic beamforming in order to realize massive MIMO. Unlike previous network planning tools, it takes a user-centric approach to planning rather than a coverage-based and capacity-based approach.

\item {\textbf{RadioPlanner2.1}} \cite{radioplanner}:
RadioPlanner2.1 is one of the few simulators that enables the investigation of air-to-ground and ground-to-air communications. While it is less advanced than other simulators in terms of propagation modeling and supporting 5G features, its low price makes it an alternative option for academic use.

\item {\textbf{RanPlan Professional 5.2}} \cite{ranplan}:
This tool enables the simulation of SA and NSA systems in accordance with 3GPP Release-15 NR specifications. Apart from cellular network planning, it also enables the implementation of diverse technologies for public safety, the Internet of Things, and smart cities. RanPlan maintains a database of the electromagnetic properties of building materials for frequencies up to 100 GHz. Furthermore, new 5G services such as virtual reality and augmented reality (VR and AR) and artificial intelligence (AI) are supported.

\item {\textbf{S\_5GConnect}} \cite{s_5g}:
This simulator deploys a propagation model called volcano ray tracing and offers improved radio prediction through enhanced vegetation discrimination. It makes use of high-quality geospatial data to realistically anticipate LOS and NLOS scenarios, including buildings and foliage barriers. Additionally, the volcano ray tracing model reduces computing time by 50\%-80\% as compared to previous models.

\item {\textbf{Samsung’s CognitiV RPO}} \cite{samsung}:
CognitiV RPO employs deep learning (DL) algorithms to create a three-dimensional semantic map that takes fine details such as trees and poles into account. The AI model extracts information such as tree shapes, building surface material, and pole or streetlamp heights from a collection of satellite and street-side photos. Ray-tracing algorithms are used to predict radio propagation using the reconstructed 3D map.

\end{itemize}

Table \ref{tab:commercial} summarizes the 5G features available in commercial simulators. It is worth noting that these commercial simulators make significant investments in incorporating mmWave propagation. With the capability for mmWave propagation, these simulators become more appealing to network operators. This is because the mmWave network’s limited coverage necessitates a higher degree of precision in site placement and parameter settings \cite{huawei2}. Additionally, it is worth noting that the majority of commercial simulators employ various variants of ray tracing-based propagation modeling by incorporating rigorous real-world measurement data, crowd-sourced information, GIS data, and high-fidelity environmental details such as clutter, buildings, and vegetation. On the other hand, the bulk of commercial simulators omit scalable numerology due to its limited utility in coverage and capacity design. Only Planet and RanPlan currently support all of the aforementioned features, including scalable numerology.

\section{Evaluation of Digital System Models Leveraging the Proposed Metrics}
\label{sec:system}

In this section, we conduct a comprehensive evaluation of various DSMs, shedding light on their individual strengths and limitations while also gauging their potential for transformation into fully-fledged DT models, thus transcending their current status as system-level simulators, as introduced earlier. We begin with the introduction of innovative and insightful evaluation metrics tailored to the intricacies of the 5G landscape. Subsequently, we proceed to an exhaustive assessment of the DSMs using these newly introduced metrics. Finally, we assess the applicability of each DSM to pertinent research domains and topics.

\subsection{New Insightful 5G-Specific Evaluation Metrics}

While the conventional metrics presented in Fig. \ref{fig:metrics} serve as a foundational means to assess 5G network DSMs at a basic level, more insightful 5G-specific metrics can provide a fuller spectrum of their capabilities and attributes, which are vital for the advancement of DTs. In this context, we introduce a novel set of metrics encompassing three fundamental attributes intrinsic to a DSM: realism, comprehensiveness, and computational efficiency. To formulate these metrics, we draw upon insights gleaned from the discussions on 5G simulator requisites articulated in Section \ref{sec:req}.

\begin{itemize}
    \item Realism:
    We define realism as the accurate representation of cellular network properties such as propagation, user mobility patterns, network elements, network layout, and handover. This aspect is essential for establishing whether DSMs are DT-ready since it directly gauges the network resemblance of the digital model. However, DSM realism is diminished when models are oversimplified and inappropriate assumptions are made during its development. Among these simplifications are the assumption of perfect hexagonal network geometry, a perfect omni-directional antenna pattern, and consideration of static users only. The need for a computationally efficient DSM or a lack of in-depth domain knowledge frequently drives this oversimplistic approach. A DSM, lacking a realistic implementation, remains confined to being a simulator and cannot attain its full potential as an effective DT.

    \item Completeness:
    The second metric, completeness, is measured by evaluating the 5G features and functions implemented within a DSM. Completeness assesses the extent to which the DSM covers all relevant aspects of the system being simulated.  We highlight eight of the most critical components of a 5G network, including (1) mmWave operation, (2) scalable subcarrier spacing, (3) massive MIMO, (4) beamforming, (5) multi-RAT dual connection, (6) carrier aggregation, (7) dynamic TDD, and (8) capacity components, as part of our completeness assessment. The completeness of a DSM quantifies the number of functions it can execute as a DT kernel.

    \item Computational Efficiency:
    Finally, even if DSMs exhibit realism and completeness, they lack the readiness for digital twin deployment if they lack computational efficiency, the third crucial metric. Generally, the computational efficiency of a DSM diminishes as the number of network elements—such as base stations (BSs), users, service types, and tested parameters—increases. Our evaluation centers on the computational efficiency of DSMs, specifically examining the strategies employed to mitigate the computational complexity inherent in simulations.

\end{itemize}

\begin{figure}[]
\centerline{\fbox{\includegraphics[width=0.45\textwidth,height=2.7in]{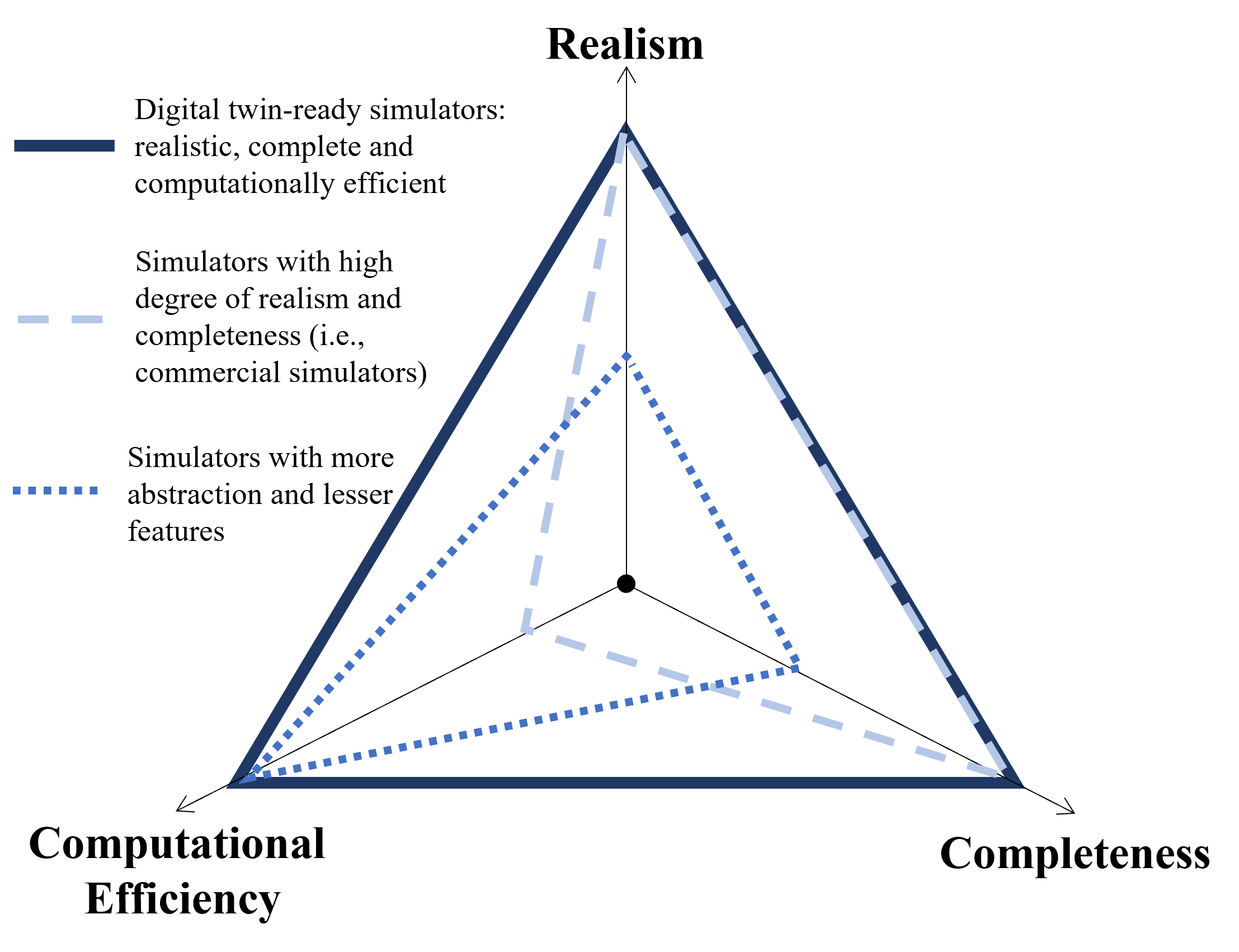}}}
\caption{Iron triangle of cellular network DSM.}
\label{fig:traingle_dt}
\vspace{-0.1in}
\end{figure}

Table \ref{tab:metric} provides a detailed description of each metric that is used to evaluate various DSMs. While certain metrics, such as the handover model, inherently possess qualitative attributes, we included them due to their significant value in developing DTs and aiding readers during the selection of a simulator. To quantify these metrics, we establish a rating system corresponding to each metric. The rating system for each DSM ranges from 1 to 3 stars, with 3 denoting the highest rating. Each DSM is given a star rating based on how well it satisfies the requirements for realism, completeness, and computing efficiency for each area. To qualify as DT-ready, a DSM must attain a formidable level of performance across all these dimensions, as depicted in the illustrative Fig. \ref{fig:traingle_dt}. This rating system offers a streamlined and intuitive means to facilitate nuanced comparisons among distinct DSMs. Moreover, it adeptly pinpoints specific domains that need enhancement, thereby guiding the developers towards refining the DSMs for optimal digital twin integration.

\subsection{Realism Evaluation}

Table \ref{tab:realism} presents a comparison of DSMs based on their degree of realism. The evaluation shows that the Vienna SL 5G Simulator and WiSE have the most practical propagation models. Both simulators incorporate the impact of blockages into their propagation models, with WiSE even considering the effect of oxygen absorption. 5G-air-simulator, 5G K-SimSys, Vienna SL 5G Simulator, and WiSE are among the DSMs with a rich selection of standard channel models. However, while the other three simulators specified performing channel model calibrations, the Vienna SL 5G Simulator has not specified any calibrations in its channel models. Meanwhile, SiMoNe is the only simulator that implements the ray-tracing propagation approach.

With regards to handover modeling, only two DSMs, SyntheticNet and SiMoNe, explicitly discuss the implementation of handover functions. While some may include handover support, it is not clearly stated in the accessible manuscripts for these simulators. Similarly, SyntheticNet and SiMoNe have the most comprehensive mobility models, including the ability to import real mobility traces. Meanwhile, 5G-air-simulator, simulator by \emph{M.Liu et al.}, and SiMoNe have the most diverse traffic models. Additionally, SiMoNe also leverages a real traffic database consisting of video streaming, web browsing, and voice traffic.

The Vienna SL 5G Simulator is by far the most extensive simulator for network geometry, while both SyntheticNet and SiMoNe support the import of user-defined topologies. Meanwhile, despite the importance of simulators for testing security aspects, none of the surveyed simulators have implemented threat models essential for evaluating potential security enhancement techniques.

\begin{table*}[]
\caption{Summary of Realism Comparison Between Different 5G System-level Simulators}
\centering
\label{tab:realism}
\begin{tabular}{|c|c|c|c|c|c|c|c|}
\hline

\rowcolor[HTML]{EFEFEF} 
\textbf{\begin{tabular}[c]{@{}c@{}}System-level\\ Simulators\end{tabular}} & \multicolumn{1}{c|}{\cellcolor[HTML]{EFEFEF}\textbf{Propagation}} & \multicolumn{1}{c|}{\cellcolor[HTML]{EFEFEF}\textbf{\begin{tabular}[c]{@{}c@{}}Channel Model\\ and Calibration\end{tabular}}} & \multicolumn{1}{c|}{\cellcolor[HTML]{EFEFEF}\textbf{\begin{tabular}[c]{@{}c@{}}Handover \\ Model\end{tabular}}} & \multicolumn{1}{c|}{\cellcolor[HTML]{EFEFEF}\textbf{\begin{tabular}[c]{@{}c@{}}User Mobility \\ Model\end{tabular}}} & \multicolumn{1}{c|}{\cellcolor[HTML]{EFEFEF}\textbf{\begin{tabular}[c]{@{}c@{}}Traffic Demand \\ Model\end{tabular}}} & \multicolumn{1}{c|}{\cellcolor[HTML]{EFEFEF}\textbf{\begin{tabular}[c]{@{}c@{}}Network \\ Geometry\end{tabular}}} & \multicolumn{1}{c|}{\cellcolor[HTML]{EFEFEF}\textbf{\begin{tabular}[c]{@{}c@{}}Threat \\ Model\end{tabular}}} \\ \hline

\rowcolor[HTML]{DAE8FC} 
\textbf{\begin{tabular}[c]{@{}c@{}}5G-air-\\ simulator \cite{MARTIRADONNA2020107314}\end{tabular}} & \begin{tabular}[c]{@{}c@{}}Path loss, \\ shadowing, \\ penetration \\ loss, \\ and fast fading\end{tabular} & \begin{tabular}[c]{@{}c@{}}UMa, SMa, RMa, \\ UMi, UFe,\\ WINNER  DL, \\ Basic DL\\ and IMT models \\(Calibrated \\ based on \\industry model)\end{tabular} & None & \begin{tabular}[c]{@{}c@{}}Standstill,  RWP, \\ linear  movement, \\ random walk, and \\ Manhattan\end{tabular} & \begin{tabular}[c]{@{}c@{}}Video streaming, \\ voice traffic, \\ web browsing, \\ and constant\\ bitrate\end{tabular} & \begin{tabular}[c]{@{}c@{}}Hexagonal \\ grid\end{tabular} & None \\ \hline

\rowcolor[HTML]{FFFFFF} 
\textbf{\begin{tabular}[c]{@{}c@{}}5G K-SimSys \\ \cite{han20185g}\end{tabular}}  & Not specified & \begin{tabular}[c]{@{}c@{}}InH, Dense\\ urban, UMa, and \\ RMa (Calibrated\\ based on \\ industry models)\end{tabular} & None & Not specified & \begin{tabular}[c]{@{}c@{}}Full buffer, and \\ non-full buffer \\ model\end{tabular} & \begin{tabular}[c]{@{}c@{}}Hexagonal \\ grid\end{tabular} & None \\ \hline

\rowcolor[HTML]{DAE8FC} 
\textbf{\begin{tabular}[c]{@{}c@{}}I.Belikaidis\\ et al. \cite{8802028} \end{tabular}} & Not specified & Not specified & None & \begin{tabular}[c]{@{}c@{}}Random walk, \\ and linear motion\end{tabular} & \begin{tabular}[c]{@{}c@{}}FTP traffic, \\ full buffer\\ and HTTP traffic \\model\end{tabular}  & Not specified & None \\ \hline

\rowcolor[HTML]{FFFFFF}
\textbf{\begin{tabular}[c]{@{}c@{}}K.Bakowski \\ et al. \cite{7454442}\end{tabular}} & \begin{tabular}[c]{@{}c@{}}Large-scale \\ fading,\\ and shadowing\end{tabular} &  \begin{tabular}[c]{@{}c@{}}O2O, O2I, \\ and I2I \end{tabular} & None & \begin{tabular}[c]{@{}c@{}}Shortest path, \\ and random movement \\ following the street\end{tabular} & Not specified & \begin{tabular}[c]{@{}c@{}}Madrid grid \\ model\\ (MGM)\end{tabular} & None \\ \hline

\rowcolor[HTML]{DAE8FC} 
\textbf{\begin{tabular}[c]{@{}c@{}}M.Liu et al. \\ \cite{liu2016design}\end{tabular}}
 & \begin{tabular}[c]{@{}c@{}}Path loss, \\ shadowing,\\ antenna gain, \\ and \\ penetration loss\end{tabular} & \begin{tabular}[c]{@{}c@{}}UMa, UMi, \\ RMa, and SMa \end{tabular} & None & \begin{tabular}[c]{@{}c@{}}Random walk, \\ standstill, and \\ fix-track\end{tabular} & \begin{tabular}[c]{@{}c@{}}FTP traffic model, \\ full buffer traffic \\ model, and \\ HTTP traffic model\end{tabular} & Not specified & None \\ \hline

\rowcolor[HTML]{FFFFFF}
\textbf{\begin{tabular}[c]{@{}c@{}}S. Cho et al. \\ \cite{7993797}\end{tabular}} & Not specified & Not specified & None & Not specified & Not specified & \begin{tabular}[c]{@{}c@{}}Hexagonal \\ grid\end{tabular} & None \\ \hline

\rowcolor[HTML]{DAE8FC} 
\textbf{SiMoNe \cite{7146084}} & \begin{tabular}[c]{@{}c@{}}Antenna pattern, \\ and \\ building data\end{tabular} & \begin{tabular}[c]{@{}c@{}}Macro-predictor, \\3D ray-\\ tracer for \\outdoor cells, \\ and analytical \\3D ray-\\ launcher\end{tabular} & \begin{tabular}[c]{@{}c@{}}3GPP X2-\\ based \\ handover\end{tabular} & \begin{tabular}[c]{@{}c@{}}RW, real  trajectories, \\ customized 3D, RWP,  \\ vehicular, pedestrian \\ and cyclist, and 3D \\ indoor\end{tabular} & \begin{tabular}[c]{@{}c@{}}Real traffic \\ database (video \\ streaming, \\ web browsing, \\ and voice traffic)\end{tabular} & \begin{tabular}[c]{@{}c@{}}User-defined \\ topology, \\ and \\ hexagonal grid\end{tabular} & None \\ \hline

\rowcolor[HTML]{FFFFFF}
\textbf{\begin{tabular}[c]{@{}c@{}}Synthetic-\\ NET \cite{9084113}\end{tabular}} & \begin{tabular}[c]{@{}c@{}}Shadowing, \\ realistic\\ antenna \\ pattern, \\ and influence of \\ blockage objects\end{tabular} & \begin{tabular}[c]{@{}c@{}}FSPL, \\2-slope pathloss,\\ and \\ other empirical \\models\end{tabular} & \begin{tabular}[c]{@{}c@{}}3GPP X2-\\ based \\ handover\end{tabular} & \begin{tabular}[c]{@{}c@{}}RWP, manhattan, \\ SLAW,  traces from\\ mobility generator \\ (SUMO, BonnMotion), \\ and real traces\end{tabular} & Not specified & \begin{tabular}[c]{@{}c@{}}User-defined\\ topology\end{tabular} & None \\ \hline

\rowcolor[HTML]{DAE8FC} 
\textbf{\begin{tabular}[c]{@{}c@{}}Vienna SL \\ 5G \\ Simulator \cite{Vienna5GSLS}\end{tabular}} & \begin{tabular}[c]{@{}c@{}}Path loss, shadow \\ fading, antenna \\ pattern, small \\ scale fading, and \\ influence of \\ blockage objects\end{tabular} & \begin{tabular}[c]{@{}c@{}}Fixed, FSPL, \\ InH, RMa, \\ SMa, UMa,\\ 3D-UMa, UMi, \\ and 3D-Umi\end{tabular} & None & \begin{tabular}[c]{@{}c@{}}RWP, and pre-defined \\ trajectories\end{tabular} & Not specified & \begin{tabular}[c]{@{}c@{}}Gauss cluster, \\ hexagonal grid, \\ hexagonal ring, \\ Manhattan grid, \\ predefined, \\ uniform cluster, \\ and uniform \\Poisson\\ Point Process\end{tabular} & None \\ \hline

\rowcolor[HTML]{FFFFFF}
\textbf{Wise \cite{8352614}} & \begin{tabular}[c]{@{}c@{}}Influence of \\ blockage objects, \\ and oxygen \\ absorption effects\end{tabular} & \begin{tabular}[c]{@{}c@{}}InH, \\Dense Urban,\\ RMa,  UMa, \\and 3D-UMa \\ (Calibrated\\ based on \\ 3GPP calibration\\ campaign)\end{tabular} & None & Not specified & Full buffer & \begin{tabular}[c]{@{}c@{}}Hexagonal \\ grid\end{tabular} & None \\ \hline

\rowcolor[HTML]{DAE8FC} 
\textbf{\begin{tabular}[c]{@{}c@{}}X. Wang \\ et al. \cite{wang}\end{tabular}} & \begin{tabular}[c]{@{}c@{}}Antenna pattern, \\ and shadowing\end{tabular} & \begin{tabular}[c]{@{}c@{}}ITU-R \\ Pedestrian-B \end{tabular} & None & Not specified & Not specified & \begin{tabular}[c]{@{}c@{}}Hexagonal \\ grid\end{tabular} & None \\ \hline
\end{tabular}
\end{table*}

\subsection{Completeness Evaluation}

The comparison of the completeness of several DSMs is shown in Table \ref{tab:completeness}. It is worth noting that none of the currently available simulators incorporates all the listed key features of 5G NR. While the majority of simulators offer scalable numerology, massive MIMO, and beamforming, the implementation of mmWave propagation remains scarce despite being one of the cornerstones of 5G\&B networks. In fact, only three DSMs, 5G K-simsys, Vienna SL 5G Simulator and WiSE, support mmWave. Furthermore, support for multi-RAT dual connectivity, carrier aggregation, and dynamic TDD is not commonly implemented. Lastly, only the Vienna SL 5G Simulator supports the simulation of bandwidth parts. This completeness analysis demonstrates that current DSMs still fall short of fully implementing 5G NR features.

\begin{table*}[]
\caption{Summary of 5G Features Supported by Different 5G System-level Simulators}
\label{tab:completeness}
\centering
\begin{tabular}{|c|c|c|c|c|c|c|c|c|c|}
\hline
\rowcolor[HTML]{EFEFEF} 
\textbf{System-level Simulators} & \textbf{mmWave} & \textbf{\begin{tabular}[c]{@{}c@{}}Scalable\\ Numerology\end{tabular}} & \textbf{\begin{tabular}[c]{@{}c@{}}Massive\\ MIMO\end{tabular}} & \textbf{Beamforming} & \textbf{\begin{tabular}[c]{@{}c@{}} Dual \\ Connectivity\end{tabular}} & \textbf{\begin{tabular}[c]{@{}c@{}}Carrier\\ Aggregation\end{tabular}} & \textbf{\begin{tabular}[c]{@{}c@{}}Dynamic\\ TDD\end{tabular}} & \textbf{\begin{tabular}[c]{@{}c@{}}Bandwidth\\ Parts\end{tabular}} \\ \hline

\rowcolor[HTML]{DAE8FC} 
\textbf{5G-air-simulator \cite{MARTIRADONNA2020107314}} & \xmark & \cmark & \cmark & \cmark & \xmark & \cmark & \cmark & \xmark \\ \hline

\rowcolor[HTML]{FFFFFF} 
\textbf{5G K-simsys \cite{han20185g}} & \cmark & \cmark & \cmark & \cmark & \cmark & \xmark & \xmark & \xmark \\ \hline

\rowcolor[HTML]{DAE8FC} 
\textbf{I.Belikaidis et al.\cite{8802028} } & \xmark & \xmark & \cmark & \cmark & \xmark & \cmark & \xmark & \xmark \\ \hline

\rowcolor[HTML]{FFFFFF} 
\textbf{K.Bakowski et al.\cite{7454442}} & \xmark & \xmark & \xmark & \xmark & \xmark & \xmark & \xmark & \xmark \\ \hline

\rowcolor[HTML]{DAE8FC} 
\textbf{M. Liu et al. \cite{liu2016design}} & \xmark & \xmark & \xmark & \xmark & \cmark & \xmark & \xmark & \xmark \\ \hline

\rowcolor[HTML]{FFFFFF} 
\textbf{S. Cho et al. \cite{7993797}} & \xmark & \xmark & \xmark & \xmark & \xmark & \xmark & \xmark & \xmark \\ \hline

\rowcolor[HTML]{DAE8FC} 
\textbf{SiMoNe \cite{7146084}} & \xmark & \xmark & \xmark & \xmark & \cmark & \xmark & \xmark & \xmark \\ \hline

\rowcolor[HTML]{FFFFFF} 
\textbf{SyntheticNET \cite{9084113}} & \xmark & \cmark & \xmark & \xmark & \cmark & \cmark & \cmark & \xmark \\ \hline

\rowcolor[HTML]{DAE8FC} 
\textbf{Vienna SL 5G Simulator \cite{Vienna5GSLS}} & \cmark & \cmark & \cmark & \cmark & \xmark & \xmark & \xmark & \xmark \\ \hline

\rowcolor[HTML]{FFFFFF} 
\textbf{Wise \cite{8352614}} & \cmark & \cmark & \cmark & \cmark & \xmark & \xmark & \cmark & \xmark \\ \hline

\rowcolor[HTML]{DAE8FC} 
\textbf{X. Wang et al. \cite{wang}} & \xmark & \xmark & \xmark & \xmark & \xmark & \xmark & \xmark & \xmark \\ \hline
\end{tabular}
\end{table*}


\subsection{Computational Efficiency Evaluation}
Several strategies for speeding up the simulation process for 5G networks are now being employed in several 5G DSMs, as shown in Table \ref{tab:efficiency}. For instance, \emph{S. Cho et al.}, SyntheticNET, and Vienna 5G SL Simulator all exploit parallel processing to accelerate simulation run-time. Meanwhile, SyntheticNET, and the Vienna 5G SL Simulator include pre-generated reference signal received power (RSRP) and channel traces, respectively, while SiMoNe implements an efficient uplink abstraction based on \cite{viering2010efficient}. On the other hand, 5G-air-simulator and the Vienna 5G SL Simulator optimize run time through the use of object-oriented programming (OOP)-based architecture. To increase time efficiency in MIMO-based simulations, WiSE combines pre-generation of MIMO precoding matrices and smart beam-sweeping link selection.

\begin{table}[]
\caption{Summary of the Methods Applied to Improve the Computational Efficiency of Different 5G System-level Simulators}
\centering
\label{tab:efficiency}
\begin{tabular}{|c|l|l|}
\hline
\rowcolor[HTML]{EFEFEF} 
\textbf{\begin{tabular}[c]{@{}c@{}}System-level   \\ Simulators\end{tabular}}  & \multicolumn{1}{c|}{\cellcolor[HTML]{EFEFEF} \textbf{\begin{tabular}[c]{@{}c@{}}Methods Applied   to Improve \\ Computational Efficiency\end{tabular}}} \\ \hline

\rowcolor[HTML]{DAE8FC} 
\textbf{5G-air-simulator \cite{MARTIRADONNA2020107314}} &  \begin{tabular}[c]{@{}l@{}}Object-oriented programming (OOP)-\\ based implementation\end{tabular} \\ \hline

\rowcolor[HTML]{FFFFFF} 
\textbf{5G K-simsys \cite{han20185g}} & None \\ \hline

\rowcolor[HTML]{DAE8FC} 
\textbf{I. Belikaidis et al. \cite{8802028} } & None \\ \hline

\rowcolor[HTML]{FFFFFF} 
\textbf{K.Bakowski et al. \cite{7454442}} &  None \\ \hline

\rowcolor[HTML]{DAE8FC} 
\textbf{M.Liu et al. \cite{liu2016design}} & None \\ \hline

\rowcolor[HTML]{FFFFFF} 
\textbf{S. Cho et al. \cite{7993797}}  & Parallel   processing \\ \hline

\rowcolor[HTML]{DAE8FC} 
\textbf{SiMoNe \cite{7146084}}  & Uplink abstraction based on \cite{viering2010efficient} \\ \hline

\rowcolor[HTML]{FFFFFF}
\textbf{SyntheticNET \cite{9084113}} & \begin{tabular}[c]{@{}l@{}}1. Parallel Processing \\ 2. Channel traces pre-generation \\ 3. User traces pre-generation\end{tabular} \\ \hline

\rowcolor[HTML]{DAE8FC} 
\textbf{\begin{tabular}[c]{@{}c@{}}Vienna SL 5G \\ Simulator \cite{Vienna5GSLS}\end{tabular}}  & \begin{tabular}[c]{@{}l@{}}1. Abstractions of PHY and MAC layer\\ 2. Network geometry and channel traces \\ pre-generation\\ 3. User traces pre-generation\\ 4. Object-oriented programming (OOP)-\\ based implementation\\ 5. Optimizing the aggregate interference\\ 6. Lite simulation\\ 7. ROI gets split into a set of pixels of a \\ fixed size\\ 8. Parallel processing\end{tabular} \\ \hline

\rowcolor[HTML]{FFFFFF} 
\textbf{WiSE \cite{8352614}} & \begin{tabular}[c]{@{}l@{}}1. Pre-generation of MIMO pre-coding \\ matrices\\ 2. Smart beam forming\end{tabular} \\ \hline

\rowcolor[HTML]{DAE8FC} 
\textbf{X. Wang et al. \cite{wang}}  & None \\ \hline
\end{tabular}
\end{table}

\subsection{Evaluation Summary and Simulator Applicability}

\begin{table*}[]
  \caption{Evaluation of different 5G\&B simulators based on insightful 5G-specific metrics defined in Table \ref{tab:metric}}
  \label{tab:star}
  \fbox{\includegraphics[width=0.98\textwidth, height=10.5cm]{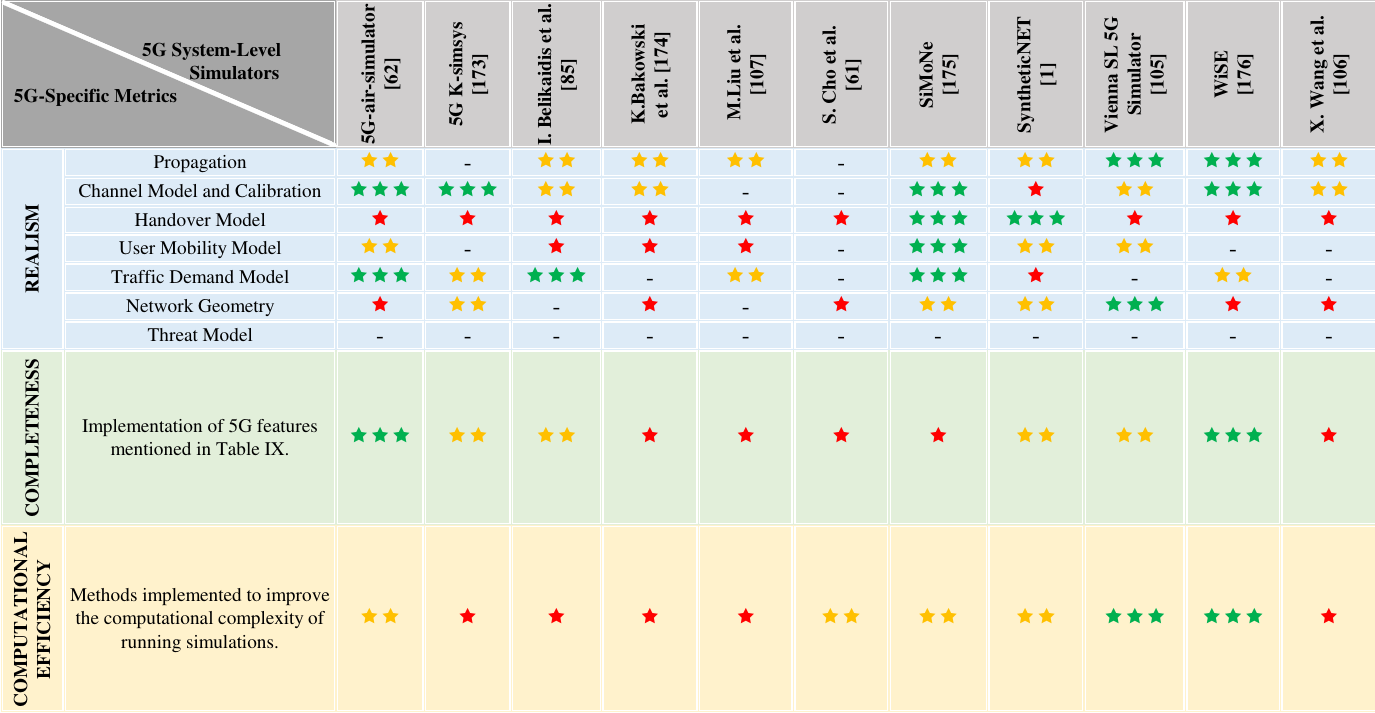}}
\end{table*}
\hspace{10pt}

Table \ref{tab:star} offers a concise overview of the progress made by the DSMs in their journey toward DT readiness. The evaluation is based on key factors including realism, completeness, and computational efficiency. This table provides a condensed representation of how each DSM aligns with these essential criteria. It is important to note that this assessment is exclusively grounded in open-source user manuals and existing literature. Any developments or enhancements not documented within these sources at the time of writing are not accounted for. For fairness, any 5G components that lack specific details in the open-source literature are indicated by the symbol (-) within the table. It's crucial to stress that the intention behind assigning rating scores is not to establish a hierarchical ranking among the DSMs. Each DSM possesses its own unique strengths and capabilities. Instead, the goal is to ascertain whether a particular DSM has the potential to evolve into a full-fledged DT or whether it is better suited as a system-level simulator. An additional advantage of this evaluation process is its utility for researchers, as it may aid them in pinpointing the most pertinent simulators for their individual research endeavors.

\begin{table}[]
  \caption{Applicability of Different 5G System-Level Simulators to Different Research Topics.}
  \label{tab:applicability}
  \centerline{\includegraphics[width=.46\textwidth,height=6.0 in]{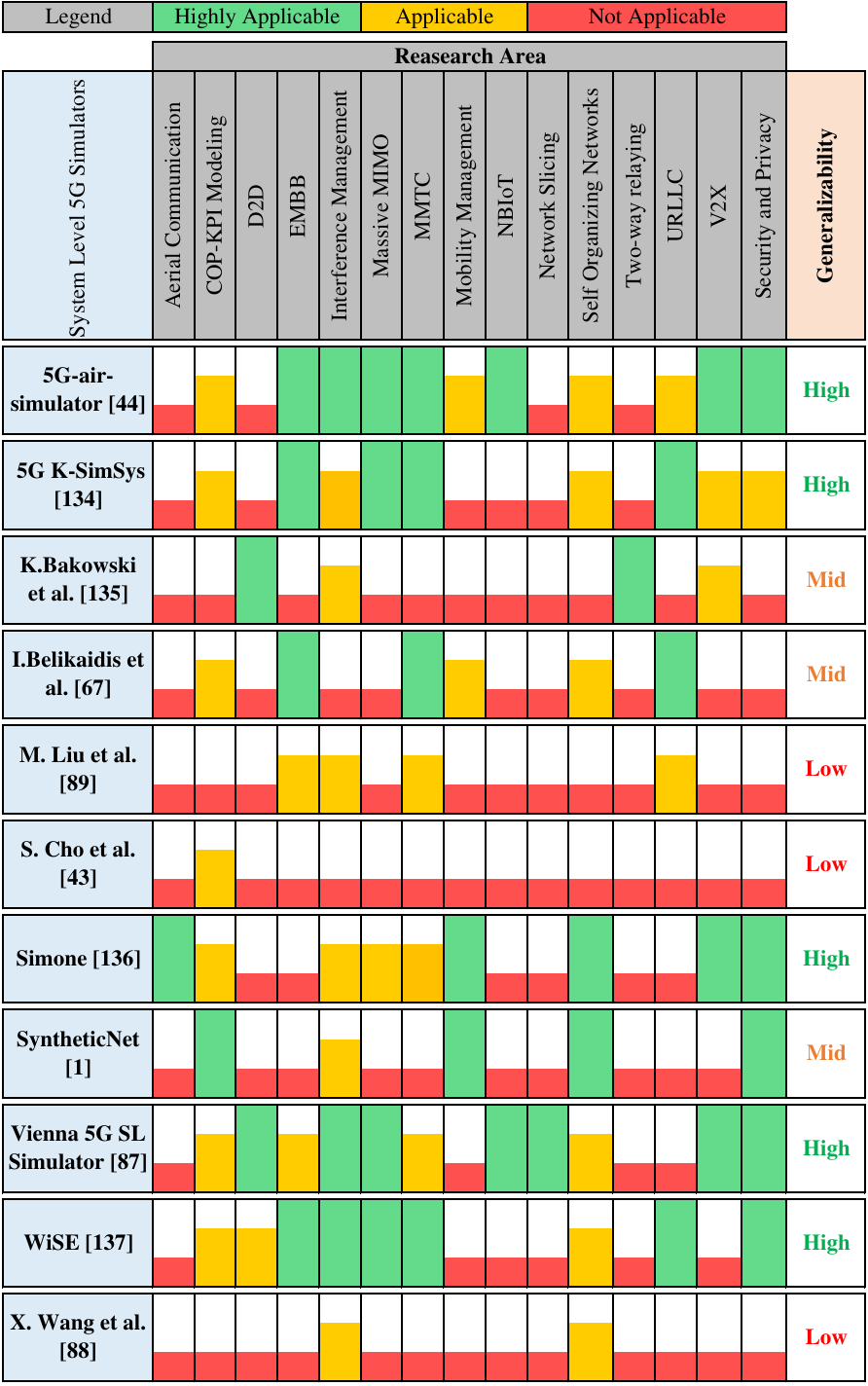}}
\end{table}
\hspace{10pt}

The evaluation shown in Table \ref{tab:star} highlights that SiMoNe has the highest degree of realism owing to the realistic modeling of the channel, handover, user mobility, and traffic demand. Likewise, 5G-air-simulator, Vienna, and WiSE are not far behind in terms of realism. These DSMs provide accurate representations of propagation and channel behavior. SyntheticNET has an in-depth implementation of handover while the simulator developed by \emph{I. Belikaidis et al.} exhibits a high traffic demand model rating. Meanwhile, 5G-air-simulator and Vienna demonstrate a high level of completeness. Both DSMs implemented five of the eight advanced 5G features listed in Table \ref{tab:metric} indicating that these DSMs are capable of closely mimicking several aspects of a real 5G network. 5G K-SimSys, \emph{I. Belikaidis et al.}, SyntheticNET, and Vienna receive two-star ratings for incorporating at least one advanced feature of 5G, while the remaining DSMs receive a single-star rating. Finally, Vienna and WiSE emerge as the most computationally efficient DSMs, utilizing multiple techniques to minimize the computational complexity associated with running simulations.

The evaluation highlights that among the presented DSMs, the Vienna, SiMoNe, WiSE, and 5G-air-simulator are poised closest to transcending the challenges of the "iron triangle" and achieving DT readiness. This implies that they require relatively minimal further development efforts before they can fully attain DT readiness. Conversely, a different scenario unfolds for 5G K-SimSys and SyntheticNET. These two DSMs demand more extensive refinements to progress toward DT readiness effectively. Finally, The remaining DSMs (i.e.,  \emph{I. Belikaidis et al., K.Bakowski et al.,M.Liu et al., and S. Cho et al.}) are better suited to remain in their current capacities as simulators. Their characteristics and capabilities may align more naturally with simulator functions rather than the complexities associated with achieving DT readiness.

To summarize the current state of these simulators, as detailed in Table \ref{tab:star}, the most significant gaps in realism within existing simulators relate to handover models, user mobility models, network geometry, and security. Table \ref{tab:completeness} further reveals that nearly all simulators lack support for bandwidth parts aspects. Additionally, only 3 out of 11 system-level simulators incorporate key features such as mmWave, dynamic TDD, and carrier aggregation, indicating that we are still far from achieving a simulator with all essential features. There is also a notable lack of methods to address computational complexity in many simulators.Sections \ref{subsec:DSM_to_DT}, \ref{sec:challenges}, \ref{sec:6G}, and \ref{sec:conclusion} outline future directions to address these shortcomings.

Table \ref{tab:applicability} summarizes the applicability of each simulator in major current and emerging research areas of cellular networks. The selection of these research topics is based on their relevance and usefulness in both the present 5G networks and the upcoming 6G networks. This applicability is based on the insights gained from the evaluation using the proposed metrics. This table is intended to provide assistance to the research community in selecting the appropriate simulator based on their research requirements. We categorize the applicability of each simulator into three levels: highly applicable, applicable, and not applicable, based on the implemented features and functionalities. For instance, simulators with a rich selection of traffic demand models, i.e., 5G-air-simulator and SiMoNe can be used to conduct experiments on different use cases such as eMBB, MMTC, and URLLC that require a variety of traffic scenarios. Similarly, the 3GPP-based handover implementation offered by SiMoNe and SyntheticNet makes them excellent choices for research on mobility management. Finally, unique features such as side-link support in Vienna, ground-to-air link support in SiMoNe, and the novel design of \emph{K.Bakowski et al.} make them highly applicable in D2D, aerial communication, and two-way relaying scenarios, respectively.

To effectively test security threats, the accuracy and completeness of the simulation platform are paramount, as emphasized in \cite{10.1093/cybsec/tyab005}. Realism enables researchers to simulate complex attack scenarios and evaluate security measures under conditions that closely resemble real-world networks. Simplifying protocols and features, on the other hand, may lead to an underestimation of potential security risks. Additionally, the efficiency of the simulator is crucial, as noted in \cite{9921600}. Consequently, simulators that excel in these three metrics, such as 5G-air-simulator, SiMoNe, Vienna, Wize, and SyntheticNET, prove invaluable in simulating network attacks and devising countermeasures. In a recent study, authors in \cite{10082863} leveraged SyntheticNET to evaluate a solution against the sophisticated Minimization of Drive Test (MDT) report attack. SyntheticNET's ability to generate MDT reports and seamlessly integrate with ML-based solutions facilitated the evaluation and implementation of effective countermeasures.

While some simulators meet the basic requirements for simulating security aspects, none have been specifically developed with a focus on security. In particular, most simulators lack advanced security-specific features, such as comprehensive threat models.

\subsection{DSM to Digital Twin Evolution}
\label{subsec:DSM_to_DT}

Simulators that meet the introduced criteria can already be considered high-fidelity and of good quality, thus qualifying as DT-ready. However, a DSM linked to the real network is what constitutes a DT. To become a DT, a DSM necessitates continuous and meticulous calibration against real-world measurements. This calibration is facilitated through a link between the real network and the DSM, serving as the fourth criterion for converting a DSM into an effective DT. This link serves as the conduit for transferring data between the real network and the DT, enabling various functionalities such as data-driven and AI-based model training and sharing, establishing feedback loops, facilitating the exchange of control instructions, among others. However, the details of this link are beyond the scope of this survey paper. Interested readers are encouraged to explore the following articles for more in-depth information regarding the connection between digital models and real networks: \cite{9540090,gao2023digital,villa2024colosseum,khan2022digital,9711524,9374645,kuruvatti2022empowering}, \cite{mihai2022digital,xu2023survey,sheraz2024comprehensive,han2023digital,minerva2020digital,boltsi2024digital,arrano2022modular,wen2023review,wu2024microgrids,nardini2024enabling}. It is paramount to recognize that a DSM failing to meet the criteria of completeness, realism, and computational efficiency might not achieve DT status, even when integrated with a real network. Attempting to link an unrealistic, incomplete, or computationally inefficient simulator would prove ineffective.

\begin{figure*}[t!]
	\centering
	\fbox{\includegraphics[width=.98\textwidth, height=6.5cm]{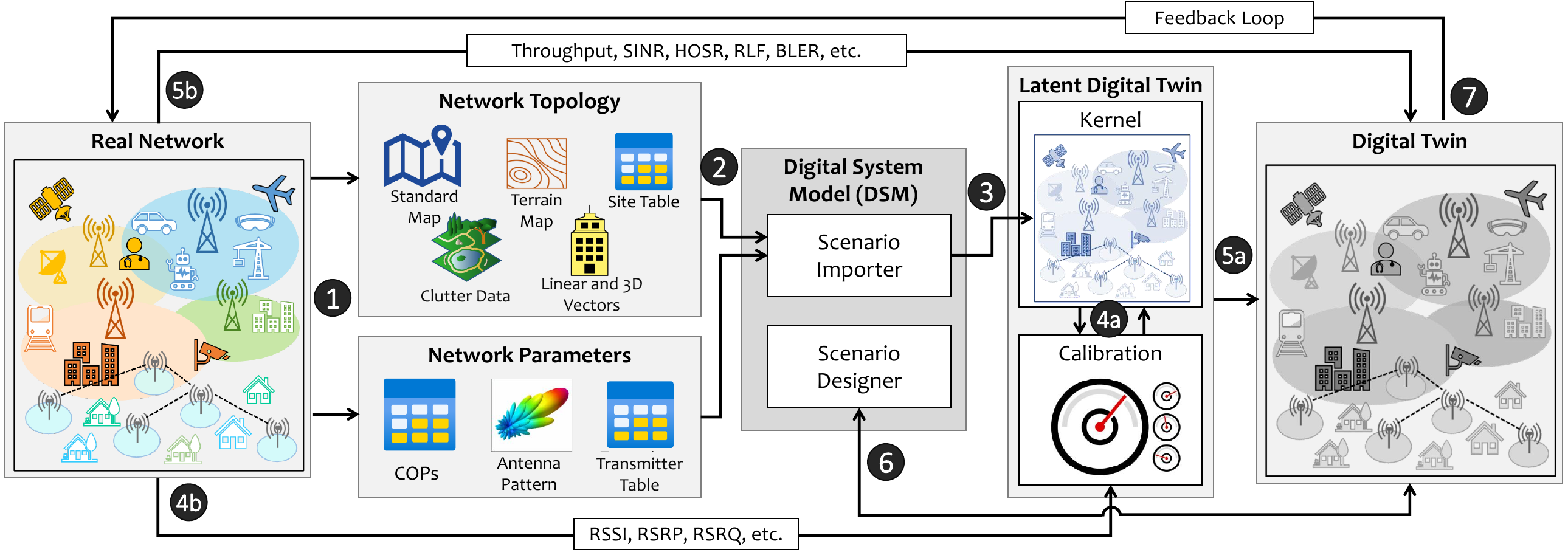}}
	\caption[]{{Framework to create digital twin using DSM}} 
	\label{fig:DT}
\end{figure*}

In Fig. \ref{fig:DT}, we present a framework outlining the step-by-step process of integrating DSM with a real network to enable DT. This illustration is a more detailed representation of Fig. \ref{fig:simulator_dt}. By leveraging network deployment insights and real data obtained from real networks (i.e., testbeds and network operators), the DSM can be tailored to create a customized DT. This DT can serve as a dynamic and accurate representation of the actual network, facilitating its monitoring, maintenance and optimization. The following steps are involved in the generation of a DT using DSM:

\begin{enumerate}
    \item \textbf{Data Acquisition:} The initial step in creating a DT through a DSM involves collecting data on the cellular network's physical attributes and configurations, including network topology and parameter settings. This data may encompass site maps, clutter types (e.g., buildings, roads, vegetation), site coordinates, terrain specifics, antenna specifications, azimuth and tilt angles, tower heights, transmit power, base station types (small, macro, or micro cells), operating frequencies, and bandwidth. These details can be sourced from testbed administrators and cellular network operators.

    \item \textbf{Data Import:} The next step is importing this data into the DSM using a scenario importer, a tool designed to facilitate the import of network topology and parameters for simulation. The importer supports various data formats (structured, unstructured) and is equipped with parsing, validation, and error-handling capabilities to ensure data integrity and consistency.

    \item \textbf{Data Aggregation and Initial Modeling:} After importing, the DSM aggregates the data to create a detailed network model, representing the static and semi-static characteristics, such as propagation and channel models, beamforming techniques, and handover protocols. This initial model, referred to as the DT kernel, serves as a static representation of the real network, based on the combined data and DSM features. At this stage, the kernel is uncalibrated.

    \item \textbf{Calibration:} The kernel is then calibrated to improve the fidelity of the DT by integrating real network data, such as Received Signal Strength Indicator (RSSI), RSRP and Reference Signal Received Quality (RSRQ) measurements. Calibration fine-tunes DSM models (e.g., channel, propagation) to closely match real-world conditions, enhancing realism.

    \item \textbf{Dynamic Integration:} Post-calibration, a baseline DT is established, accurately representing the network's current state. Dynamic attributes, like throughput, SINR, Handover Success Rate (HOSR), Radio Link Failures (RLF), and BLER, are integrated using live data feeds, allowing real-time monitoring and analysis of network performance.

    \item \textbf{Scenario Exploration:} With the DT established, users can explore various scenarios using a scenario designer. This tool allows manipulation of parameters (e.g., user numbers, antenna settings, transmit power, cellular network parameters) to simulate different scenarios and assess potential impacts, optimizing network performance.

    \item \textbf{Feedback and Integration:} Finally, stakeholders can integrate the DT into their systems by establishing a feedback loop, allowing direct parameter adjustments based on DT insights. The simulated data generated by the DT becomes valuable for tasks like network optimization and training ML/DL models, emphasizing the importance of computational efficiency in the DSM.

\end{enumerate}

Once a basic DT with functionalities presented in the above discussion is established, it should be able to handle several advanced capabilities compared to conventional simulators. These advanced capabilities are listed below:

\begin{itemize}
    \item \textbf{Autonomous Operation:}  Unlike conventional simulators, which require manual input and intervention, a DT should operate as an autonomous agent \cite{9420037}. It should independently simulates the behavior of a system, allowing it to continuously monitor, analyze, and predict the system's future state without human intervention.
    
    \item \textbf{Predictive Modeling:} DTs should be equipped with advanced algorithms and machine learning models that enable them to predict future states of the system with high accuracy \cite{10336798}. By analyzing real-time data and historical patterns, a DT should be able to forecast potential issues or performance bottlenecks before they occur, enabling proactive management and maintenance.

    \item \textbf{Performance Optimization:} While traditional simulators focus primarily on replicating past and current conditions, DTs should actively search for optimal solutions to enhance system performance \cite{9420037,dong2019deep}. They should have the capability to evaluate various scenarios and parameters in real-time to identify the best course of action, providing actionable insights that drive performance improvements.

    \item \textbf{Dynamic Feedback Loop:} DTs should be able to establish a continuous feedback loop with the physical system. They will provide real-time feedback based on simulation results, allowing for immediate adjustments to be made in the physical system's control parameters \cite{9899718}. This dynamic interaction helps in refining operational strategies, enhancing overall system efficiency and effectiveness.

    \item \textbf{Adaptive Learning:} DTs should leverage machine learning and artificial intelligence to continuously learn from the data they collect and the scenarios they simulate \cite{cronrath2019enhancing, sun2020adaptive}. This adaptive learning capability will allow them to improve their predictive accuracy and optimization strategies over time, unlike traditional simulators, which typically operate on static models.

    \item \textbf{Real-Time Decision Support:} By continuously monitoring the physical system and simulating potential future states, DTs should provide real-time decision support to operators \cite{10355584,9899718,9440709}. This capability will enable quicker and more informed decision-making, reducing downtime and improving operational resilience.

    \item \textbf{Enhanced Scalability:} DTs should have the capacity to scale more effectively than conventional simulators. They need to handle complex, interconnected systems across different domains and scales, from small-scale testbeds to large, real-world deployments, offering flexibility and adaptability in diverse operational environments \cite{khan2022digital}.
\end{itemize}

\subsection{DSMs Preferences: Simulators vs DT}
While the importance of DTs cannot be overstated, simulators still hold their unique value. Simulators are ideal for initial testing of new technologies, providing a cost-effective, controlled environment to demonstrate feasibility and basic functionality without the need for extensive infrastructure. While useful for early testing, simulators lack the ability to fully replicate the dynamic, real-world conditions and complex interactions of a live network, which are critical for thorough validation. DTs are more suitable for the following tasks:

\begin{enumerate}
    \item \textbf{Necessity of DT Testing:} After proof-of-concept, it is crucial to use DTs for advanced testing. DTs offer a high-fidelity, real-time simulation of the network, capturing detailed behaviors, user interactions, and unexpected conditions that simulators cannot. Rigorous testing within a DT is essential to ensure new technologies integrate smoothly, perform reliably under real-world conditions, and do not disrupt existing network operations.

    \item \textbf{Risk Mitigation and Continuous Improvement:} DTs allow safe experimentation, minimizing risks before deployment. The feedback loop from DT testing supports continuous refinement and optimization of new technologies.

    \item \textbf{Standardization and Readiness:} Thorough DT testing is a prerequisite for new technologies to be considered for network standards or management. This process ensures the technology is mature, stable, and scalable for full deployment.
\end{enumerate}

In summary, while simulators are useful for initial testing, DTs are necessary for comprehensive validation before new technologies are standardized or deployed in real networks.

\section{Developing Digital System Models for 5G\&B Networks: Challenges and Potential Solutions} 
\label{sec:challenges}

\begin{figure*}[]

	\centering
	\fbox{\includegraphics[width=.98\textwidth, height=9cm]{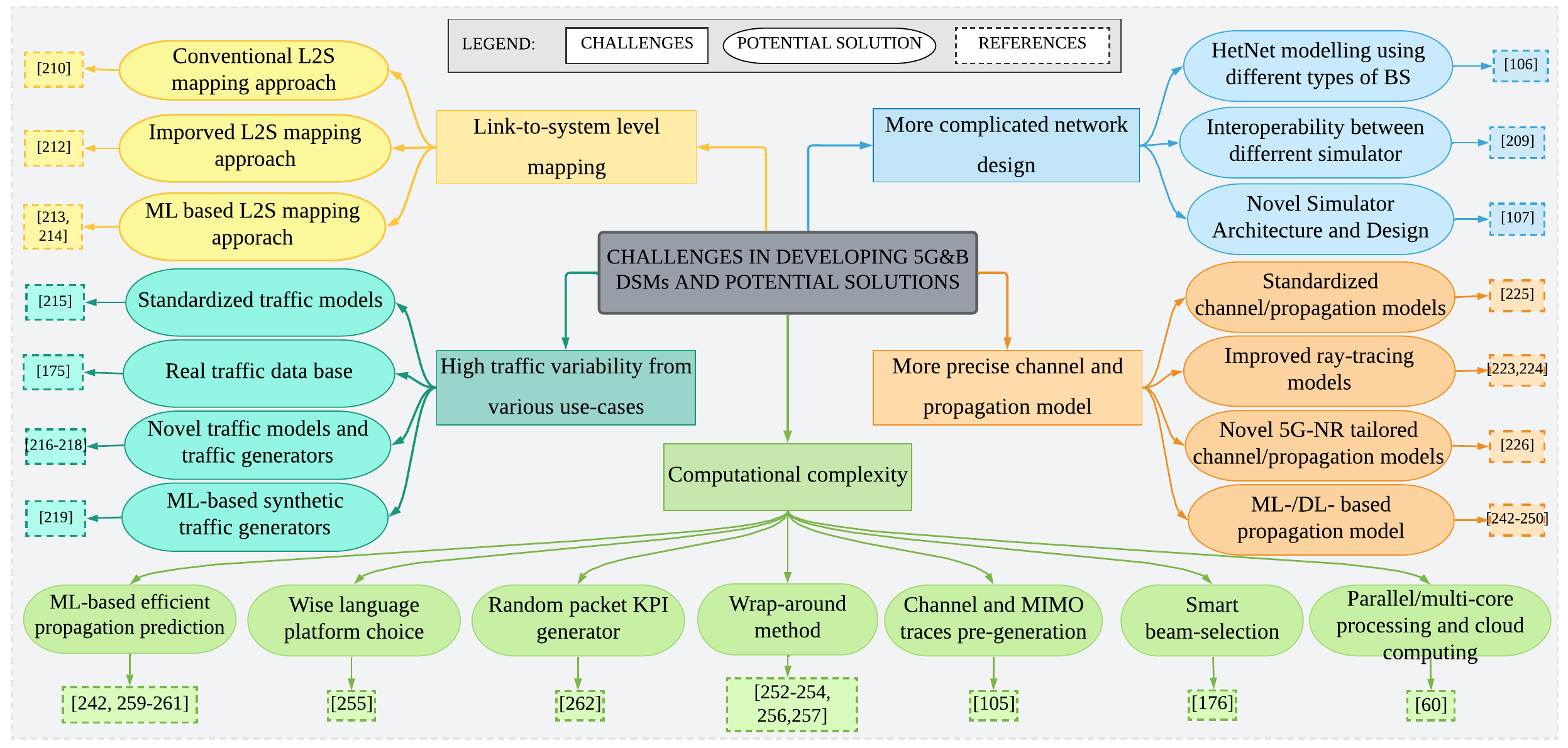}}
	\caption[]{Challenges with the corresponding potential solutions in 5G network DSM development.}
	\label{fig:challenges_sol}
\end{figure*}

This section discusses the challenges associated with the development of 5G\&B DSMs. These challenges are motivated by 5G\&B’s innovative design, diverse use cases, and enabling technologies. We then present the different approaches that can be leveraged to address these challenges, drawing on the various relevant literature summarized in Fig. \ref{fig:challenges_sol}.

\subsection{Advanced network design brought by HetNets} 

\subsubsection{Challenges}

The convergence of diverse types of BSs (e.g., macro, micro, and pico cells), along with a mix of radio access technologies like LTE-A, Wi-Fi, and 5G NR within heterogeneous 5G networks, presents notable hurdles in the evolution of DSMs. For instance, the presence of HetNets introduces intricacies in modeling processes such as cell selection, cell discovery, and handover management \cite{9206546}. Furthermore, the differences in the protocols inherent in these distinct technologies introduce an additional layer of intricacy in the development of these DSMs.

\subsubsection{Potential Solutions}
Various simulators have incorporated intelligent mechanisms to address issues posed by the variability of 5G networks. Vienna 5G SL Simulator models heterogeneity by allowing the users to deploy different types of BS (i.e., pico cell, femto cell, macro cell) with distinct propagation behavior and users (i.e., static, mobile) with different service requirements placed arbitrarily depending on the scenario. Meanwhile, SyntheticNET offers leeway to import a highly customizable file in comma-separated values (CSV) format containing individual BS characteristics such as location, type, operating frequency, transmission power, tilt, azimuth angle, and MIMO configuration, among others.

Some heterogeneous network scenarios, including support for macro- and femto-cells, are also available in 5G-air-simulator. To simulate a heterogeneous network, authors in \cite{wang} proposed a new design for a system-level simulator by decoupling the RRHs from eNodeB, making them two separate entities. However, the heterogeneity scope of the aforementioned simulators is limited to single radio access technology (RAT).

The more challenging task is the integration of multiple RATs in a single simulation environment. To address this issue, authors in \cite{liu2016design} proposed to achieve multi-RAT heterogeneity by equipping the users with two air interfaces, enabling them to concurrently communicate with two different technologies. Although the proposed design involves LTE and WiFi, this approach can easily be extended to 5G NR. To achieve multi-RAT connectivity, 5G K-SimSys employs the channel model for the sub-6GHz (frequency used for LTE) and above 6GHz bands (frequency used for 5G mmWave) in a single simulation environment.

In contrast to proposing novel architectures, another way to address the challenge of complexity brought by the inherent heterogeneity of 5G networks is to allow interoperability between different DSMs. Authors in \cite{pham2019enabling} explored this strategy by building an adapter that allows the simultaneous use of different simulators such as ns-3, Mininet-WiFi, Omnet++, and OpenAirInterface5G. The authors in \cite{pham2019enabling} successfully built a 5G network simulator capable of simulating a large-scale network with various types of networking technologies. This technique can be adopted for future DSMs by allowing inter-operation between currently available simulators to achieve the desired level of heterogeneity.


\subsection{Comprehensive link-to-system-level mapping}

\subsubsection{Challenges}
As previously mentioned, PHY layer or link-level abstraction is a common practice to reduce the time complexity of running simulations. While the authors in \cite{L2S_2} determined that L2S is insensitive to different numerologies, additional 5G NR factors necessitate the use of more sophisticated techniques for coupling link-level and system-level simulations. For instance, compared to the smaller number of transport block sizes (TBS) for LTE as defined in Table 7.1.7.1-1 and Table 7.1.7.2.1-1 from 3GPP 36.213 \cite{3GPP36.213}, 5G NR supports a larger number of TBS due to a significantly wider range of bandwidths, a broad range of transmission durations, slot-based resource allocation, and variations in the overhead size. In addition, the code block segmentation procedure has also become more complex with the introduction of low-density parity-check code (LDPC). Moreover, 5G NR has at least three MCS index tables depending on the downlink control information (DCI), cell radio network temporary identifier (C-RNTI), and supports modulation orders of up to 256 QAM. These challenges ultimately impact link-to-system (L2S) model mapping, which is essential for developing DT-ready DSMs.

\subsubsection{Potential Solutions}

The state-of-the-art approach to link abstraction is through the use of L2S mapping models, which entails two significant procedures. The first step is to calculate the effective SINR from the various post-processing SINRs received from each subcarrier, and the second step is to calculate the associated transport BLER using an SINR-BLER lookup table \cite{L2S_2}. To estimate the overall quality of the channel represented by the effective SINR mapping (ESM), some of the commonly used model functions include the mutual information ESM model (MIESM) and the exponential ESM model (EESM). For example, 5G-air-simulator utilized MIESM, 5G K-SimSys incorporates EESM, while \cite{liu2016design} supports both MIESM and EESM L2S mapping.

While the current link-to-system (L2S) mapping models used in system-level simulators are practical, they do not explicitly depict the aforementioned PHY layer updates in 5G NR. To address this issue, authors in \cite{L2S_2} presented a 5G NR-based PHY abstraction model. Based on the EESM method, the developed model supports two standard MCS table settings covering up to 256-QAM, HARQ combining mechanisms, code block segmentation, and link adaptation procedures. Meanwhile, in \cite{LS2_8}, authors proposed an enhanced EESM that accounts for the variation in channel conditions in networks supporting multi-connectivity.

Recently, machine learning (ML)-based link abstraction has gained traction for L2S mapping. The authors in \cite{LS2_9} proposed a BLER prediction model using logistic regression by exploiting the mean and standard deviation of SINR. Meanwhile, authors in \cite{L2S_1} leveraged deep neural networks (DNN) to develop a novel L2S mapping technique for 5G and IoT networks. The authors developed a model to predict the BLER using a DNN-based regression. Both studies \cite{LS2_9, L2S_1} have demonstrated that the ML/DL-based approach provides better accuracy and time complexity than analytical-based link abstraction.

\subsection{Large distributed system with high traffic variability}

\subsubsection{Challenges}

5G\&B network is intended to enable an extensive array of vertical applications and services. This means that performance evaluations are no longer restricted to a single service type. Vertical applications are frequently dispersed across a vast geographic area, resulting in diverse traffic patterns. This adds extra complexity to the task of developing DSMs capable of evaluating a large, distributed 5G system with considerable traffic variability.

\subsubsection{Potential Solutions}
Current simulators such as WiSe, 5G-air-simulator, 5G K-simsys, and SiMoNe incorporate a variety of network traffic models to capture 5G traffic heterogeneity. Some of the most common traffic models include full buffer, non-full buffer, web browsing, constant bitrate, video streaming, and voice traffic, as shown in Table \ref{tab:realism}. SiMoNe also models traffic demand using traffic intensity maps in addition to a real-world traffic database to account for heterogeneous traffic. While the current traffic modeling approaches are practical, they only address a limited subset of the traffic scenarios that could occur in a real 5G network.

Several studies in the literature have identified a diverse range of traffic scenarios for 5G networks. Authors in \cite{8985528} conducted a comprehensive survey of 5G network use case scenarios and traffic models. The survey includes an in-depth discussion of the various standardized test environments and traffic models available from IMT-2020, 3GPP, the Telecommunications Industry Association (TIA), and the 5G Infrastructure Public Private Partnership (5G PPP), which may all be incorporated into 5G system-level simulators. While the study in \cite{8985528} outlined current standardized traffic models, other studies have offered novel traffic models and traffic generators for certain service types in 5G \cite{zvei,6899230,7930654,9110314}. Authors in \cite{zvei} discussed several 5G user data traffic models tailored to industrial use cases that adhere to the 3GPP radio network standard. Meanwhile, authors in \cite{6899230} devised a traffic generator for LAN networks through a tool called “SourcesOnOff” that can generate realistic internet-like traffic. In \cite{7930654}, the authors introduced a data flow model for M2M communication. Finally, authors in \cite{9110314} described a process for generating a large volume of synthetic traffic data by leveraging live or recorded network traffic.

\subsection{Highly precise channel and propagation model}

\subsubsection{Challenges}
The 3GPP TR 38.900 (Release-14) \cite{3gpp_channel} standardized five channel models: Urban micro-cell (UMi), Urban macro-cell (UMa), rural macro-cell (RMa), suburban macro-cell (SMa), and indoor hot-spot (InH), keeping in view the diverse 5G environments. Recent developments include the introduction of 3D-UMa and 3D-UMi, which capture both the horizontal and vertical spatial properties of a channel \cite{7060514}. Additionally, the effects of large-scale path loss, small-scale fading, shadowing, and obstruction should be taken into account in order to offer a more precise and accurate propagation model. While incorporating various channel models in a DSM is already challenging enough, this is exacerbated with the introduction of massive MIMO and ultra-dense antenna deployments. The channel behavior usually changes considerably with a large array of antennas compared to a traditional small antenna array deployment. As a more detailed resolution of the environment is considered to achieve a more precise propagation model, the computational complexity also increases significantly. To achieve a realistic propagation model, detailed, precise, and time-efficient modeling of the environment is necessary. Furthermore, new applications such as V2V necessitate the development of new channel models, which should also be taken into account when developing DSMs.

\subsubsection{Potential Solutions}
Commercial 5G simulators make significant investments in enhancing the accuracy of the propagation model in order to be competitive. As illustrated in Table \ref{tab:commercial}, the vast majority of these commercial simulators employ ray tracing as their principal propagation model. Additionally, these simulations integrate real-world environmental parameters, such as buildings and vegetation. While this effort resulted in a more accurate radio propagation modeling, the applicability of the ray tracing-based approach is overshadowed by computational inefficiency and the requirement for exhaustive geographical data \cite{7152831}. Nonetheless, various studies have been conducted with the goal of optimizing the performance of ray tracing models. In \cite{8546760}, the authors enhanced the accuracy of the usual ray tracing approach by considering roadside trees as electromagnetic wave transmitting, reflecting, and diffracting objects. The authors in \cite{9090309} investigated the effect of the human body on ray tracing in a tunnel scenario and used their results to improve the performance.

In system-level simulators, empirical propagation models are a more computationally efficient technique to represent the propagation. A detailed and exhaustive survey presented by the authors in \cite{8424015} covers an expansive list of channel models for the 5G networks. The paper discussed channel models for massive MIMO, V2V, high-speed users, and mmWave scenarios. The survey also listed 5G channel models from various standardization bodies, such as IMT-2020, IEEE 802.11ay, and COST 2100, to name a few. Although the majority of existing system-level simulators already incorporate multiple 3GPP channel models, they can be further enhanced by including the 5G channel models from other standardization bodies, as highlighted by the survey.

Apart from the more traditional empirical and ray tracing-based approaches, numerous models have been developed that can be utilized to enhance future DSMs. The authors in \cite{9318511} presented a general 3D wireless channel model framework applicable to most of the 5G network scenarios (i.e., massive MIMO, V2V, high-speed users, and mmWave). The proposed model captures key channel properties of 5G\&B networks such as space-time-frequency (STF), non-stationarity, and spherical wavefront (SWF), to name a few. Several other new models are developed for specific scenarios like massive MIMO \cite{9019645, 9309503,9187740,8974233,rao2021massive}, mmWave \cite{9014521,s20143880,xu20213d,3418431,9230975}, V2V \cite{8974233,9316947}, HST \cite{9187740,xu20213d}, for UAV \cite{3418431,9146923,9085356}, and outdoor-to-indoor (O2I) \cite{9166323,Muttair_2020}.

ML/DL-based propagation models provide another promising direction for designing a precise and time-efficient library of propagation models for DSMs. A preliminary study in \cite{9014187} demonstrated the feasibility of ML-based propagation modeling by comparing it with the highly accurate ray tracing approach. Authors in \cite{thrane2020deep,8950164,1444345} presented deep learning-based propagation models leveraging geographical images, satellite imagery, position indicators, and drive test measurements, respectively. Similar to \cite{9014187}, studies in \cite{thrane2020deep}, \cite{8950164}, and \cite{1444345} showed improved accuracy compared to empirical models and lesser complexity than ray tracing-based models. Meanwhile, the effectiveness of path loss estimation using neural networks and random forest-based propagation models for NBIoT was analyzed by the authors in \cite{8741751}. The authors in \cite{8865025} investigated the optimal parameters of a neural network that affect propagation prediction. Although the study focuses on frequencies of 189.25 MHz and 479.25 MHz, the proposed approach is adaptable to higher frequencies of 5G networks as the approach used is frequency band agnostic. Meanwhile, work done in \cite{8949757, 9037126} proposed deep-learning based massive MIMO channel estimation, while a study in \cite{9136601} leveraged ML to devise map-based mmWave channel models. Current as well as future DSMs can leverage these efficient ML-based propagation models.

\subsection{Extreme computational complexity}

\subsubsection{Challenges}

Ideally, a DSM that is DT-ready should incorporate a high degree of realism and completeness while utilizing a small amount of computational resources. However, in reality, with the current approach to developing these DSMs, computational efficiency deteriorates with the increase in realism and completeness. As a compensation approach, most DSM developers tend to employ abstractions and oversimplifications of computationally intensive tasks and incorporate limited features and functionalities. Moreover, a recent article \cite{scalable} demonstrates a non-linear surge in the simulation run time as the network size increases, showing an increase from 100 seconds for 1000 nodes to more than 2000 seconds for a 3000-node scenario. With dense BS deployment and billions of connected devices, simulators with insufficient computational efficiency will fail to replicate the 5G ecosystem.

In addition to the network scale discussed previously, another critical factor contributing to computational complexity is the size of the simulation area. To ensure realism and applicability across various scenarios, simulation areas often need to encompass vast regions with numerous nodes and UEs. This challenge is compounded by mobility, as users moving within the simulation area may traverse long distances throughout the course of the simulation \cite {muller2018flexible, 7974289, 8727204}.

\subsubsection{Potential Solutions}
As shown in Table \ref{tab:efficiency}, several system-level simulators have implemented strategies to increase their computational efficiency, including parallel processing, channel pre-generation, and smart beam selection. On the other hand, several commercially available simulators also exploit schemes to reduce computational complexity. For instance, Samsung’s CognitiV RPO increases the simulation speed using a proprietary acceleration algorithm that can cut the analysis speed from half-day to just minutes compared to conventional ray tracing tools. In \cite{6994948}, the authors presented three potential solutions to improve the simulation efficiency. These include multi-machine and multi-core parallel simulation, hardware acceleration for high-dimensional matrix computing, and cloud computing. Similarly, authors in \cite{manalastastowards} improved the computational efficiency of a simulator using methods such as binning, COP-KPI pre-generation, and parallel and distributed processing, among others.

To mitigate the challenge of computational complexity brought by the large size of the network, the wrap-around method can be utilized. This technique establishes cyclic boundary conditions within the simulation environment, crucial for accurately representing signal propagation across the boundaries of the simulation area. By interconnecting the boundaries, signals reaching one side wrap around to the opposite side, ensuring continuous connectivity and enabling precise simulation of wireless signal behavior within confined spaces. Implementing the wrap-around technique eliminates the need for additional computations to handle boundary interactions, as signals reaching one side of the simulation area simply wrap around to the opposite side. Moreover, this technique significantly mitigates computational complexity in simulations by eliminating the need to create an excessively large networking area for vehicle users by allowing continuous movement within a relatively smaller and confined simulation area. Notably, the Vienna simulator is among the few simulation tools implementing the wrap-around technique \cite {muller2018flexible}. Meanwhile, authors in \cite {7974289} implemented a wrap-around mechanism for system level simulation of LTE cellular networks in ns-3. In another work, strategies for handling border effects, including the use of wrap-around techniques and dummy interferers outside the network, have been discussed in the literature \cite{8727204}. Here, the authors evaluate the effectiveness of wrap-around method in approximating an infinitely stretched network while considering network size and path loss exponent. Finally, the authors in \cite{7127464} evaluate different wrap-around techniques within the context of 3D channel models in system-level simulations. Their analysis revealed that a radio distance-based wrap-around scheme provides a more accurate representation of various system-level simulation calibration metrics, such as SINR compared to geographic distance-based wrap-around scheme.

However, despite its benefits, implementing wrap-around techniques is not always straightforward in all cases. For example, wrap-around techniques should be considered when generating shadow fading value (SFV) maps for system-level simulations, especially as the simulation area expands. However, as noted by the authors in \cite{7248659}, new methods are needed to generate wrap-around SFV maps to prevent discontinuities in shadow fading between the simulated area and the wrap-around area. To address this issue, the authors employed a novel 3-tier approach that decomposes the simulation into multiple string loops, followed by SFV generation for both the outermost and inner string loops. The results demonstrate the accuracy of the proposed scheme, showing that the maximum absolute correlation error is significantly lower compared to other schemes (e.g., \cite{1651489}).

Apart from the aforementioned approaches, further strategies for addressing the computational complexity associated with efficient propagation modeling can be applied. For instance, results from the recent study in \cite{9014187} demonstrate the ability of ML-based propagation models to outperform empirical models in terms of accuracy while also being 12x faster than the ray tracing approach. In \cite{9459462}, authors examined the trade-off between accuracy and complexity for mmWave ray tracing and proposed a solution to balance the trade-off. Their approach involved simplification of the model by eliminating some of the multi-path components. In \cite{8815888}, authors proposed a practical and accurate channel estimation for cell-free mmWave massive MIMO framework based on the fast and flexible denoising convolutional neural network (FFDNet). Meanwhile, authors in \cite{8863648} presented a model cross-application wherein a propagation model calibrated for a specific area is applied to locations with no prior measurement data. Although the efficacy of the proposed approach depends on the similarity of the areas, this solution can reduce the computational complexity brought about by calibrating multiple propagation models.

Another ML-based strategy for improving the computational efficiency of simulators is presented in \cite{manzanilla20205g}. The authors proposed a method of simulating a large-scale city-wide 5G network with massive IoT connections using a realistic random packet KPI generator. The authors devised a KPI estimator using a regression model built on the data from a detailed cell-level 5G simulator.

The choice of language platform is also critical to guaranteeing an accelerated simulator run time \cite{manalastastowards}. The majority of existing DSMs are built using either C++ or MATLAB. However, a C++-based DSM necessitates a highly specified C++ skill set. This hinders the use of C++-based DSMs to some extent, despite their ability to utilize the high-performance computing power of C++. On the other hand, MATLAB-based DSMs are relatively easier to learn, but open-source MATLAB lacks the high-performance computing power to fully exploit the available resources. Due to this constraint, MATLAB-based DSMs are relatively slow. Python, on the other hand, combines the advantages of C++ and MATLAB. Python is an open-source programming language that makes use of high-performance computing technologies. Additionally, Python syntaxes are substantially easier to comprehend and remember than those of other programming languages.

\subsection{Roles of AI in DSM Advancements}

A recurring theme among the challenges discussed is the reliance on AI-based (i.e., ML and DL models) techniques as solutions. For instance, AI is utilized in ML-based L2S mapping approach to tackle the Link-to-system level mapping challenge, while ML-based synthetic traffic generators were implemented to mitigate high traffic variability across diverse use-cases. Additionally, the integration of ML/DL-based propagation models is crucial for achieving greater precision in channel and propagation modeling, effectively addressing computational complexity challenges. These examples underscore the pivotal role of AI in both DSM development and simulation execution.

In addition to the aforementioned roles, AI holds further potential for advancing DSM development and facilitating simulation execution. For example, leveraging AI techniques can aid in generating realistic mobility models that closely resemble real data, while also addressing privacy concerns and legal considerations. Various techniques for generating mobility patterns have been explored in the literature, as discussed by the authors in \cite{9061335}. This survey reviews and summarizes recent advancements in user mobility synthesis schemes utilizing Generative Adversarial Networks (GANs), including variations such as Social GAN \cite{gupta2018social}, non-parametric trajectory generators \cite{ouyang2018non}, GAN-Based Location Density Matrix Generators \cite{yin2018gans}, and TrajGANS \cite{liu2018trajgans}. Meanwhile, in \cite{luca2020deep}, the authors have compiled various methods aimed at generating mobility patterns using a deep learning approach. Specifically, this survey focuses on how deep learning can be utilized to address challenges related to next-location prediction, crowd flow prediction, and trajectory generation.

Moreover, AI is poised to assume a crucial role in the transformation of DSMs into fully realized DTs. Specifically, it can facilitate effective and efficient data collection, fusion, and analysis, that are key components in this evolution. Authors in \cite{gao2023digital} discussed how AI can reduce the complexity of running 6G radio testing is a DT environment. For instance, AI can optimize simulation parameters, including the number of nodes and network topology, to streamline the testing process complexity and enhance result accuracy. This strategic approach mitigates the need for excessive data gathering and redundant modeling, resulting in significant conservation of network resources.

Furthermore, in addition to enhancing the realism, completeness, and computational efficiency of DSMs, AI can also be leveraged to improve user experience. This can be achieved by employing ML/DL algorithms to analyze user feedback and simulation results, thereby identifying areas for enhancement and optimizing the user experience. While literature specifically focusing on enhancing user experience in running simulations within cellular networks using AI is limited, a more general discussions on this topic can be found, such as the work presented in \cite{stige2023artificial}. Additionally, AI techniques are increasingly utilized for code generation and bug detection \cite{10.1145/3383458,10.1145/3505243,fan2023large}. These methods are applicable to ensure the accuracy and reliability of DSM code implementation, thereby minimizing errors and ensuring bug-free operation.

Another important use of AI for DTs can be in the form of Artificial Intelligence-Generated Content (AIGC) \cite{du2023user,ho2020denoising,ho2022video,li2022diffusion,reid2023diffuser,huang2022prodiff,kong2020diffwave,niu2020permutation,ketata2023diffdock}, which has the potential to significantly enhance the diversity and accuracy of network simulations. AIGC can automatically generate realistic traffic patterns, user behaviors, and environmental factors, which are crucial for creating varied and comprehensive simulation scenarios. For example, AI can create synthetic datasets that replicate real-world conditions with high fidelity, such as simulating the effects of different weather conditions, user mobility patterns, and device interactions in a network. This capability allows for more nuanced and realistic modeling of network dynamics, leading to more accurate predictions of network performance under various conditions. Additionally, AIGC can continuously update these datasets based on real-time data feeds, ensuring that simulations remain current and reflective of the latest network trends and behaviors.

AI algorithms can also identify and incorporate complex dependencies and interactions within the network that may not be immediately apparent to human designers. This includes understanding how various network parameters influence each other and predicting the potential impact of changes in one part of the network on overall performance. By doing so, AI can provide a more holistic and detailed representation of network behavior, improving the simulation’s ability to capture intricate details of network operations. AI-driven models can also simulate rare or extreme events that are difficult to observe or replicate in real-world testing. These could include sudden network outages, unexpected surges in traffic, or the impact of hardware failures. By training on large datasets, AI can generate these scenarios and provide valuable insights into how networks might behave under stress, helping operators and researchers develop more resilient network designs and strategies. In summary, leveraging AIGC within network simulations enhances the breadth and depth of the scenarios that can be modeled, leading to more robust and comprehensive testing and evaluation of network technologies and strategies.

\section{Paving the Way Towards 6G: Future Requirements and Challenges of the Next Generation Digital System Models}
\label{sec:6G}

\begin{figure*}[t]
	\centering
	\fbox{\includegraphics[width=.9\textwidth, height=13.0cm]{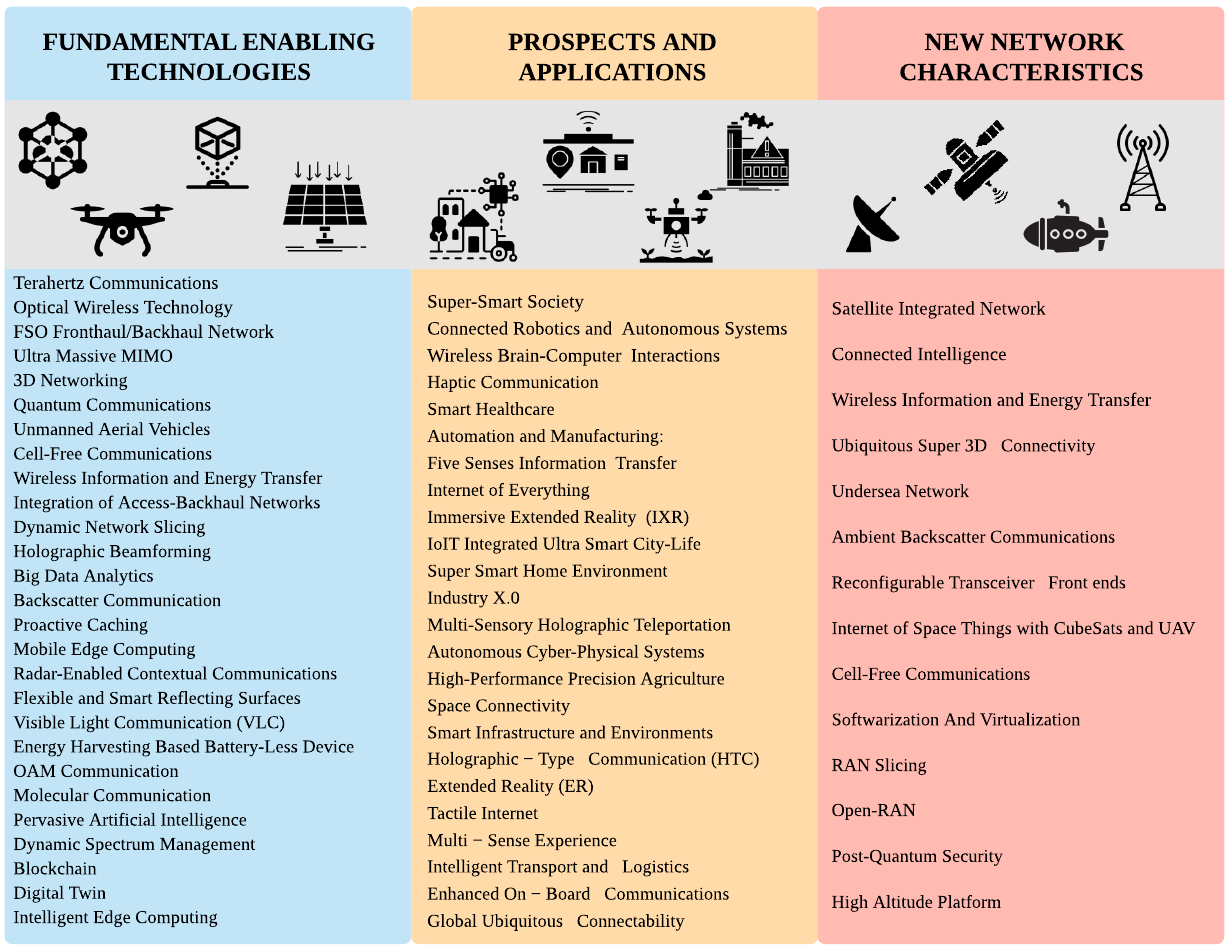}}
	\caption[]{Summary of enablers, prospect applications and novel architecture design of the incoming 6G network \cite{chowdhury20206g,8922617,tariq2020speculative,akyildiz20206g,jiang2021road}.}
	\label{fig:6G}
\end{figure*}

Although no standard for 6G has been developed to date, it is not difficult to predict what it will look like based on current trends. At this early stage, several studies are available in the literature ranging from the anticipated architecture, potential applications and use cases, and the expected enabling 6G technologies summarized in Fig. \ref{fig:6G} \cite{chowdhury20206g,8922617,tariq2020speculative,akyildiz20206g,jiang2021road}. This list provides an idea of the mounting challenges in developing future DSMs that can act as a backbone of DT.

\subsection{Challenges brought by new network characteristics}

While 5G anchors on the heterogeneity of BS types, it exclusively exploits terrestrial network deployment. However, 6G is likely to expand beyond this traditional approach towards a 3D network deployment. This form of deployment will result in the convergence of space, aerial, and underwater networks coexisting with the terrestrial network. Up until 5G, one side of the network has always remained static, i.e., the base stations, while serving mobile users. However, with 6G, the network nodes are anticipated to become mobile as well, i.e., satellites and drones, bringing a different level of complexity in modeling the network \cite{manalastastowards}. In addition, 3D network deployment will give rise to 3D mobility, which further aggravates the challenge of developing 6G DSMs. With this novel network deployment, existing analytical modeling techniques such as point processes and stochastic geometry that are applicable to 5G and other legacy technologies with static elements might not remain appreciable anymore and eventually collapse. Similarly, the modeling of interference management is exacerbated with the introduction of 3D networking.

Several simulators have been developed in response to this new architecture. Authors in \cite{9469494} presented a simulator for a non-terrestrial network, while authors in \cite{9843098} presented a satellite-terrestrial integrated network simulator.

\subsection{Challenges brought by novel prospects and applications}

6G is likely to sustain a broader range of use cases and applications compared to 5G. Some of the most notable novel applications include urban air mobility, the Internet of Everything, multi-sensory XR, and wireless brain-computer interaction \cite{akyildiz20206g,jiang2021road}. Although some have already been validated with 5G, such as self-driving cars, smart homes, smart cities, and smart health care, to name a few, full realization and utilization of these applications is anticipated to be achieved in 6G. These new use cases and applications bring further challenges in terms of widely distributed and heterogeneous traffic models and more complex mobility management.

\subsection{Challenges brought by fundamental enabling technologies}
The 6G enablers possess a massive challenge in developing future simulators. For instance, the challenge of sophisticated propagation modeling arises due to the anticipated utilization of the THz band. Compared to the GHz band used in 5G, the THz band is more demanding with respect to the Line of Sight (LOS) requirements and is highly susceptible even to the slightest obstruction (i.e., penetration even to a sheet of paper is a challenge) \cite{singh2019beyond}. To boost the minute coverage and increase the spectral efficiency when operating on very high frequency bands, one promising solution is the use of reconfigurable intelligent surfaces. This potential was investigated by authors in \cite{sihlbom2022reconfigurable} using a system-level simulator called Coffee Grinder Simulator. Meanwhile, the multiplicity of the antennas utilized to realize MIMO brings additional complexity due to an increase in the precoding matrices needed to be generated during the simulation. The concept of ultra-massive MIMO is proposed for 6G, wherein a plasmonic nanoantenna array of size 1024x1024 is envisioned \cite{chowdhury20206g}. With this huge antenna configuration, the generation of precoding matrices will be daunting for the simulators. These new enablers are anticipated to bring more computational complexity to simulators for emerging networks. Finally, the choice of language to develop a DSM also becomes more crucial due to the introduction of pervasive AI.

\subsection{Financial Challenges}
While DTs have not yet become prevalent in 5G networks, they are expected to play a significant role in 6G. Therefore, it is essential to examine the financial challenges associated with their development and operation. These costs encompass computing resources, storage, software development, and system integration and maintenance. Currently, there is no precise estimate of these expenses, though some studies, such as the one presented in \cite{OETTL2023318}, attempt to provide cost estimates for DT development. Despite the potential substantial initial investment, a well-implemented digital twin has the potential to deliver significant long-term savings in network management, optimization, and troubleshooting.

\section{Conclusion and Future Directions}
\label{sec:conclusion}

Simulators continue to play a crucial role in the development of cellular network technologies, and with the advent of DTs, their utility proliferates. In this paper, we thoroughly analyze the different simulators developed for 5G\&B networks. Our comprehensive literature review includes over 35 existing 5G\&B simulators for academia and industry, ranging from open-source link-level, system-level, network-level simulators to commercial and industry-grade simulators. We provide a brief discussion of each simulator and present a comparison to raise awareness of their features and capabilities. Our in-depth examination of the peculiarities of 5G\&B networks enables the identification of several requirements that have to be considered to facilitate an effective and efficient method of simulator development. Furthermore, a thorough examination of the current metrics used in evaluating simulators shows that they are insufficient in capturing their full utility. To address this issue, we present a novel 5G-specific metric for evaluating digital system models that focuses on realism, completeness, and computational efficiency. This new metric allows us to evaluate the suitability of these DSMs for use in developing DTs. Insights from this evaluation are also used to generate an applicability matrix that can assist the research community in selecting the simulator most suited to their particular use cases. While the development of DSMs presents several challenges, we demonstrate that there are also solutions and strategies for overcoming them. Finally, findings from examining the enablers, prospective new use cases, and unique architecture design of the future 6G network emphasize the upcoming challenges in the development of the next generation of DSMs.

As technology advances, the evaluation metrics may also need to evolve to better align with the specific nuances of emerging technologies. For example, in 6G, the metrics for realism, completeness, and computational efficiency might involve different parameters than those discussed in Section IV, given the unique enablers, prospective applications, and novel architectural designs of the upcoming 6G network, as detailed in Section VII.

This paper also reveals a gap in simulators capable of addressing security enhancements within the network. Given the critical importance of security, this is an area that warrants further exploration.

The role of simulators is poised to evolve alongside technological advancements, with DSMs emerging as crucial components in this progression, as highlighted in the manuscript. One notable example of this trend is exemplified in the recent study conducted by \cite{NARDINI2023101320}, which employed the Simu5G simulator to generate datasets used for training and testing AI models tailored for 5G/6G network applications. However, significant research endeavors are still required to fully harness their potential. One pressing area is the transformation of DSMs into comprehensive digital twins by establishing seamless connectivity between these models and the real-world networks they represent. More in-depth research into the requirements, architecture, and components needed to establish this linkage is imperative to facilitate effective integration and utilization of DSMs to realize the framework to create digital twin using DSM presented in Section V. Aspects such as data accuracy, integrity, security and privacy will need to be studied. Standardization efforts may prove essential in this endeavor to ensure uniformity and compatibility across various DSM implementations, thereby streamlining development efforts and promoting interoperability. By establishing standardized protocols and frameworks, developers can streamline development efforts and promote interoperability among different simulators. This not only enhances collaboration but also enables researchers to leverage existing tools and resources more efficiently, fostering accelerated progress and innovation within the field.

Another avenue for exploration involves developing solutions to the challenges in creating digital system models, as discussed in Section VI. While several solutions already exist, new approaches that consider the constraints of the iron triangle are needed. Additionally, there is significant potential to explore AI/ML-based approaches, as currently, only a few solutions leverage these technologies to address the identified challenges. Given the growing significance of AI in molding cellular networks, it is imperative that future research and development efforts in DSMs prioritize the exploration of AI techniques. This exploration aims to harness AI's potential in crafting simulators that are not only more realistic and comprehensive but also computationally efficient. Through the integration of AI capabilities into DSMs, researchers can unlock novel insights, enhance performance, and pave the way for the development of advanced network management and optimization strategies tailored to the evolving requirements of next-generation cellular networks.

\section*{Acknowledgment}
This work is supported by the National Science Foundation under Grant Number 1923669 and NASA under Proposal Number 20-2020EPSCoR-0014. The statements made herein are solely the responsibility of the authors. For more details about these projects, please visit: http://www.ai4networks.com

\ifCLASSOPTIONcaptionsoff
  \newpage
\fi

\bibliography{References}
\bibliographystyle{ieeetr}

\end{document}